\documentclass[fleqn,usenatbib]{mnras}

\usepackage{mathptmx}

\usepackage[T1]{fontenc}
\usepackage{ae,aecompl}

\usepackage{graphicx}	
\usepackage{amsmath}	
\usepackage{amssymb}	
\usepackage{textcomp}
\usepackage{color}

\setcitestyle{notesep={  }}

\let\oldAA\AA
\renewcommand{\AA}{\text{\normalfont\oldAA}}
\title[SHiZELS Survey: $\sim$kpc-scale spectroscopy of `typical' star-forming galaxies at $z=0.8-2.2$]{SINFONI-HiZELS: The dynamics, 
merger rates \& metallicity gradients of `typical' star-forming galaxies at $z$\,=\,0.8--2.2}

\author[J.~Molina et al.]{
J.~Molina,$^{1}$\thanks{E-Mail: \texttt{jumolina@das.uchile.cl}}
Edo~Ibar,$^{2}$
A.~M.~Swinbank,$^{3,4}$
D.~Sobral,$^{5,6}$
P.~N.~Best,$^{7}$
I.~Smail,$^{3,4}$
\newauthor{ A.~Escala$^{1}$ \& M.~Cirasuolo$^{7,8}$}\\
$^{1}$Departamento de Astronom\'ia, Universidad de Chile, Casilla 36-D, Santiago, Chile\\
$^{2}$Instituto de F\'isica y Astronom\'ia, Universidad de Valpara\'iso, Avda. Gran Breta\~na 1111, Valpara\'iso, Chile\\
$^{3}$Centre for Extragalactic Astronomy, Department of Physics, Durham University, South Road, Durham DH1 3LE, UK\\
$^{4}$Institute for Computational Cosmology, Durham University, South Road, Durham DH1 3LE, UK\\
$^{5}$Department of Physics, Lancaster University, Lancaster, LA1 4YB, UK \\
$^{6}$Leiden Observatory, Leiden University, P.O.\ Box 9513, NL-2300 RA Leiden, The Netherlands \\
$^{7}$SUPA, Institute for Astronomy, Royal Observatory, Blackford Hill, Edinburgh, EH9 3HJ, UK\\
$^{8}$European Southern Observatory, Karl-Schwarzschild-Strasse 2, D-85748 Garching bei Muenchen, Germany\\
}

\pubyear{2016}

\begin{document}
\label{firstpage}
\pagerange{\pageref{firstpage}--\pageref{lastpage}}
\maketitle

\begin{abstract}
We present adaptive optics (AO) assisted SINFONI integral field unit
(IFU) spectroscopy of eleven H$\alpha$ emitting galaxies selected
from the High-$Z$ Emission Line Survey (HiZELS). We obtain spatially
resolved dynamics on $\sim$kpc-scales of star-forming galaxies (stellar
mass M$_\star=10^{9.5-10.5}$\,M$_\odot$ and star-formation rate
SFR\,$=$\,2--30\,M$_\odot$\,yr${^{-1}}$) near the peak of the cosmic
star-formation rate history. Combining these observations with our
previous SINFONI-HiZELS campaign, we construct a sample of 
twenty homogeneously selected galaxies with IFU AO-aided observations 
-- the `SHiZELS' survey, with roughly equal number of galaxies per 
redshift slice, at $z=0.8$, $1.47$, and $2.23$. We measure the dynamics
and identify the major kinematic axis by modelling their velocity fields 
to extract rotational curves and infer their inclination-corrected rotational 
velocities. We explore the stellar mass Tully-Fisher relationship, finding that
galaxies with higher velocity dispersions tend to deviate 
from this relation. Using kinemetry analyses we find that galaxy
interactions might be the dominant mechanism controlling the star-formation
activity at $z=2.23$ but they become gradually less important down to
$z=0.8$. Metallicity gradients derived from the [N\,{\sc ii}]/H$\alpha$ 
emission line ratio show a median negative gradient for the SHiZELS 
survey of $\Delta$log(O/H)/$\Delta$R=$-$0.026$\pm$0.008\,dex\,kpc$^{-1}$.
We find that metal-rich galaxies tend to show negative gradients, whereas
metal-poor galaxies tend to exhibit positive metallicity gradients.
This result suggests that the accretion of pristine gas in the periphery of galaxies
plays an important role in replenishing the gas in `typical' star-forming galaxies.


\end{abstract}




\section{INTRODUCTION}


Determining the physical processes that control star-formation and mass
assembly at high redshift is an area of intense debate.
At $z=1-2$, galaxies were actively forming stars and rapidly
growing their stellar mass content \citep[e.g.][]{Madau1996, Sobral2009a}.
However, studies also found a strong decline in star formation rate (SFR) from that epoch to the present 
day: the cosmic star-formation rate density of the Universe has dropped by 
more than an order of magnitude \citep[e.g.][]{Karim2011, Gilbank2011, Rodighiero2011,
  Sobral2013b}. The primary causes of the subsequent decline of the
star-formation rate activity since $z=1-2$ is still under
debate. 

Two main explanations have emerged to explain how galaxies
maintained such high levels of star formation at those redshifts: (1) the rate of
mergers and tidal interactions may have been higher at that epoch,
driving quiescent disks into bursts of star-formation
\citep[e.g.][]{Bridge2007, Conselice2009}; and (2) galaxies were continuously
fed gas from the inter-galactic medium (IGM), promoting and
maintaining star-formation activity driven by internal dynamical
processes within the inter-stellar medium (ISM; e.g \citealt{Keres2005,
Bournaud2009, Dekel2009a}). 

To test the predictions from these galaxy 
evolution models, a method for distinguishing between mergers and galaxy disks 
needs to be implemented. Three main methods of estimating the merger fraction are: 
counting close pairs of galaxies, assuming that they will subsequently 
merge \citep[e.g.][]{Lin2008, Bluck2009}; using a method of identifying 
galaxies with merging morphology \citep[e.g.][]{Conselice2003, Conselice2008,
Conselice2009, Lotz2008, Stott2013a}; and employing detailed integral field unit
(IFU) spectroscopy to look for dynamical merger signatures \citep[e.g.][]{Shapiro2008, 
Forster2009, Bellocchi2012, Contini2012, Swinbank2012a}.

If secular processes drive the galaxy disks evolution, we need to measure 
the internal dynamical properties of galaxies at the peak epoch of the 
volume-averaged SFR, constrain how the structural properties of galaxy 
disks have varied over cosmic time and test if the prescriptions developed
to understand the star-formation processes at $z=0$ are still valid in the 
ISM of galaxies at high-$z$.

Taking advantage of integral field unit (IFU) adaptive optics assisted 
observations, significant effort has been invested to measure the kinematics 
of the gas within star-forming galaxies at $z\sim1-2$ in order to test 
competing models for galaxy growth \citep[see review by ][]{Glazebrook2013}. 
Previous observations have shown highly turbulent, rotationally supported 
disks with clumpy star-formation and large gas fractions
\citep[$f_{\rm gas}$=20--80\%;][]{Elmegreen2009, Forster2009,
Genzel2010, Geach2011, Wisnioski2011, Swinbank2012b, Stott2016}. Higher gas fractions 
might lead the formation of massive ($\sim10^9$\,M$_\odot$) clumps by gravitational 
fragmentation of dynamically unstable gas \citep{Escala2008}. The 
typical rotation velocities of these systems are $100-300$ km\,s$^{-1}$, so 
very similar to local galaxies \citep{Cresci2009, Gnerucci2011b,
Vergani2012,Swinbank2012a}, but the typical velocity dispersion 
values range from 50 to 100 km\,s$^{-1}$ \citep{Forster2006, Genzel2006}. 
This means a circular velocity to velocity dispersion ratio (V$/\sigma$) 
range from 1 to 10 at $z\sim2$ \citep{Starkenburg2008, Law2009, Forster2009, 
Gnerucci2011b, Genzel2011,Stott2016}. By comparison, the Milky Way and other 
similar thin disks galaxies at low-$z$ have V$/\sigma\sim10-20$ \citep{Epinat2010,
Bershady2010}. This suggests that the gas dynamics of high-$z$ galaxies are not 
just dominated by ordered rotation or random motions, but by a contribution from both.
  
If the structural properties of galaxy disks have varied over cosmic
time, we would expect to see evidence in kinematic scaling relations. For
example, one potential evidence would be an evolution of the
Tully-Fisher relationship \citep{TullyFisher1977}, which describes
the interdependence of baryonic and dark matter in galaxies by studying the 
evolution of the stellar luminosity (M$_{\rm B}$) versus circular velocity. 
It traces a simple means of the build-up of galaxy disks at different epochs.
Since the $B$-band luminosity is sensitive to recent star formation, attempts 
have also been made to measure the evolution of the stellar mass (M$_{\star}$) 
Tully-Fisher relation which reflects the relation between the past-average 
star-formation history and halo mass. In particular, hydrodynamic models suggest 
that the zero-point of the stellar mass Tully-Fisher relationship should 
evolve by $\sim-$1.1 dex at fixed circular velocity between
$z=0-2$ \citep{McCarthy2012}. At a given rotational velocity, the stellar mass in 
a high-$z$ disk galaxy should be smaller than a low-$z$ disk galaxy as star-formation builds it up.
Substantial efforts have been made in order to measure the Tully-Fisher relationship at
redshift $z=1-2$ \citep{Cresci2009, Forster2009, Gnerucci2011b,
 Miller2011, Miller2012, Swinbank2012a, Sobral2013b, Diteodoro2016,Tiley2016}. 
Recently \citet{Tiley2016} have measured a stellar-mass TFR zero-point evolution
of $-$0.41$\pm$0.08 dex for rotationally supported galaxies defined with V/$\sigma$\,$>$\,3
from the `KMOS Redshift One Spectroscopic Survey' (KROSS; \citealt{Stott2016}). However, they measure no significant offset in the
absolute rest-frame $K$-band TFR (M$_K$-TFR) over the same period. This excess of $K$-band luminosity 
at fixed stellar mass measured from the high-$z$ galaxies could be explained by considering their higher SFRs
in comparison with their local Universe counterparts at same stellar mass. The excess of light that comes from 
young stars decreases the mass-to-light ratio in high-$z$ galaxies decoupling the evolution of both, the M$_K$-TFR
and the M$_\star$-TFR.

If the Tully-Fisher relationship evolves with redshift, then 
it would be expected that the galaxy size-velocity
relation also evolves \citep{Dutton2011b}. In a $\Lambda$CDM
cosmology, the sizes of galaxy disks and their rotational velocities
should be proportional to their parent dark matter halos, and since
the halos are denser at high-$z$ for a fixed circular velocity, then
disk sizes should scale inversely with Hubble time
\citep{Mo1998}. The evolution of the size-velocity relation has been
observed \citep{Swinbank2012a}, but increasing the number statistics
should be helpful in order to overcome random errors due to
different methods and conversions of size measurements.

A third potential observational tool to constrain galaxy evolution
models is the measure of the chemical abundance within galaxies using a 
simple disk model. If the gas accretion in high-redshift galaxies is via 
accretion of pristine gas from the IGM along filaments onto the galaxy 
disk at 10--20\,kpc from the galaxy centre, then the inner disks of 
galaxies should be enriched by star-formation and supernovae whilst the 
outer-disk is continually diluted by pristine material, leaving strong 
negative abundance gradients \citep{Dekel2009a,Dekel2009b}. This gradient 
would flatten if the IGM gas is redistributed, e.g.\ via merger interactions.

To chart the evolution of star-forming galaxies with cosmic time,
we exploit the panoramic (degree-scale) High-$Z$ Emission
Line Survey (HiZELS). This survey targets H$\alpha$ emitting galaxies
in four precise ($\delta$z=0.03) redshift slices: $z=0.4$, 0.8, 1.47
and 2.23 \citep{Geach2008, Sobral2009a, Sobral2010, Sobral2011,
  Sobral2012, Sobral2013b}. This survey provides a large
luminosity-limited sample of homogeneously selected H$\alpha$ emitters
at the cosmic star-formation density peak epoch, and provides a
powerful resource for studying the properties of starburst galaxies and 
the star-forming galaxies that shows a tight dependence of SFR on stellar 
mass, the so-called `main-sequence' of star-forming galaxies \citep{Noeske2007, Pannella2009,
  Elbaz2011}. Most of the HiZELS galaxies will likely evolve into
$\sim$L$^*$ galaxies by $z=0$ \citep{Sobral2011}, but are seen at a time when they are
assembling most of their stellar mass, and thus are seen at a critical
stage in their evolutionary history.

In this paper, we present adaptive optics assisted integral field
spectroscopy with SINFONI, yielding $\sim0\farcs15$ resolution
($\sim$kpc scale), of eleven star-forming galaxies selected from the
HiZELS survey in three redshift slices, $z=$0.8, 1.47 and 2.23 \citep{Sobral2013a}.
The HiZELS survey is based on observations obtained using the Wide Field Camera 
(WFCAM) on the 3.8-m United Kingdom Infrared Telescope (UKIRT; \citealt{Geach2008,Sobral2009a}).
Combined with nine targets from a previous similar SINFONI campaign \citep{Swinbank2012a,Swinbank2012b}
our study present one of the largest samples of homogeneously selected high-redshift
star-forming galaxies with AO-aided resolved dynamics, star-formation and chemical 
properties. Throughout the paper, we assume a
$\Lambda$CDM cosmology with $\Omega_{\Lambda}$=0.73, $\Omega\rm_m$=0.27,
and H$_0$=72 km\,s$^{-1}$Mpc$^{-1}$, so at redshift $z=0.8$, 1.47 and
2.23, a spatial resolution of $0\farcs1$ corresponds to a physical
scale of 0.74, 0.84 and 0.82\,kpc respectively.

\section{SAMPLE SELECTION, OBSERVATIONS \& DATA REDUCTION}

\begin{figure}
 \centering
 \includegraphics[width=0.95\columnwidth]{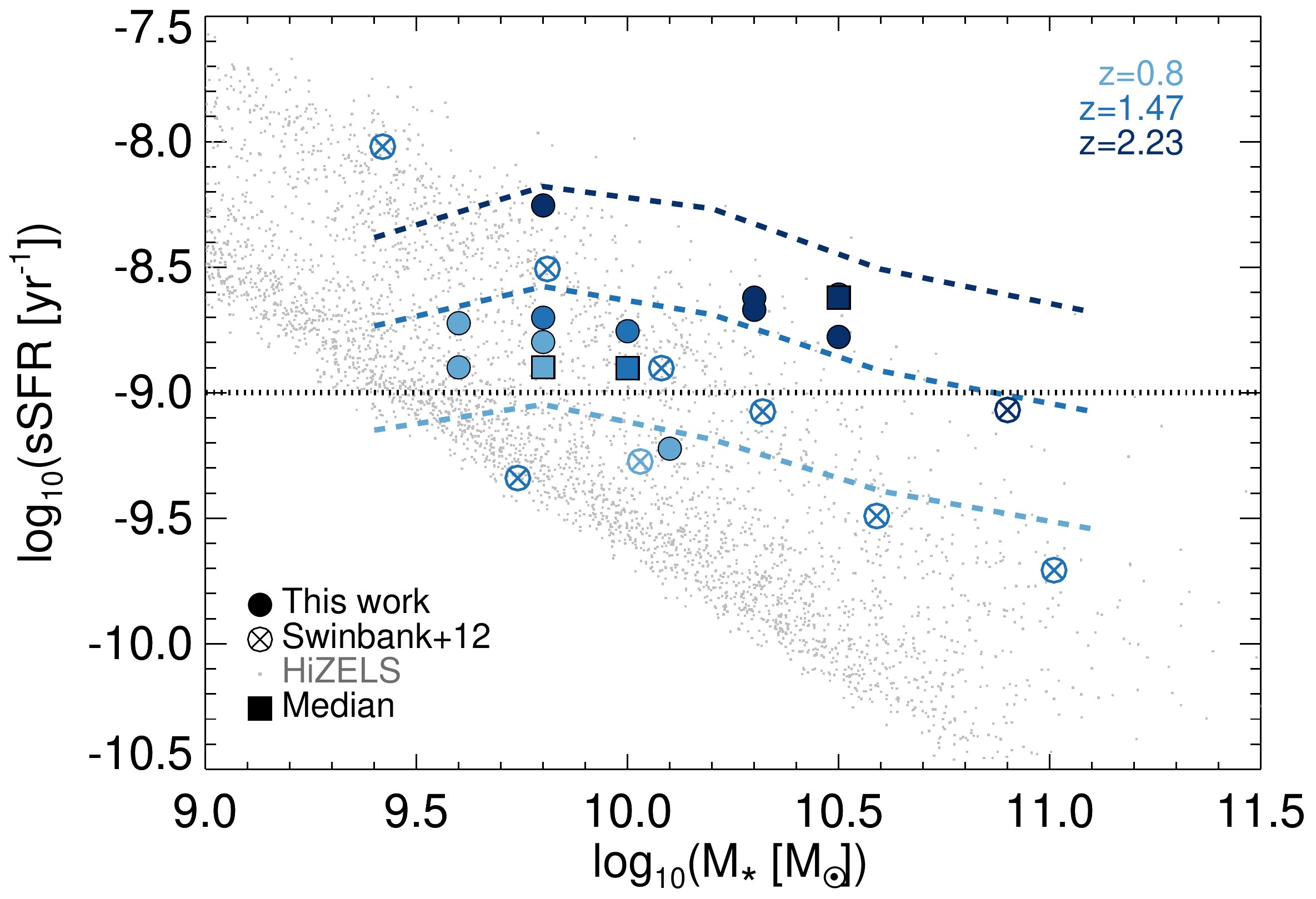}
 \caption{\label{fig:main_sequence}
	  The relation between specific star formation rate (sSFR) and stellar mass for the HiZELS survey (grey dots; 
	  \citealt{Sobral2013a,Sobral2014}), \citet{Swinbank2012a}'s sample (circles with X) and our sample (filled circles). We colour-coded 
	  our sample and the \citet{Swinbank2012a} sample by redshift. The sky blue, blue and dark blue colours represent the sources at 
	  $z=0.8$, 1.47, 2.23 respectively. The filled squares represents the median values per redshift. The black dotted line shows the 
	  sSFR$=10^{-9}$\,yr$^{-1}$ value. The colour-coded dashed lines represent the location of the `main sequence' of star-forming galaxies 
	  at each redshift slice from \citet{Karim2011} demonstrating that our sample and \citet{Swinbank2012a}'s sample are `typical' for 
	  each epoch.}
\end{figure}

\begin{table*}
	\centering
	\setlength\tabcolsep{4pt}
	TABLE 1: INTEGRATED GALAXY PROPERTIES\\
	\begin{tabular}{lccccccccccc} 
		\\
		\hline
		\hline
		ID &   RA &  Dec & Strehl & $z_{\rm H\alpha}$ & f$_{\rm H\alpha}$ & [N\,{\sc ii}]/H$\alpha$ & SFR$^{\rm obs}_{\rm H\alpha}$ & log$_{10}$(M$_*$) &  $r_{1/2}$ & E(B$-$V) &$\Delta$log(O/H)/$\Delta$R\\
		   & (J2000) & (J2000) & & & ($\times$\,10$^{-16}$ & &(M$_{\odot}$\,yr$^{-1}$) & (M$_{\odot}$) & (kpc) & & (dex\,kpc$^{-1}$)\\
		   & & & & & erg s$^{-1}$cm$^{-2}$) & & & & & & \\
		\hline
		SA22-17 & 22 19 36.1 & +00 34 07.8   & 34\% & 0.8114 & 1.7$\pm$0.1 & <0.1          & 2  & 9.6$\pm$0.1  & 4.2$\pm$0.3 & 0.5$\pm$0.2 & ... \\
		SA22-26 & 22 18 22.9 & +01 00 22.1   & 34\% & 0.8150 & 2.3$\pm$0.2 & 0.26$\pm$0.05 & 3  & 9.6$\pm$0.2  & 3.1$\pm$0.4 & 0.2$\pm$0.2 & $-$0.05$\pm$0.02 \\
		SA22-28 & 22 15 36.3 & +00 41 08.8   & 37\% & 0.8130 & 2.6$\pm$0.2 & 0.30$\pm$0.06 & 4  & 9.8$\pm$0.3  & 3.1$\pm$0.3 & 0.5$\pm$0.1 & $-$0.03$\pm$0.02 \\
		SA22-54 & 22 22 23.0 & +00 47 33.0   & 21\% & 0.8093 & 2.3$\pm$0.1 & 0.12$\pm$0.07 & 3  & 10.1$\pm$0.2 & 2.4$\pm$0.3 & 0.2$\pm$0.1 & ... \\
		\noalign{\smallskip}
		COS-16  & 10 00 49.0 & +02 44 41.1   & 32\% & 1.3598 & 1.0$\pm$0.1 & 0.10$\pm$0.04 & 5  & 9.8$\pm$0.3  & 1.5$\pm$0.4 & 0.0$\pm$0.1 & +0.08$\pm$0.02 \\
		COS-30  & 09 59 11.5 & +02 23 24.3   & 21\% & 1.4861 & 1.1$\pm$0.1 & 0.43$\pm$0.03 & 7  & 10.0$\pm$0.1 & 3.5$\pm$0.3 & 0.5$\pm$0.1 & $-$0.014$\pm$0.005 \\
		\noalign{\smallskip}
		SA22-01 & 22 19 16.0 & +00 40 36.1   & 25\% & 2.2390 & 1.0$\pm$0.1 & 0.42$\pm$0.13 & 17 & 10.3$\pm$0.4 & 2.0$\pm$0.2 & 0.1$\pm$0.1 & ... \\
		SA22-02 & 22 18 58.9 & +00 05 58.3   & 35\% & 2.2526 & 1.2$\pm$0.1 & 0.27$\pm$0.07 & 21 & 10.5$\pm$0.4 & 3.8$\pm$0.3 & 0.0$\pm$0.1 & $-$0.005$\pm$0.009 \\
		UDS-10  & 02 16 45.8 & $-$05 02 44.7 & 33\% & 2.2382 & 1.1$\pm$0.1 & 0.23$\pm$0.04 & 19 & 10.3$\pm$0.1 & 1.6$\pm$0.1 & 0.2$\pm$0.1 & ... \\
		UDS-17  & 02 16 55.3 & $-$05 23 35.5 & 12\% & 2.2395 & 1.8$\pm$0.2 & <0.1          & 31 & 10.5$\pm$0.1 & 1.5$\pm$0.3 & 0.3$\pm$0.1 & ... \\
		UDS-21  & 02 16 49.0 & $-$05 03 20.6 & 33\% & 2.2391 & 0.8$\pm$0.1 & <0.1          & 14 & 9.8$\pm$0.2  & 1.0$\pm$0.3 & 0.1$\pm$0.1 & ... \\
		\hline
		Median  & ... & ... & 33\% & ... & 1.2$\pm$0.03 & 0.27$\pm$0.02 & 12$\pm$3 & 10.1$\pm$0.2 & 2.4$\pm$0.1 & 0.2$\pm$0.1 &$-$0.014$\pm$0.009\\
		\hline
	\end{tabular}
	\caption{\label{tab:table1}
          Flux densities (f$_{\rm H\alpha}$) are taken from narrow-band photometry and include contamination
          by [N\,{\sc ii}]. SFR$^{\rm obs}_{\rm H\alpha}$ are not corrected for extinction.  $r_{1/2}$ is the H$\alpha$
          half-light radius and has been deconvolved by the PSF.}
\end{table*}

\subsection{HiZELS}
\label{sec:hizels} %

To select the targets for IFU observations, we exploited the large
sample of sources from the HiZELS imaging of the COSMOS, SA22 and UDS
fields \citep{Best2013, Sobral2013b, Sobral2015, Sobral2016a} to select 
H$\alpha$ emitters sampling the so called ``main-sequence'' at $z=0.8-2.23$ 
(Fig.~\ref{fig:main_sequence}). Taking the advantage of the large sample 
we could select galaxies which lie close ($<30''$) to bright (\textit{R}<15.0) stars, such that 
natural guide star adaptive optics correction (NGS correction) could be applied 
to achieve high spatial resolution. For this programme we selected eighteen
galaxies with stellar mass between M$_\star$= 10$^{9.5-10.5}$M$_\odot$
and H$\alpha$ fluxes greater than f$_{\rm H\alpha}\geq$
0.7\,$\times$\,10$^{-16}$erg\,s$^{-1}$\,cm$^{-2}$ to ensure that their
star-formation properties and dynamics could be mapped in a few
hours. Out of the eighteen galaxies observed with SINFONI, we
detect eleven of them with high enough signal-to-noise (S/N). Given the significant 
sky-noise in near-IR spectra, source detection was optimally performed 
by detailed visual inspection of dynamical and line width features within 
the data-cubes (using Interactive Data Language and QFitsView). Although
the rate of detection of bright H$\alpha$ emitters derived from our sample may seem modest ($\sim$60\%), it is 
comparable with the detection rate derived from the previous SINFONI campaign 
($\sim$65\%; \citealt{Swinbank2012a}). We note that both samples were drawn from the same HiZELS survey.
We note, however, that the detection rate derived from the non-AO KROSS survey (using KMOS; \citealt{Stott2016}) 
$-$which was also drawn from the HiZELS survey at $z\sim0.8-$ is nearly $\sim$92\%. This suggest that the 
modest rate of detection derived from our sample and the previous SINFONI campaign might be inherent 
to the AO-observations.

\subsection{SINFONI Observations}
\label{sec:sinfoniobs}

To measure the dynamics of our sample from the nebular H$\alpha$
emission line, we used the SINFONI IFU \citep{Eisenhauer2003} on the
ESO VLT (Project 092.A-0090(A); P.I.\,E.Ibar). We use the 
$3''\times3''$ field of view at spatial resolution of $0\farcs1$\,pixel$^{-1}$. At
$z=0.8$, 1.47 and 2.23 the H$\alpha$ emission line is redshifted to
$\sim$1.18, 1.61 and 2.12 $\mu$m and into the \textit{J}, \textit{H}
and \textit{K}-bands respectively. The spectral resolution in each
band is $\lambda/\Delta\lambda\sim3700$, and sky OH lines are
considerably narrower ($\sim4\AA$ full width half maximum -- FWHM)
compared to the galaxy line widths. We use a NGS correction since
each target is close to a bright guide star.\\

The observational setup for these 
targets was done in the same manner as in \citet{Swinbank2012a}.
To observe the targets we used ABBA chop sequences (OBs with individual 
exposures of 600 seconds), nodding $1\farcs6$ across the IFU. In order 
to achieve higher signal to noise ratios on sources at higher redshifts
we used 2, 3, 4 OB's for the $z=0.8$, 1.47, 2.23 samples implying a 
total observing time of 4.8, 7.2, 9.6 ks respectively. The 
observations were carried between 2013 October 27 and 2014
September 3 in $\sim0\farcs8$ seeing and photometric conditions. 
The median Strehl achieved for our observations is 33\%
(Table~\ref{tab:table1}).

Individual exposures were reduced using the SINFONI \texttt{ESOREX}
data reduction pipeline which extracts flat-fields, wavelength
calibrates and forms the data-cubes for each exposure. The final
data-cube was generated by aligning manually the individual OBs
on average (shifting them by $\lesssim0\farcs2\sim2\,$pix) and
then combining these using a sigma-clipping average at each pixel and
wavelength. This minimised the effect of the OH emission/absorption features
seen in the final data-cube.

\begin{figure*}
\includegraphics[width=0.5\columnwidth]{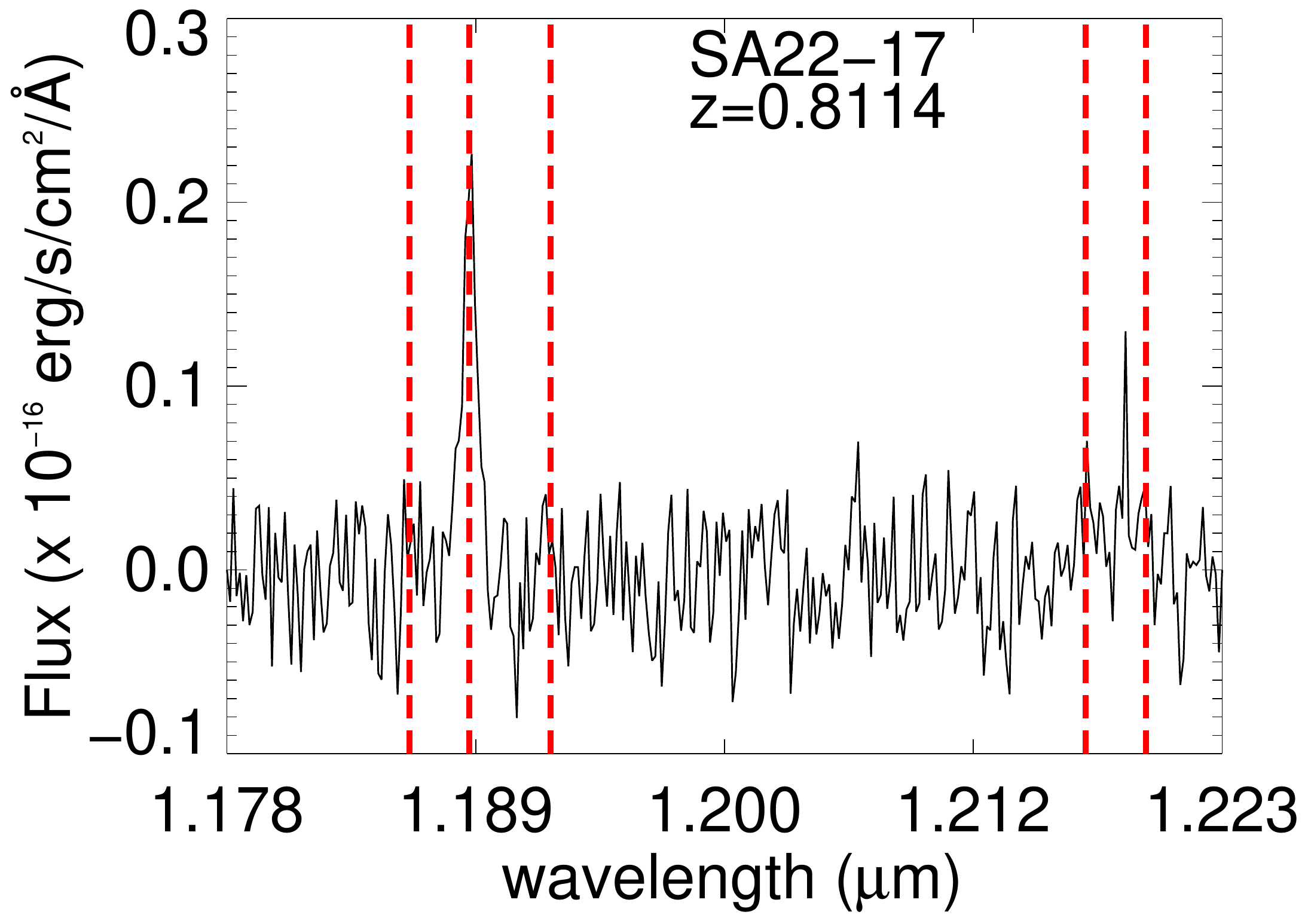}
\includegraphics[width=0.5\columnwidth]{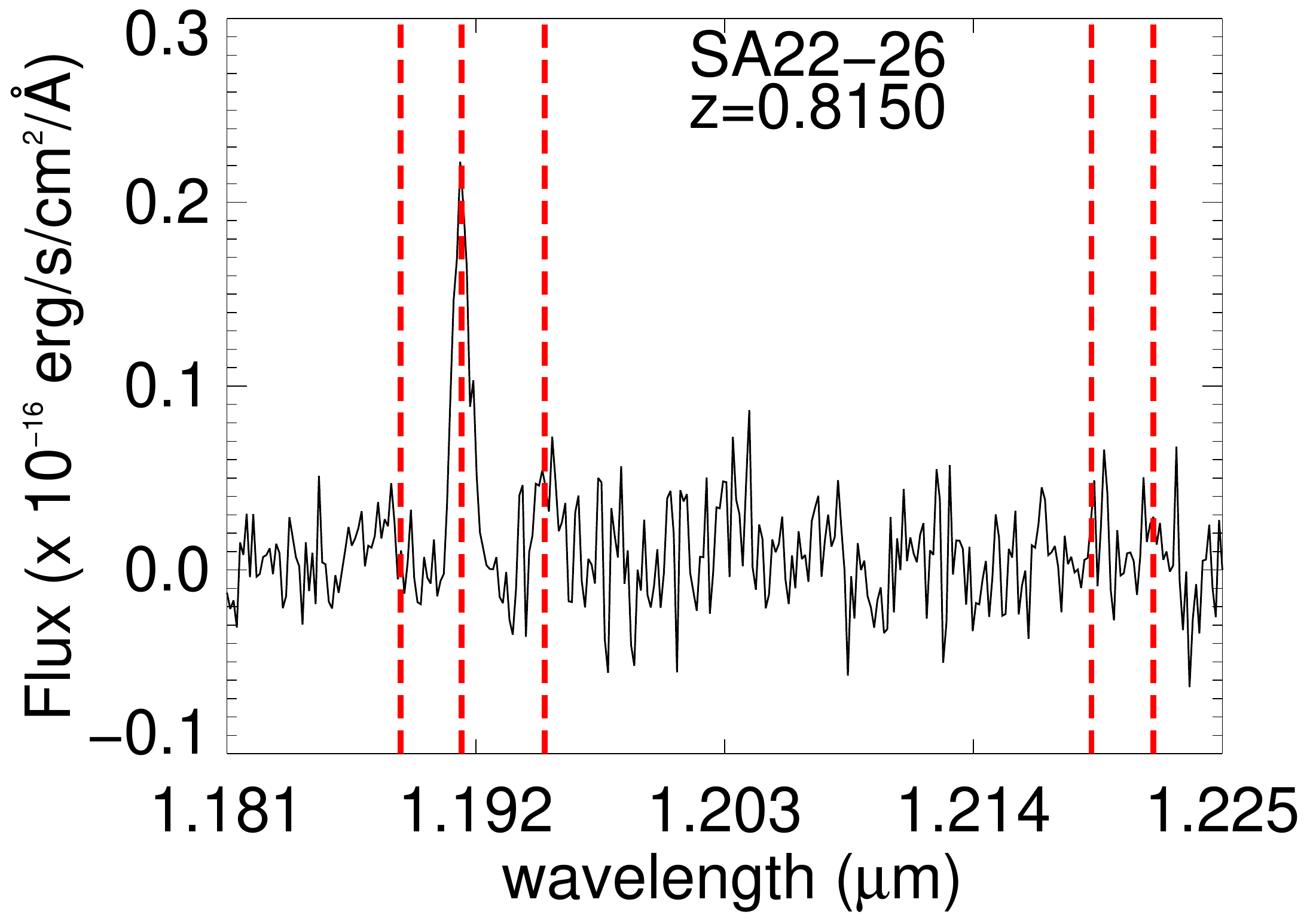}
\includegraphics[width=0.5\columnwidth]{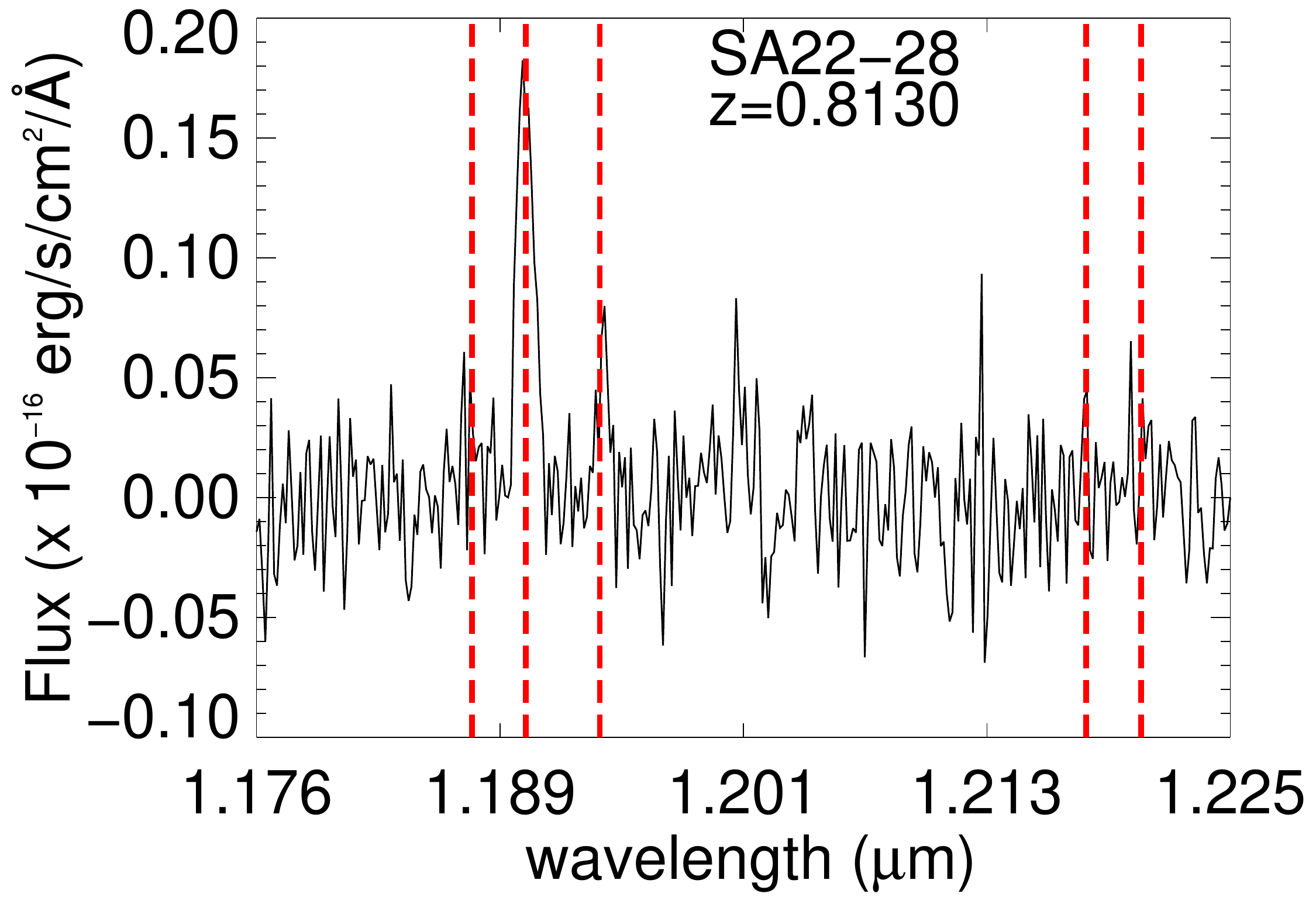}
\includegraphics[width=0.5\columnwidth]{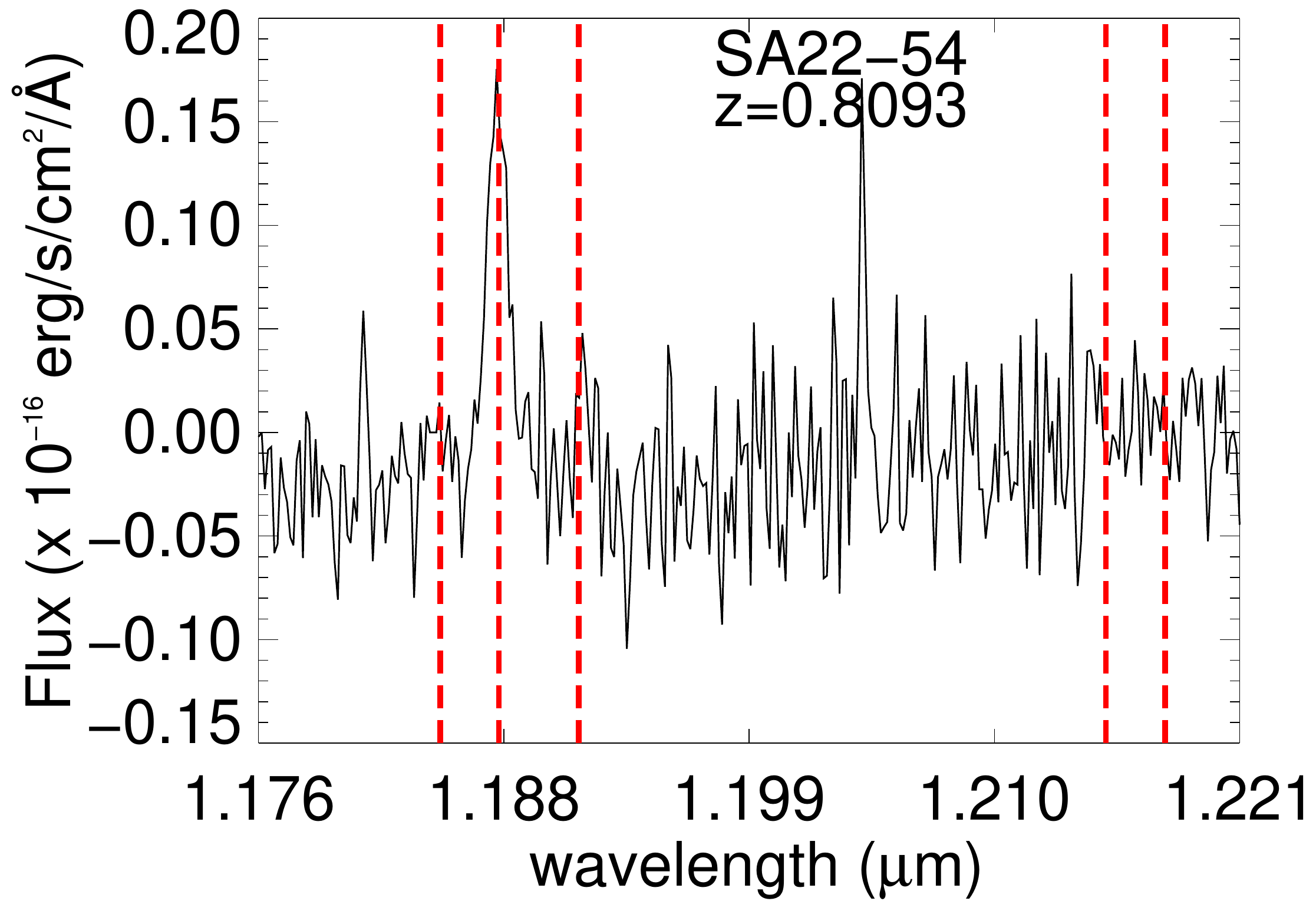}\\
\includegraphics[width=0.5\columnwidth]{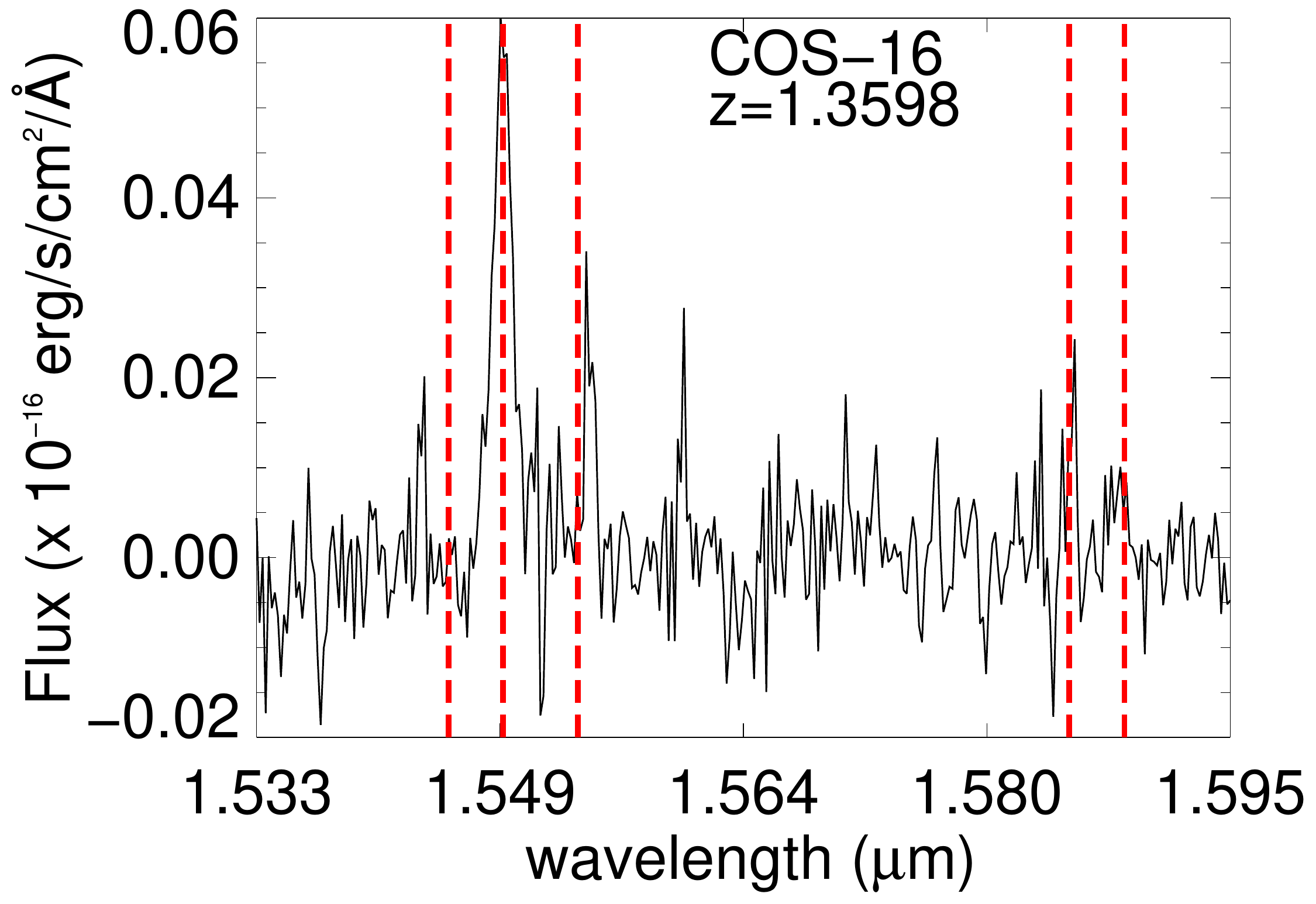}
\includegraphics[width=0.5\columnwidth]{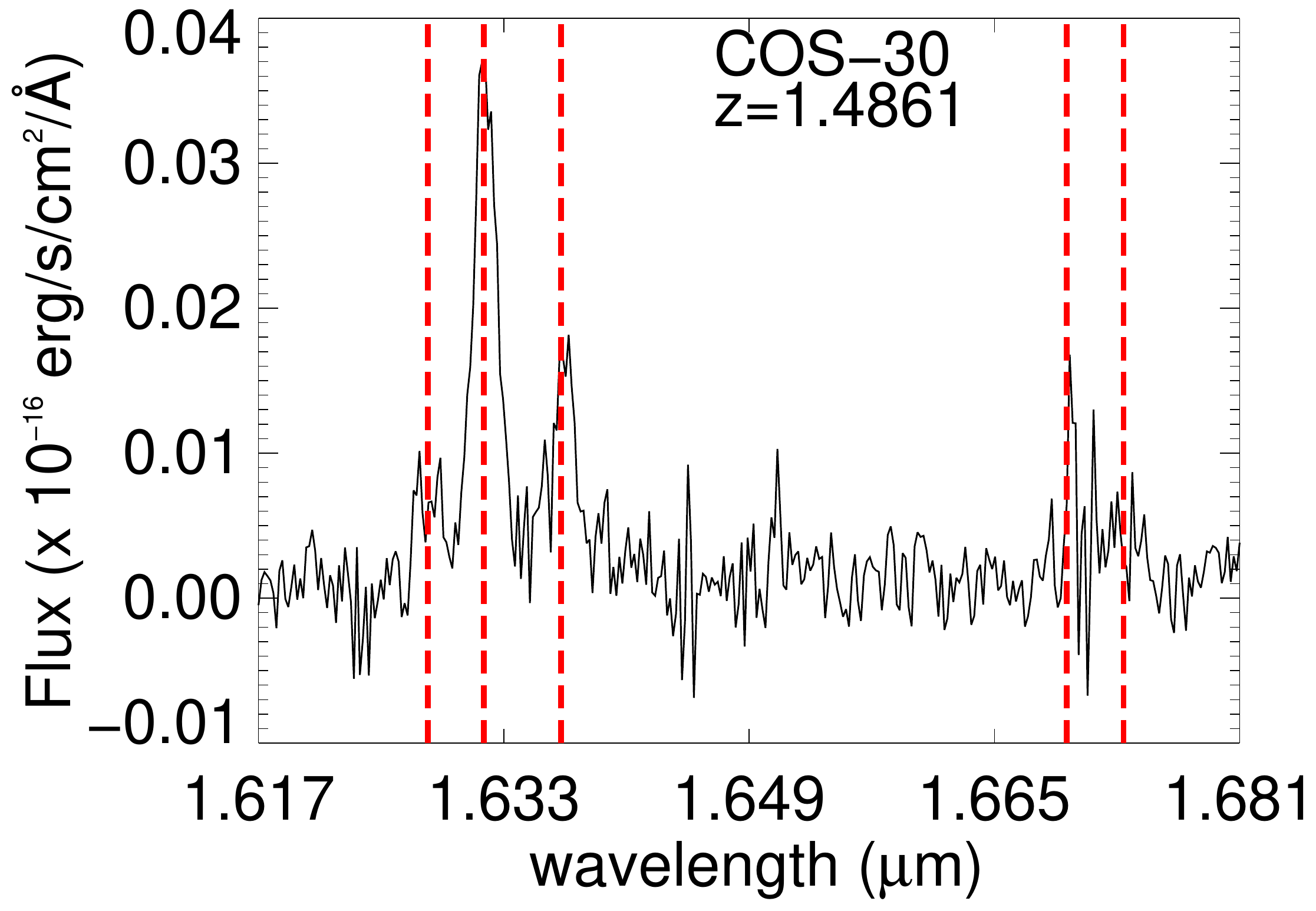}
\includegraphics[width=0.5\columnwidth]{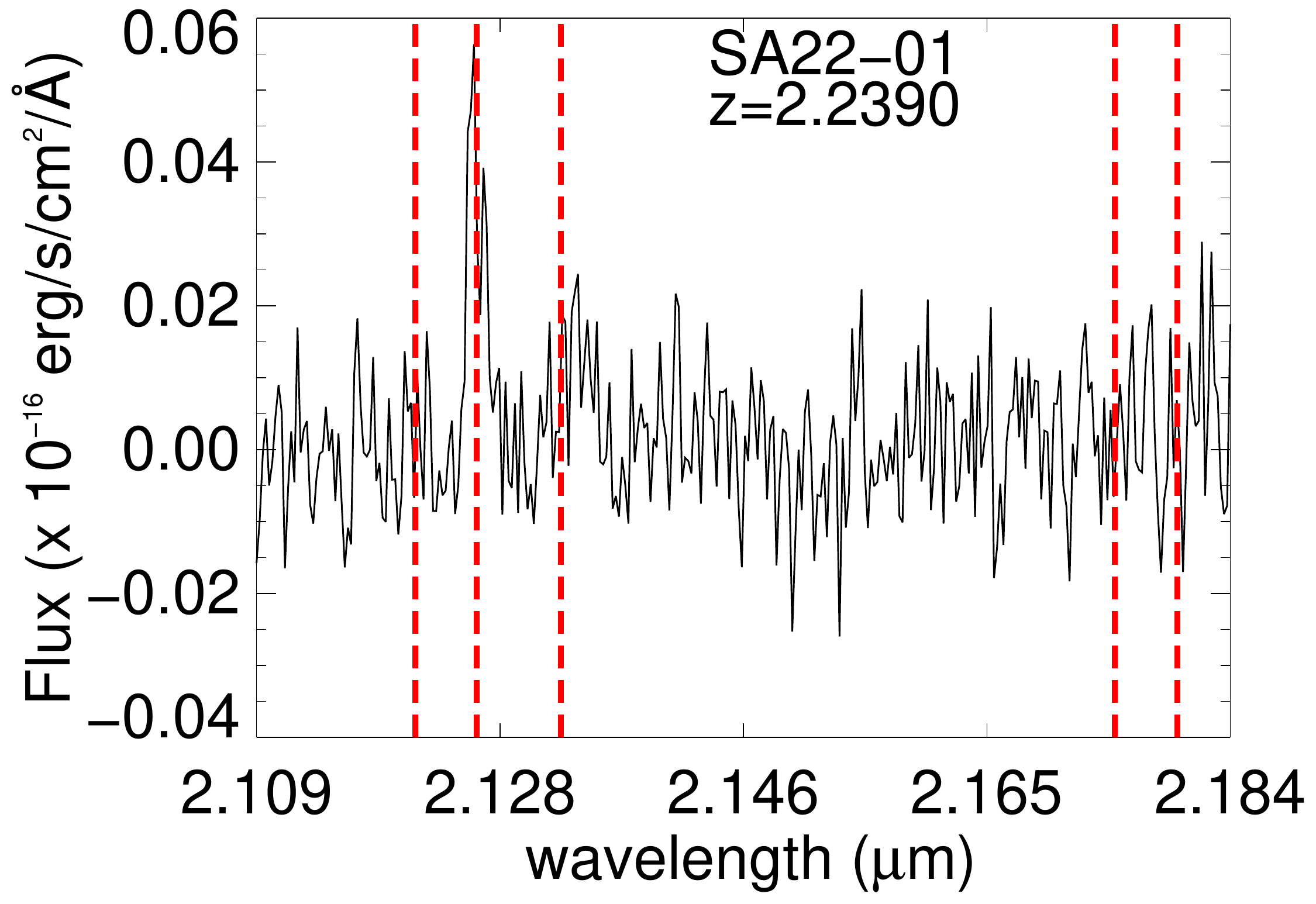}
\includegraphics[width=0.5\columnwidth]{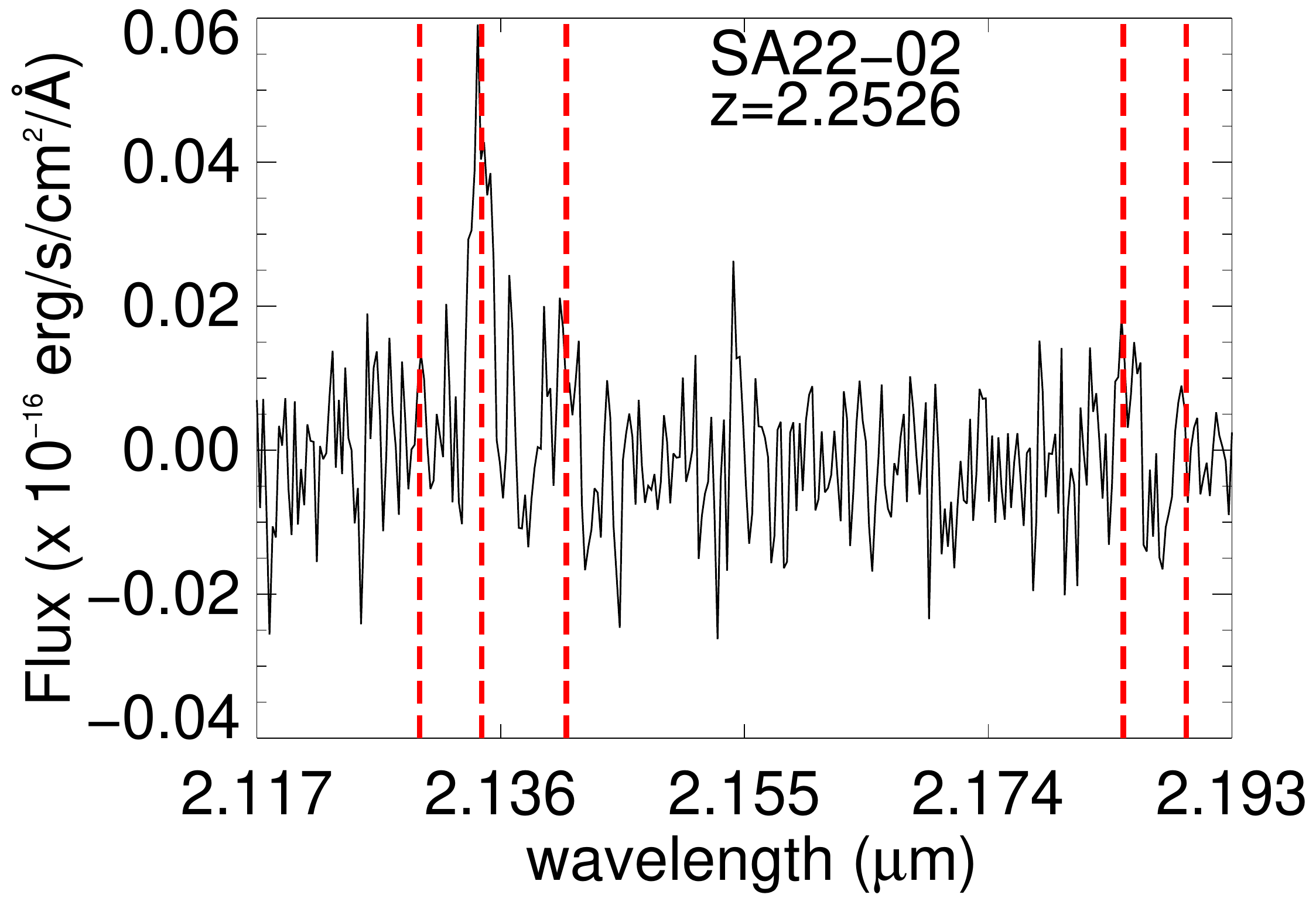}\\
\includegraphics[width=0.5\columnwidth]{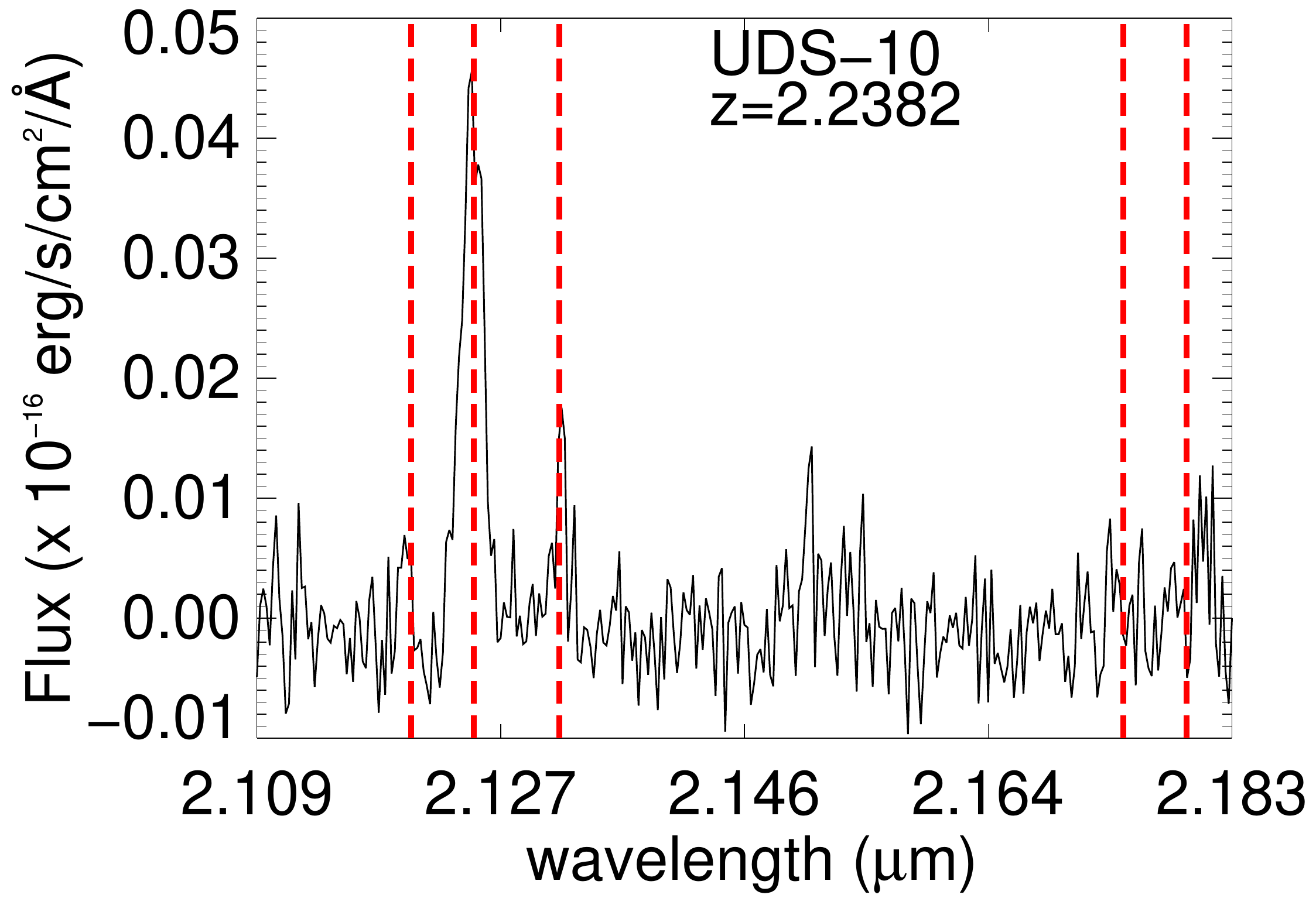}
\includegraphics[width=0.5\columnwidth]{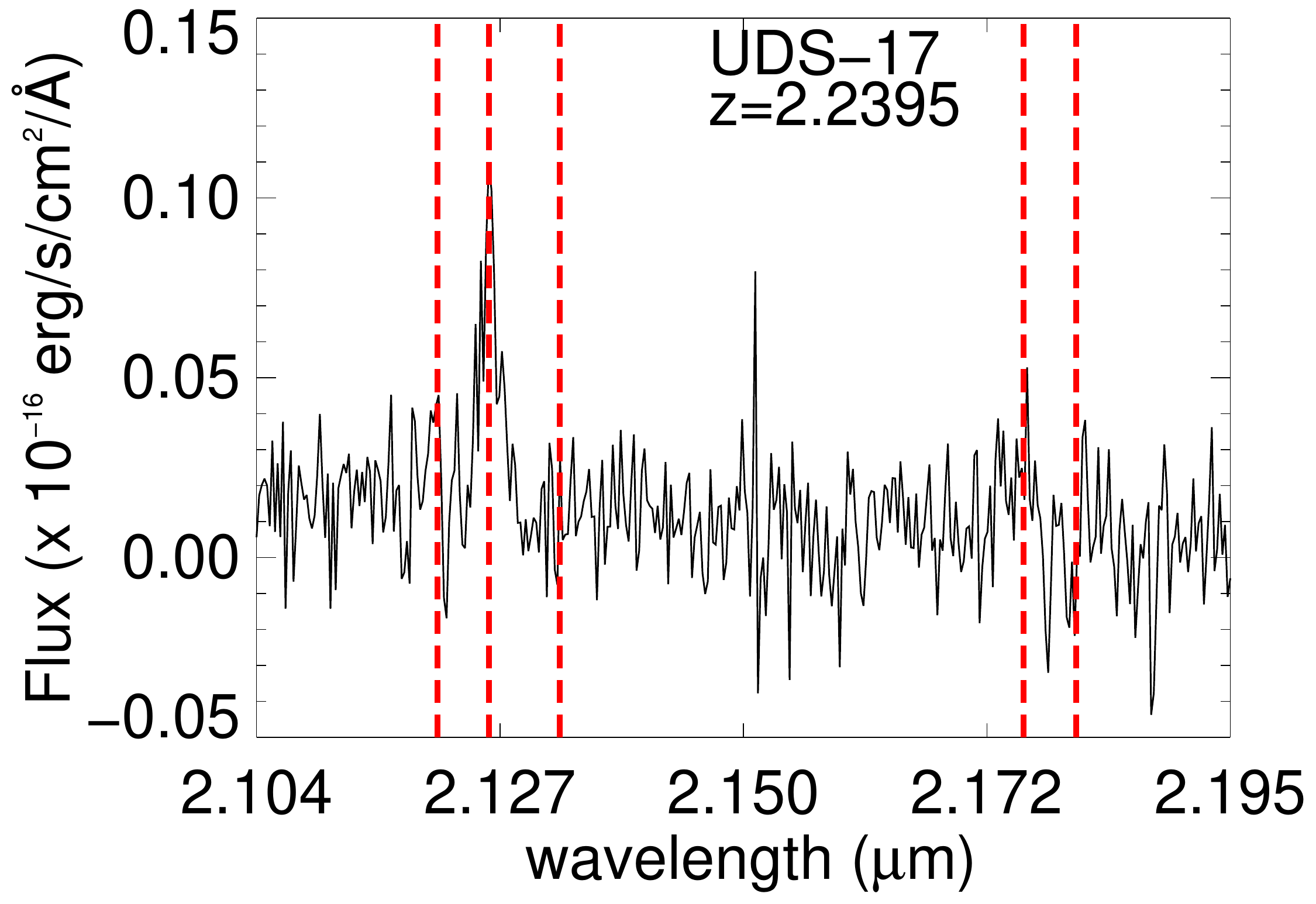}
\includegraphics[width=0.5\columnwidth]{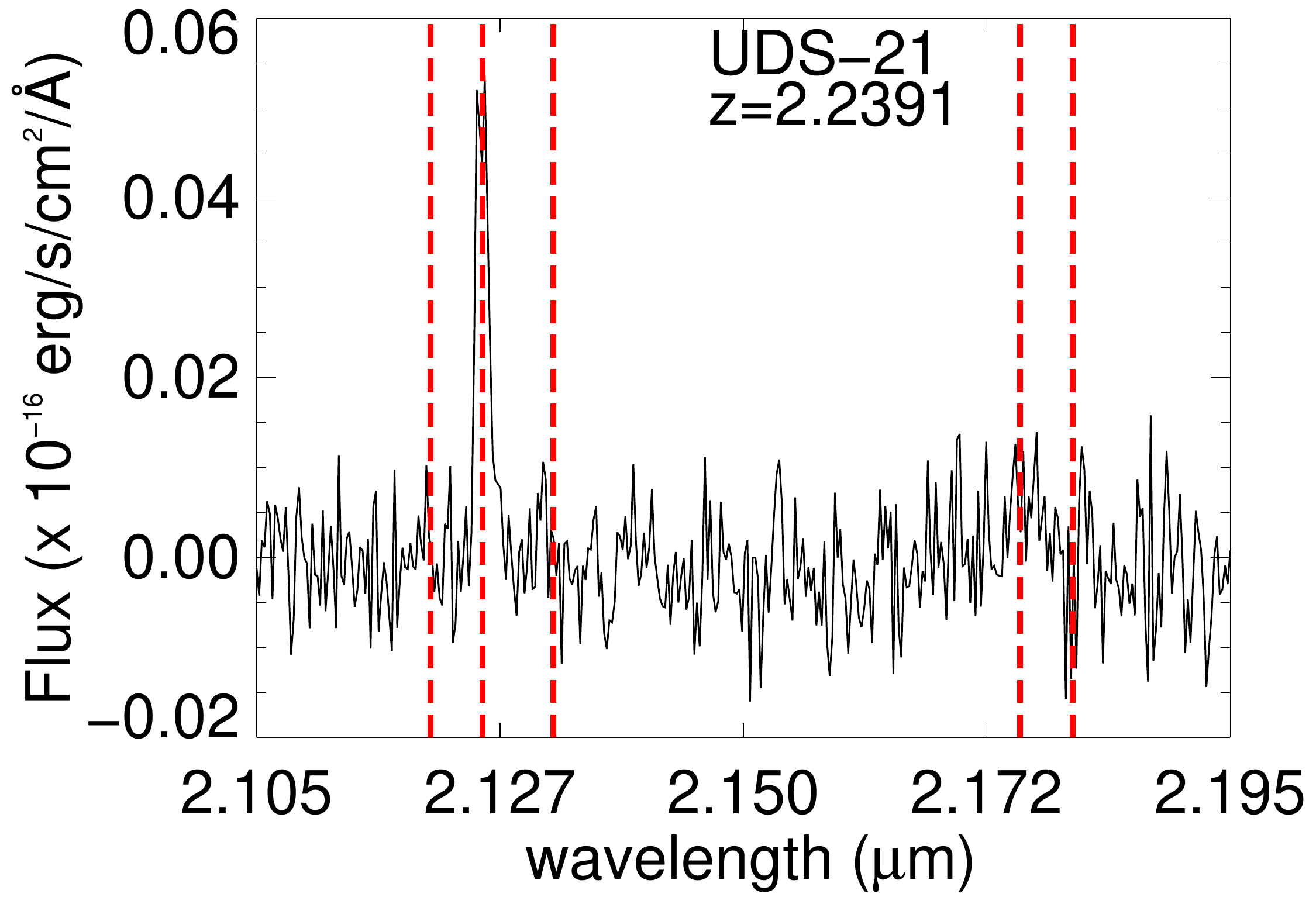}
\includegraphics[width=0.5\columnwidth]{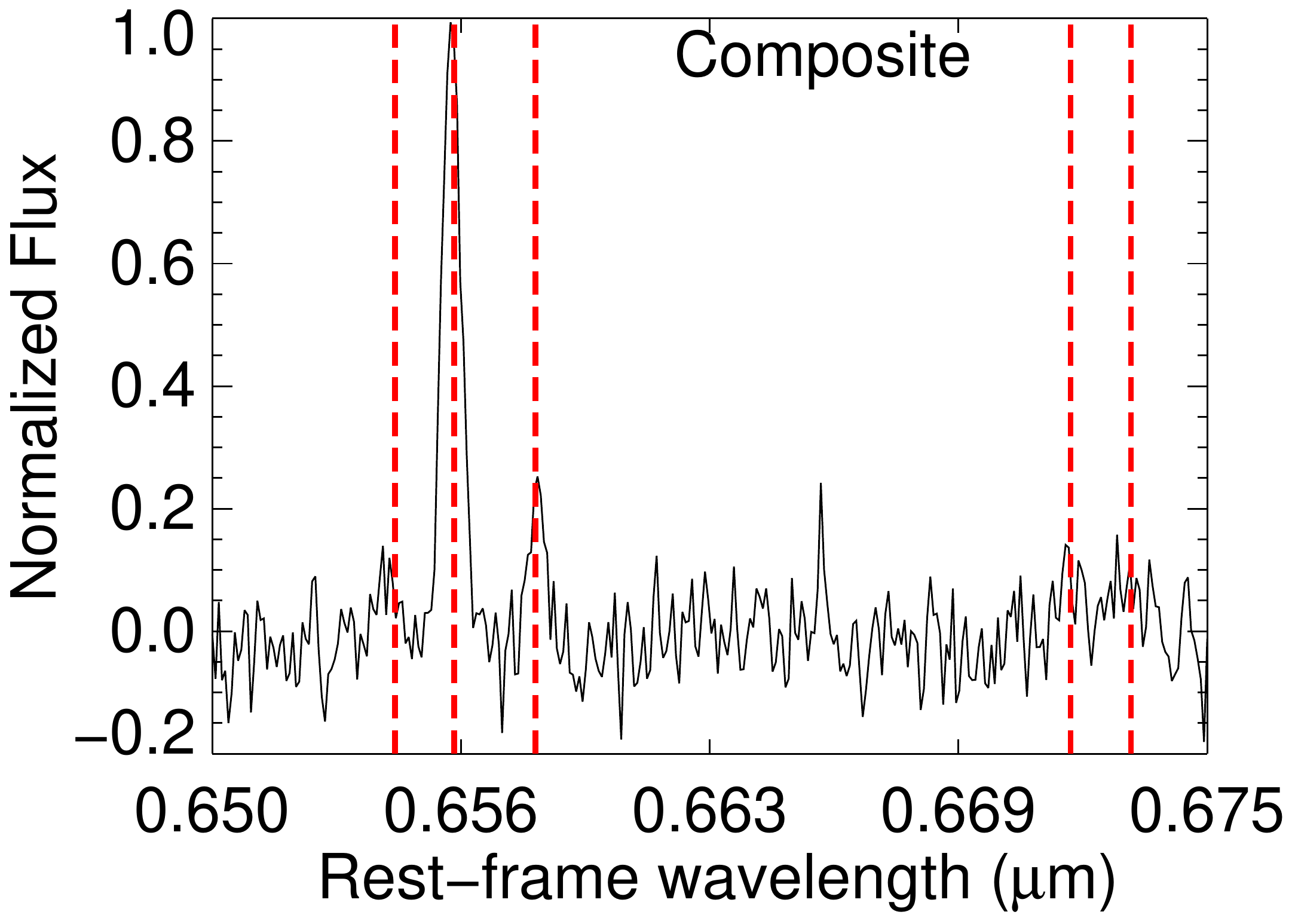}\\
	\caption{\label{fig:integrated_spectra}
	     Spatially integrated one dimensional spectra around the redshifted 
	     H$\alpha$ emission for each of the galaxies in our sample. H$\alpha$, [N\,{\sc ii}]$\lambda\lambda$6583,6548 and
	     [S\,{\sc ii}]$\lambda\lambda$6716,6731 emission lines are represented by the red-dashed lines. We detect [N\,{\sc ii}] 
	     emission in eight targets within our sample and the median [N\,{\sc ii}]/H$\alpha$ for the sample is 0.27$\pm$0.02, 
	     with a range of 0.10<[N\,{\sc ii}]/H$\alpha$<0.43. None of the galaxies display strong AGN signatures in their 
	     near-infrared spectra (e.g. broad lines or high [N\,{\sc ii}]/H$\alpha$ ratios).}
\end{figure*}

\subsection{Stellar Masses}
\label{sec:stellarmass}
Stellar masses are computed by fitting SEDs to the rest-frame UV,
optical and near-infrared data available
($FUV,NUV,U,B,g,V,R,i,I,z,Y,J,H,K,3.6,4.5,5.8$ and 8.0 $\micron$
  collated in \citealt{Sobral2014}, and references therein), following
\citet{Sobral2011}. The SED templates were generated with the
\citet{BruzualCharlot2003} package using \citet{Bruzual2007} 
models, a \citet{Chabrier2003} IMF and an exponentially
declining star formation history with the form $e^{-t/\tau}$, with
$\tau$ in the range 0.1--10\,Gyr. The SEDs were generated for a
logarithmic grid of 200 ages (from 0.1\,Myr to the maximum age at each
redshift being studied). Dust extinction was applied to the templates
using \citet{Calzetti2000} extinction law with $E(B-V)$ in the range 0
to 0.5 (in steps of 0.05) roughly corresponding to a H$\alpha$
extinction $A_{\rm H\alpha} \sim$0--2 mag. The models are generated
with different metallicities, including solar \citep{Sobral2011}. For
each source, the stellar mass and the dust extinction are computed as
the median values of the 1$\sigma$ best fits over the range of
parameters (see Table~\ref{tab:table1}).

\subsection{Star-Formation Rates}
\label{sec:SFRs}
The star-formation rates of the sample are measured from the H$\alpha$ emission line flux
calculated from the HiZELS survey. Adopting the \citet{Kennicutt1998} calibration
and assuming a Chabrier IMF, the SFRs are given by 
SFR$^{\rm obs}_{\rm H\alpha}$(M$_{\odot}$\,yr$^{-1}$)\,=\,4.6\,$\times$\,10$^{-42}$\,L$^{\rm obs}_{\rm H\alpha}$(erg\,s$^{-1}$). 
At the three redshift ranges of our sample, the average H$\alpha$
fluxes of our galaxies correspond to SFRs (uncorrected for extinction) of
SFR$^{\rm obs}_{\rm H\alpha}$(M$_{\odot}$\,yr$^{-1}$)\,$\approx$\,\,3, 6 and 21\,M$_{\odot}$\,yr$^{-1}$ 
at $z=0.8$, 1.47 and 2.23 respectively. The median E(B$-$V) for our sample is 
E(B$-$V)$=0.2\pm0.1$ (see Table~\ref{tab:table1}), which correspond to A$_{\rm H\alpha}=0.79\pm0.16$ 
(A$_{\rm V}=0.96\pm0.20$). This suggests reddening corrected star-formation rates of 
SFR$^{\rm corr}_{\rm H\alpha}$(M$_{\odot}$\,yr$^{-1}$)\,$\approx$\,6, 
13 and 43\,M$_{\odot}$\,yr$^{-1}$ at $z=0.8$, 1.47 and 2.23 respectively.
Hereafter, we use an extinction value of $A_{\rm H\alpha} = 1.0$ mag as used in previous works
based on the HiZELS survey (e.g. \citealt{Sobral2012}, \citealt{Stott2013b}, \citealt{Ibar2013}, Thomson et al.\ submitted)
in order to compare consistently.

\subsection{Spatial extent}
\label{sec:Spatial-extent}
To measure the spatial extent of the galaxy, we calculate the half-light radii ($r_{1/2}$).
Those are calculated from the collapsed continuum subtracted cubes, where the encircled H$\alpha$ flux 
decays to half its total integrated value. The total integrated value is defined as the total 
H$\alpha$ luminosity within a Petrosian radius. We adopted the `Sloan Digital Sky Survey' (SDSS)
Petrosian radius definition with R$_{\rm P,lim}=0.2$. We account for the ellipticity and position 
angle of the galaxy obtained from the best-fit disk model (see \S\ref{sec:analysis}). 
The $r_{1/2}$  errors are derived by bootstrapping via Monte-Carlo simulations the errors in measured 
emission line intensity and estimated dynamical parameters of each galaxy. The half-light radii are 
corrected for beam-smearing effects by subtracting the seeing ($~0.15''$) in quadrature. The median $r_{1/2}$
for our sample is found to be  $2.4\pm0.1$ kpc (Table~\ref{tab:table1}), which is consistent with 
previous studies at similar redshift range \citep{Swinbank2012a}.

\begin{figure*}
\flushleft
\includegraphics[width=0.343\columnwidth]{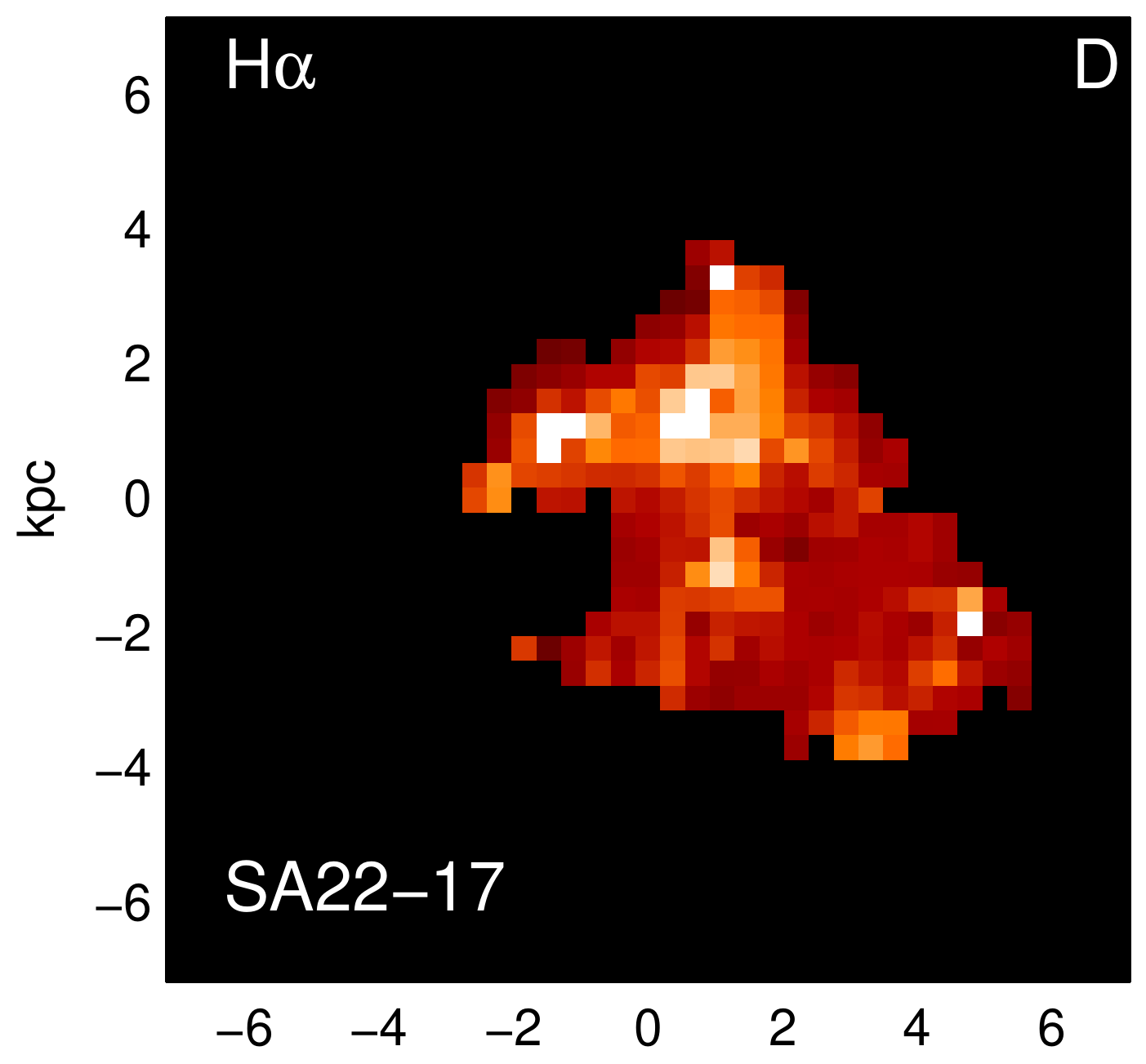}
\includegraphics[width=0.32\columnwidth]{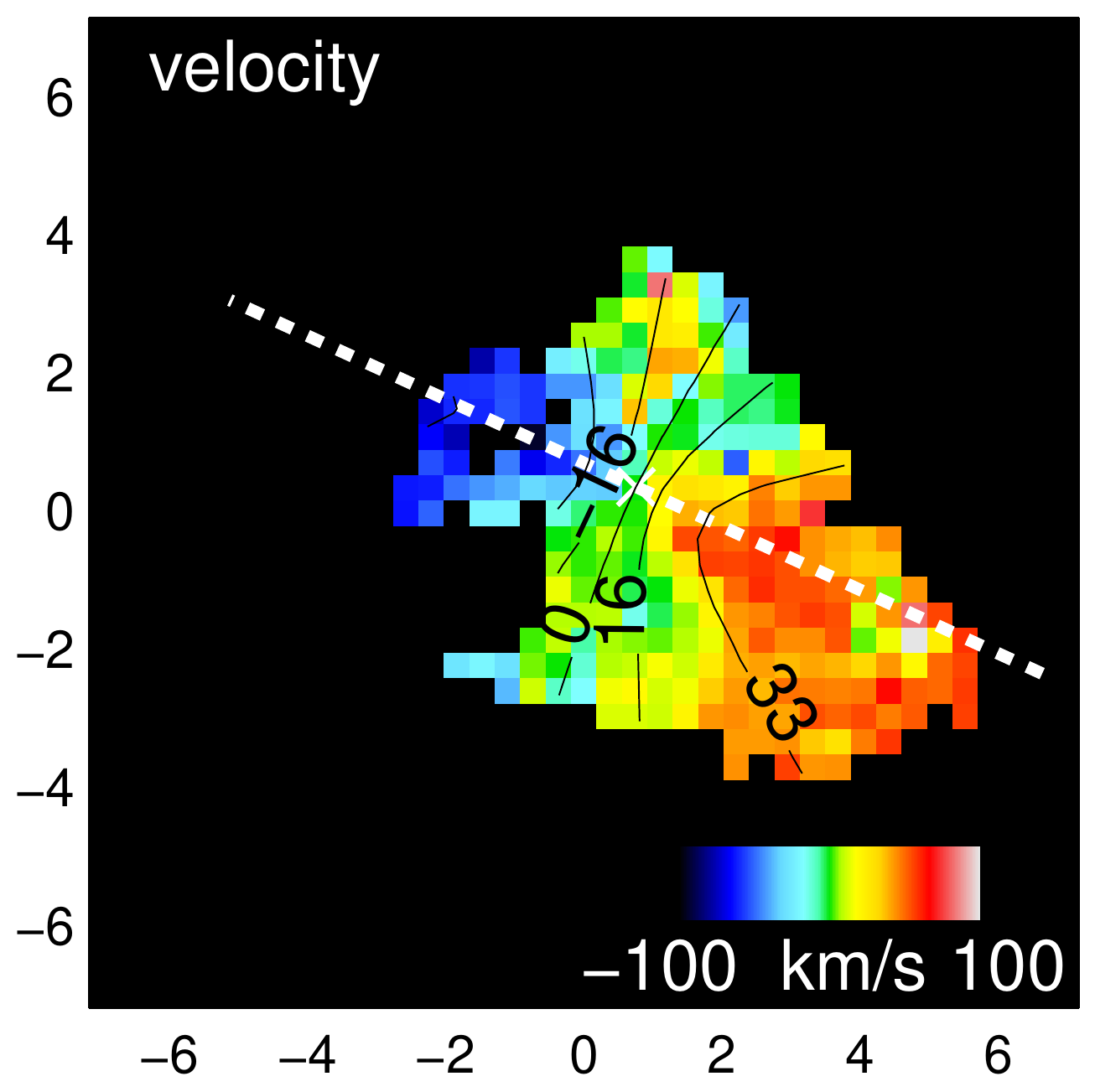}
\includegraphics[width=0.32\columnwidth]{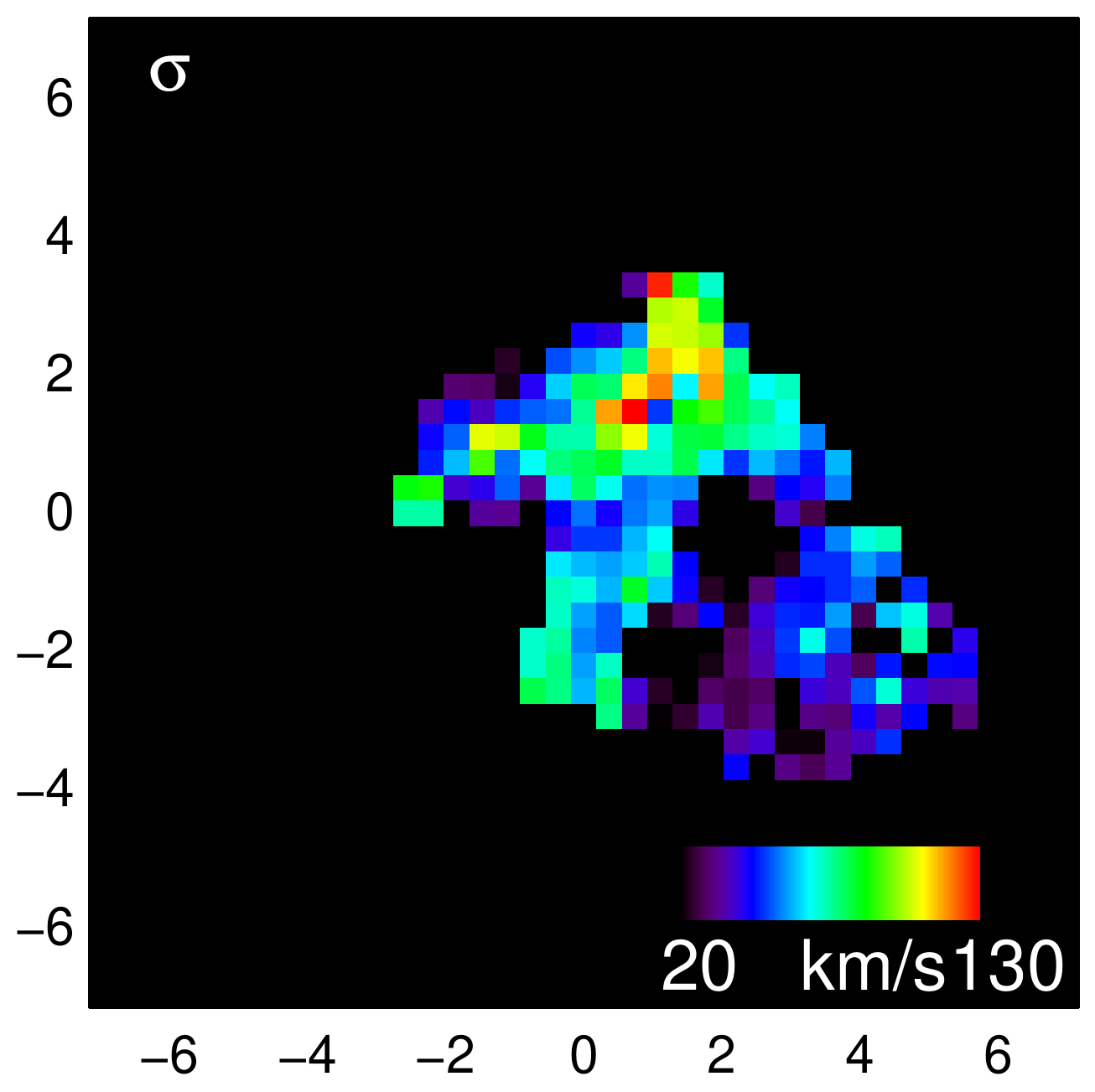}
\includegraphics[width=0.32\columnwidth]{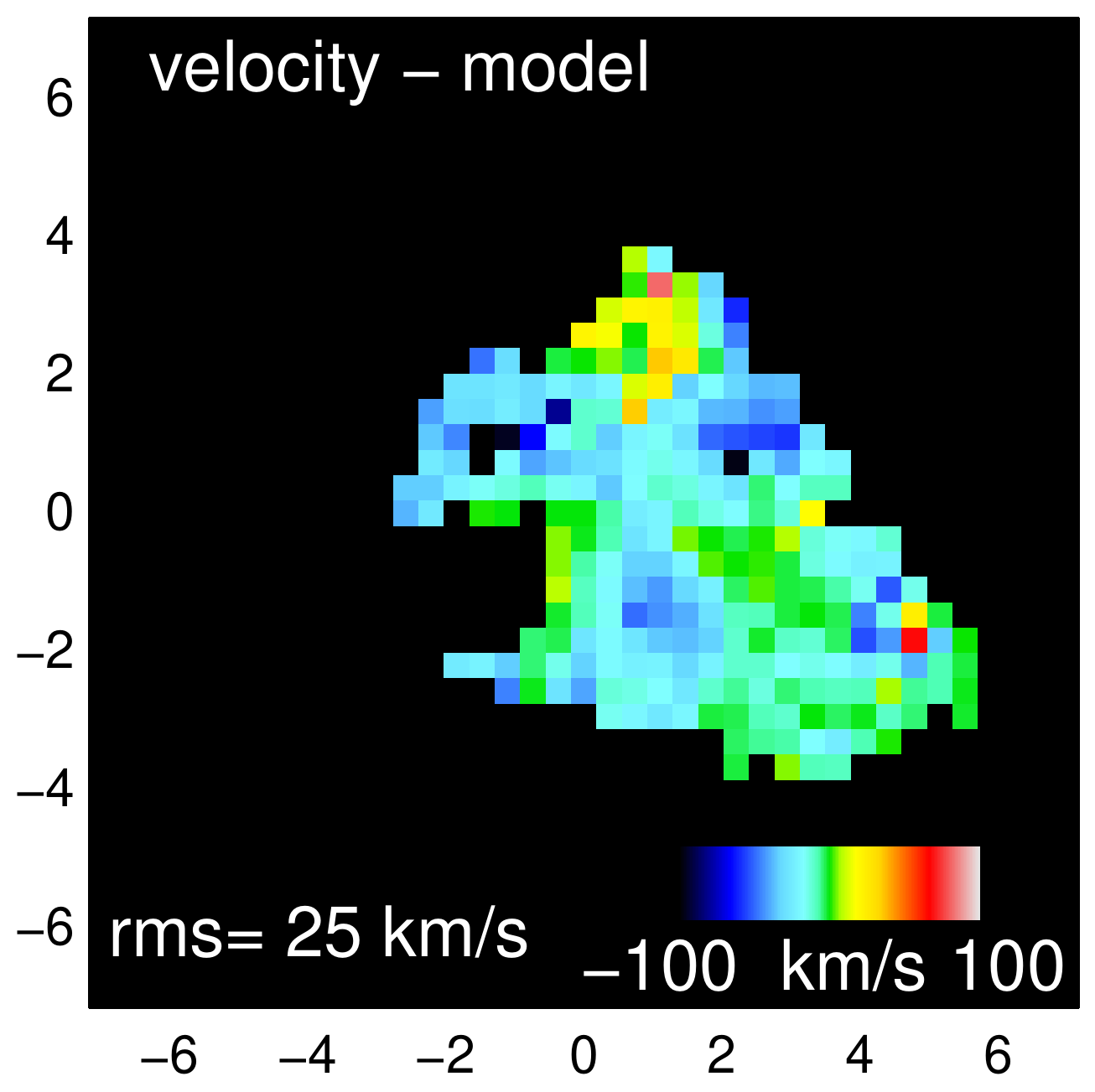}
\includegraphics[width=0.345\columnwidth]{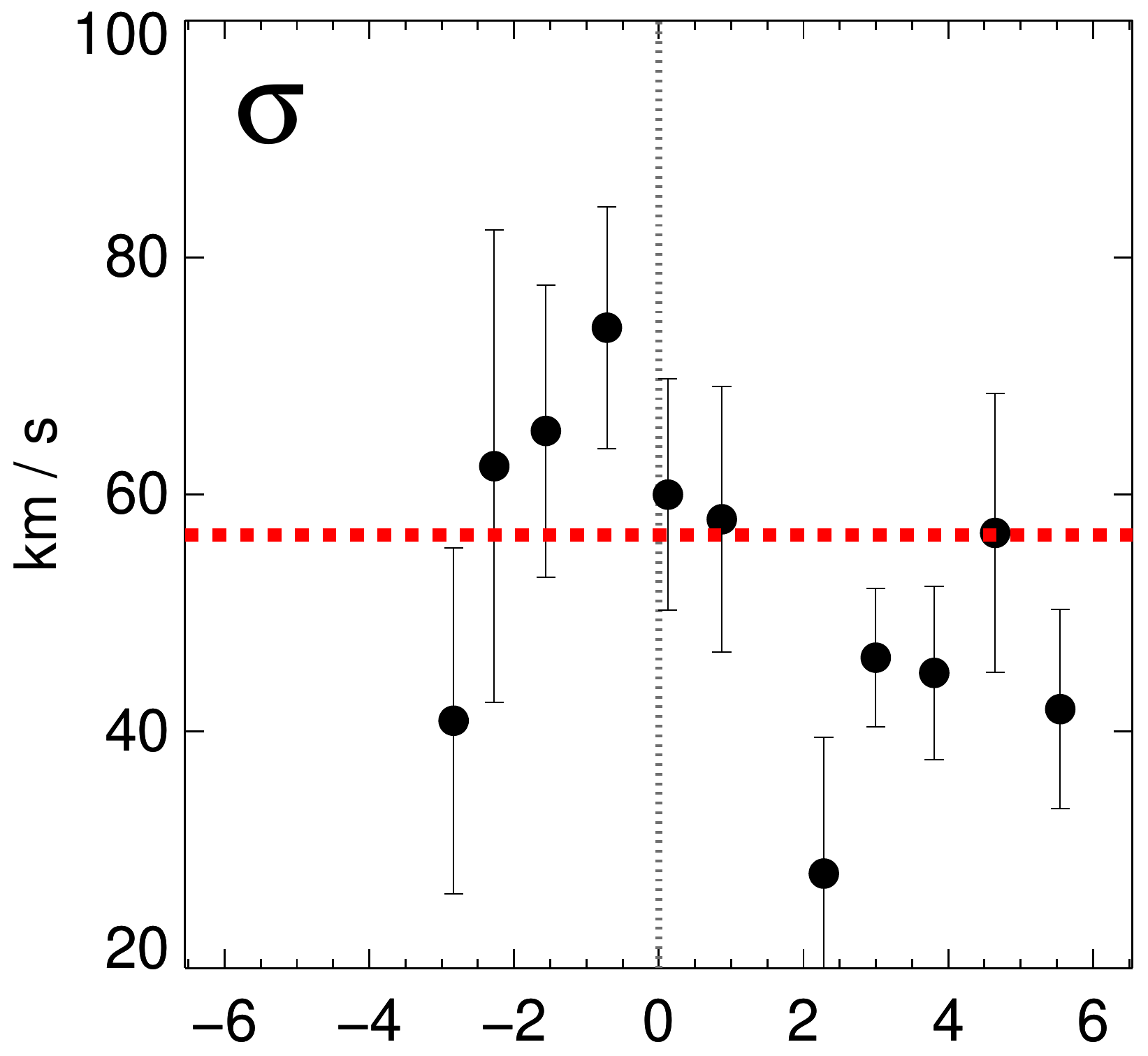}
\includegraphics[width=0.373\columnwidth]{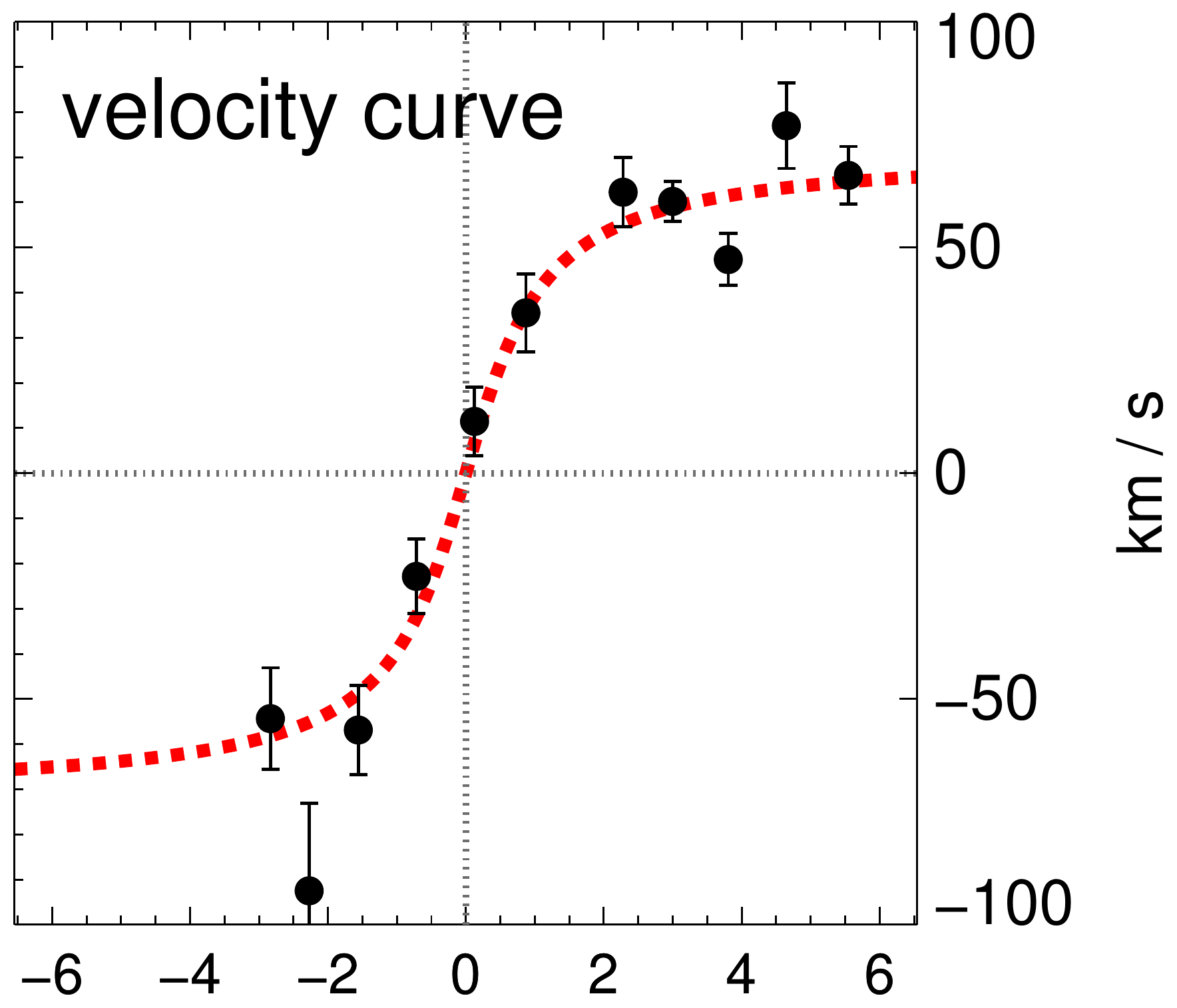}\\
\vspace{1mm}
\includegraphics[width=0.343\columnwidth]{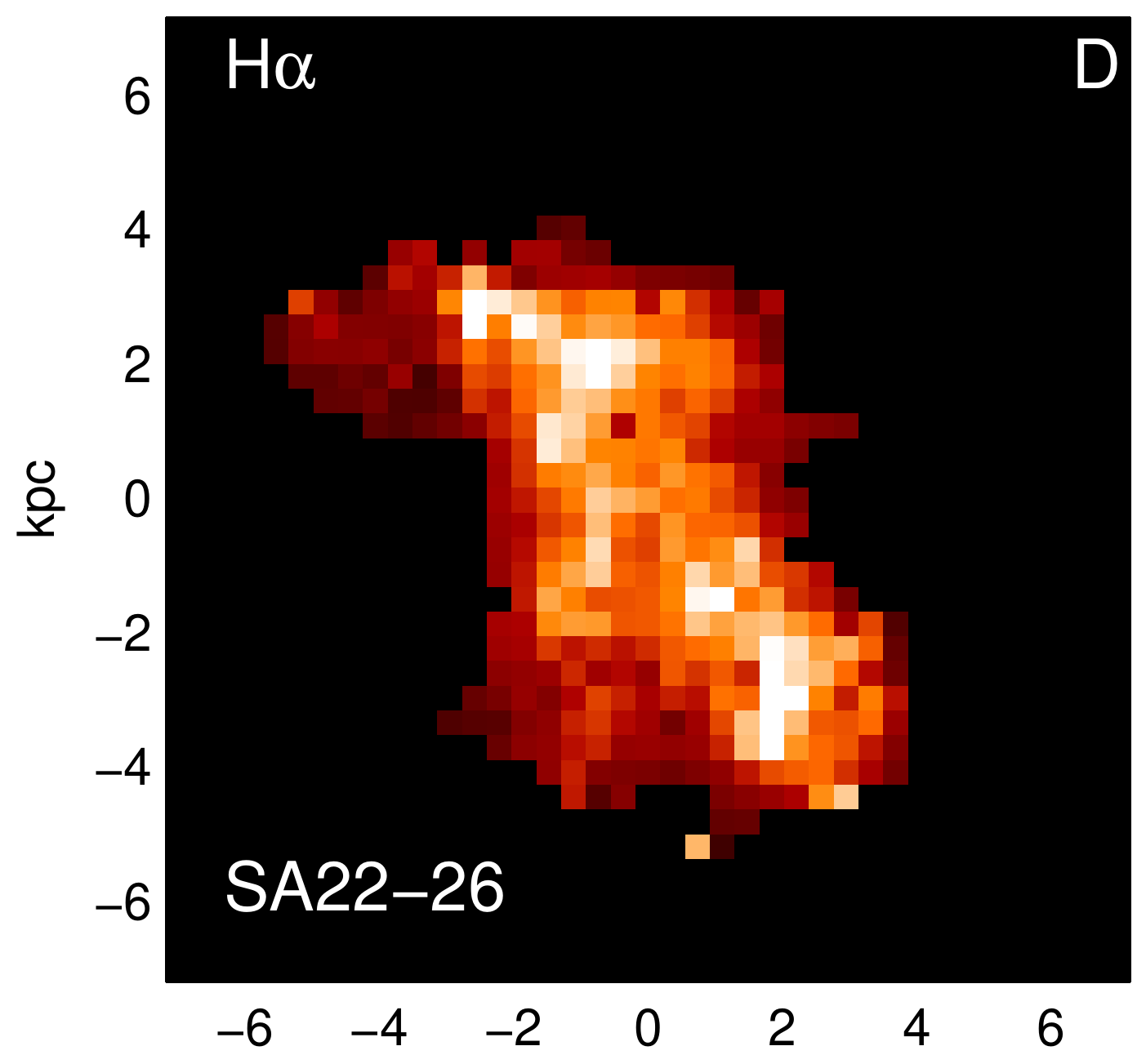}
\includegraphics[width=0.32\columnwidth]{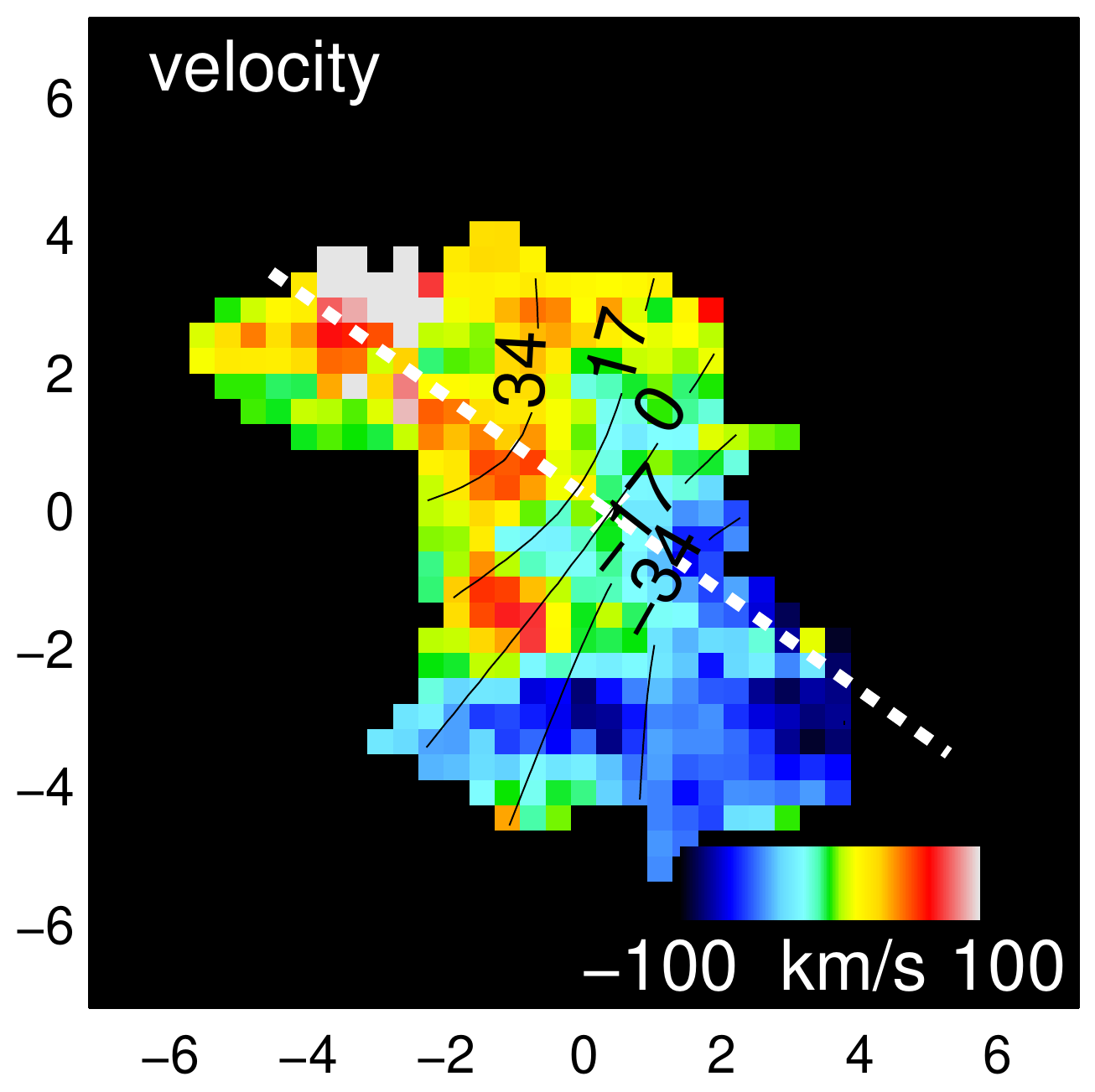}
\includegraphics[width=0.32\columnwidth]{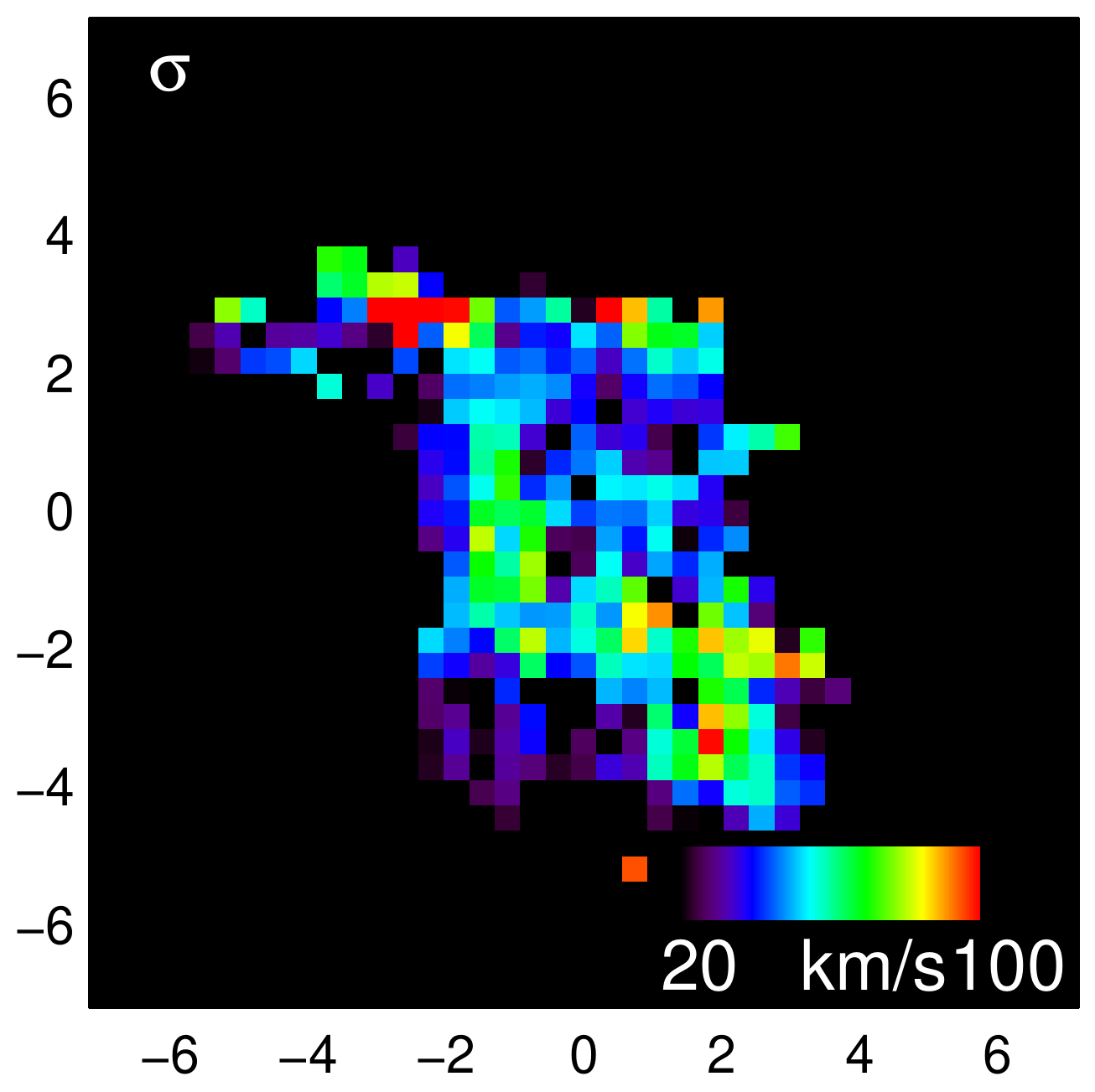}
\includegraphics[width=0.32\columnwidth]{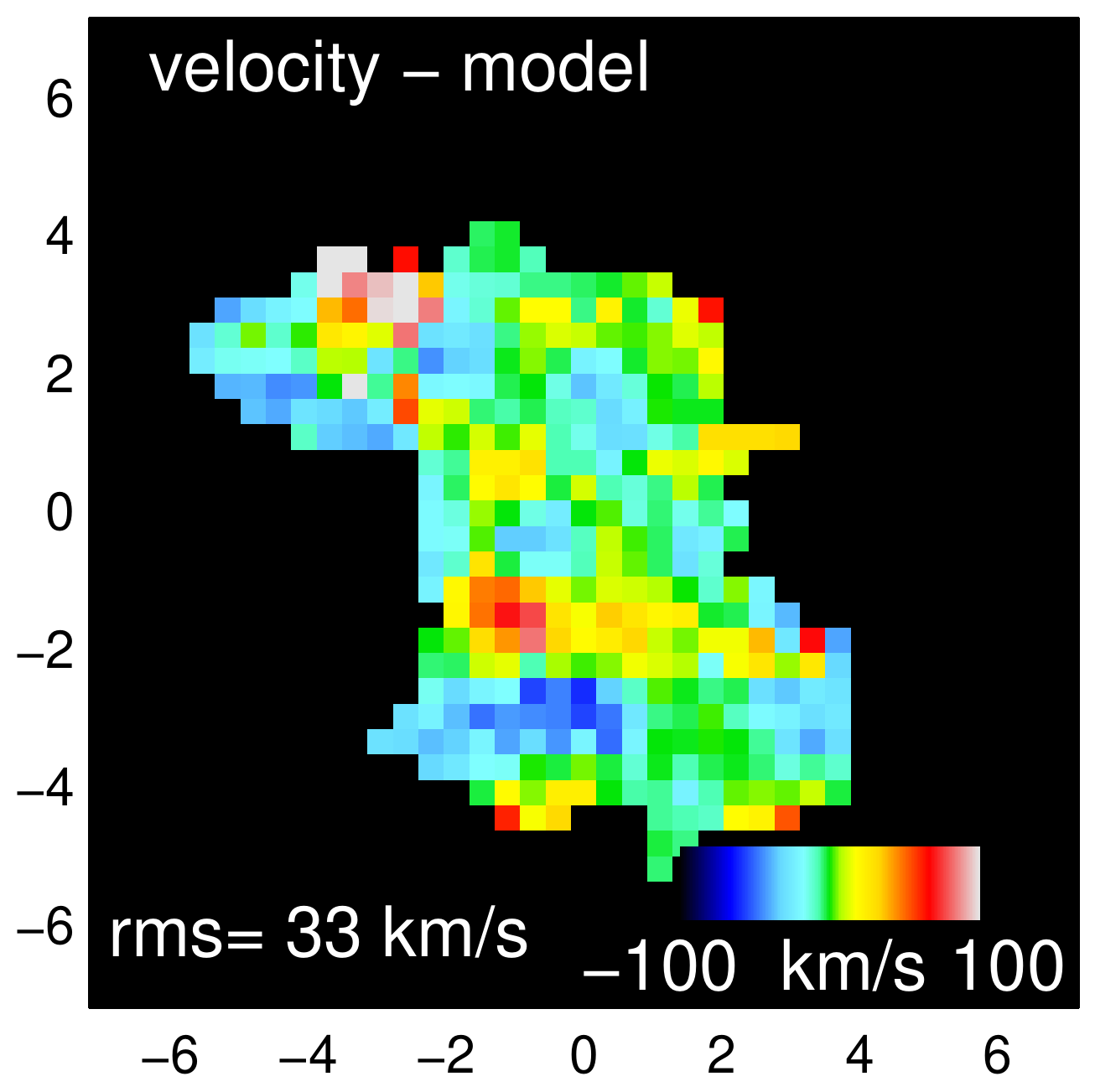}
\includegraphics[width=0.345\columnwidth]{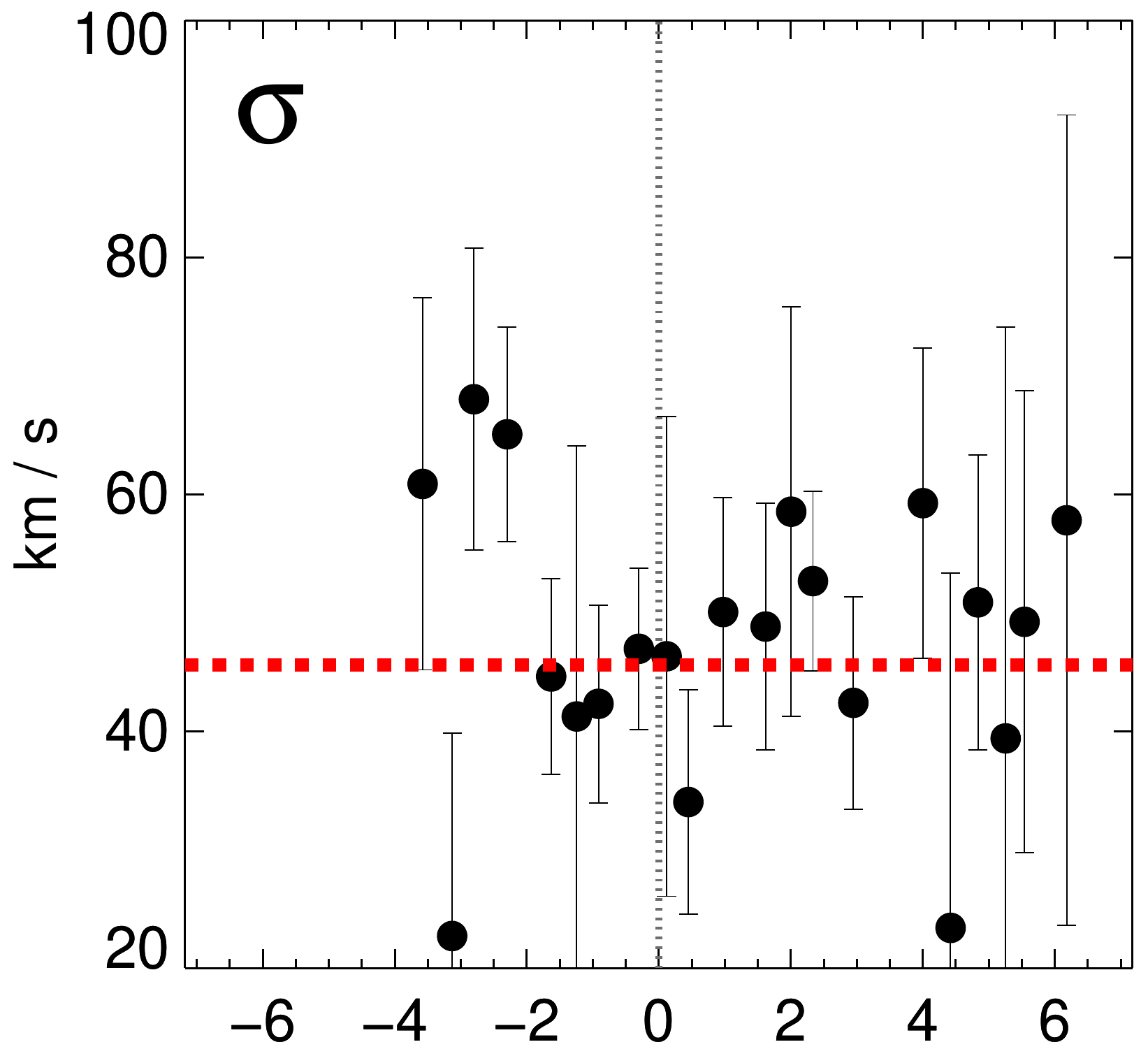}
\includegraphics[width=0.373\columnwidth]{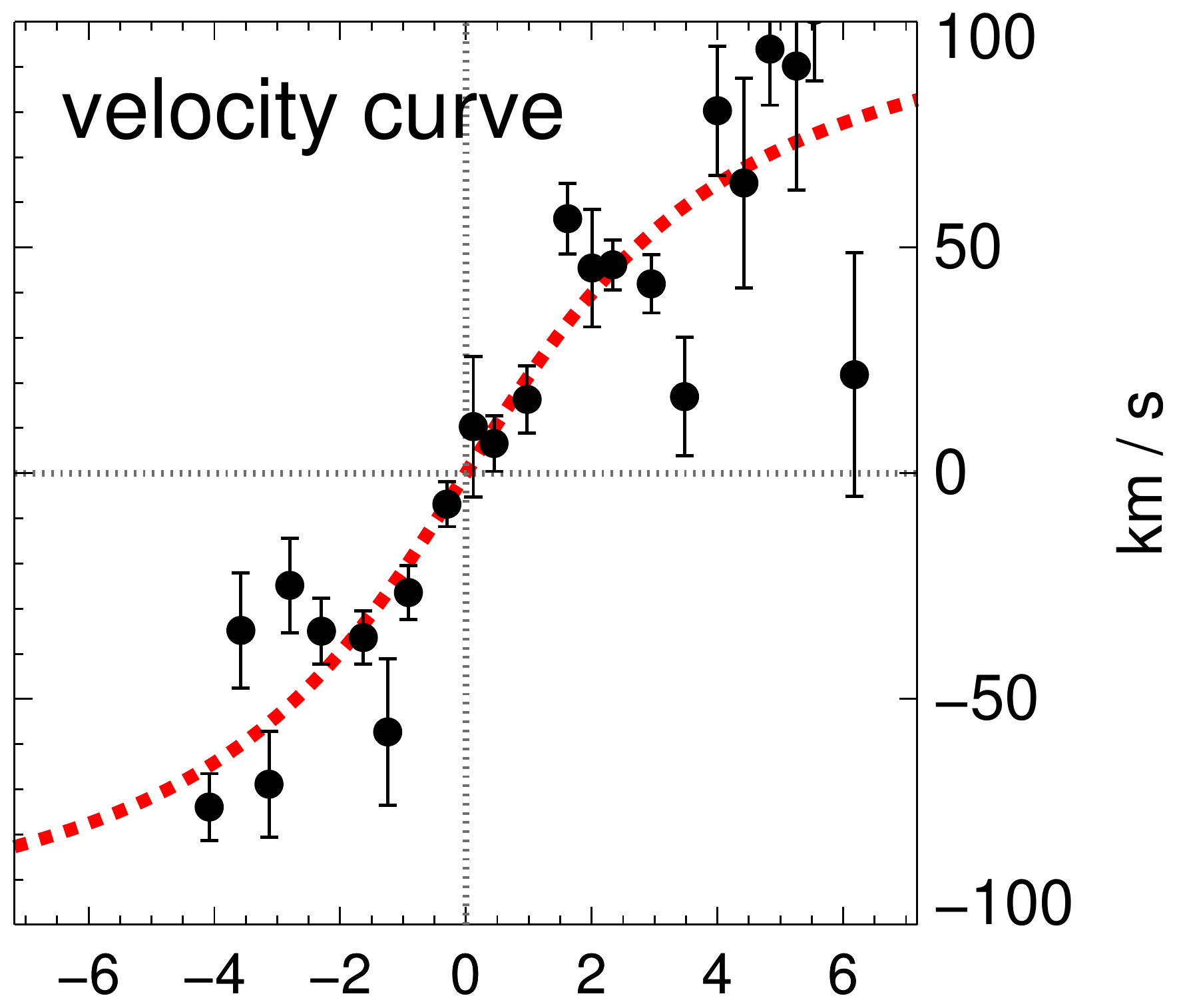}\\
\vspace{1mm}
\includegraphics[width=0.343\columnwidth]{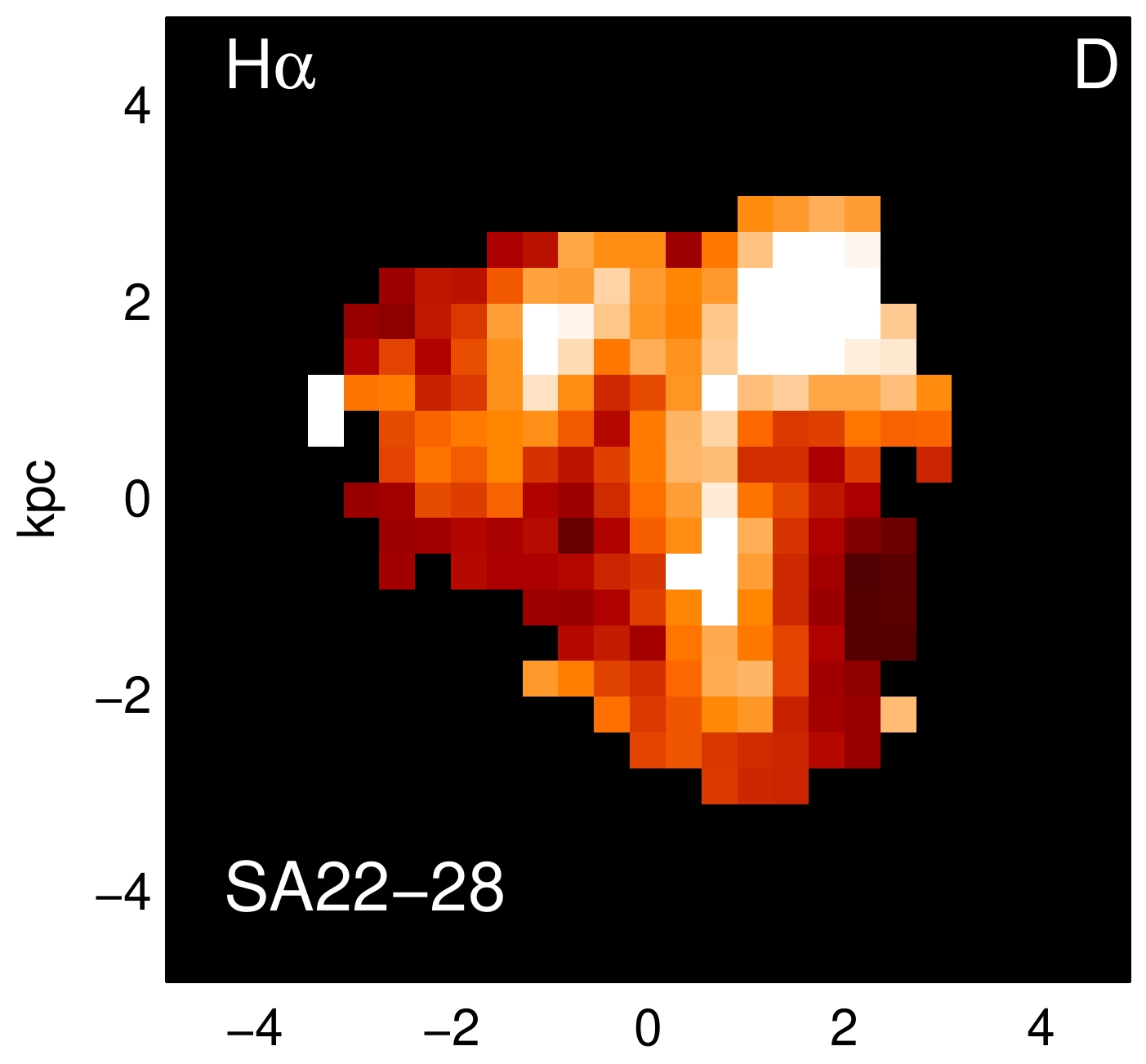}
\includegraphics[width=0.32\columnwidth]{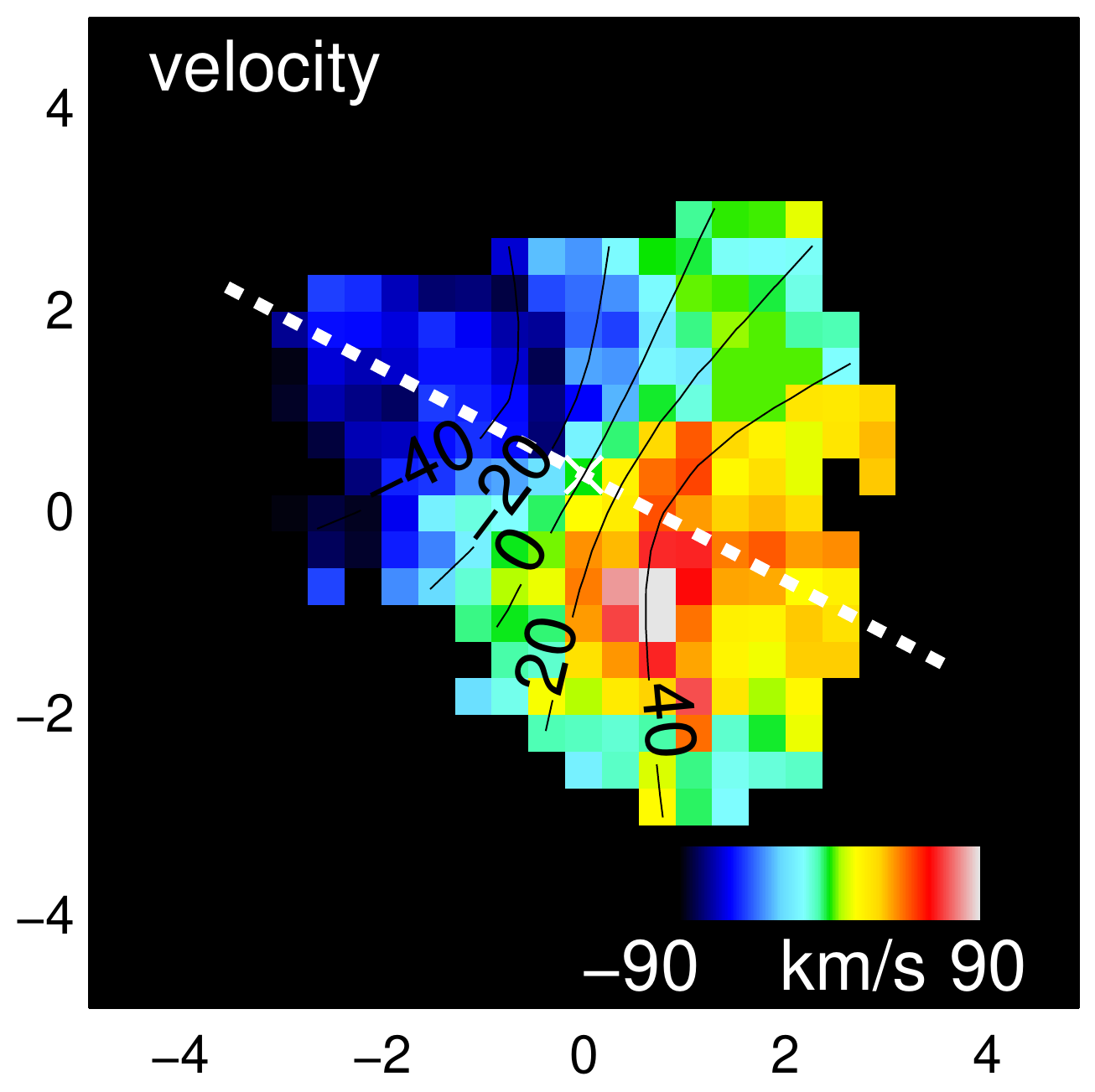}
\includegraphics[width=0.32\columnwidth]{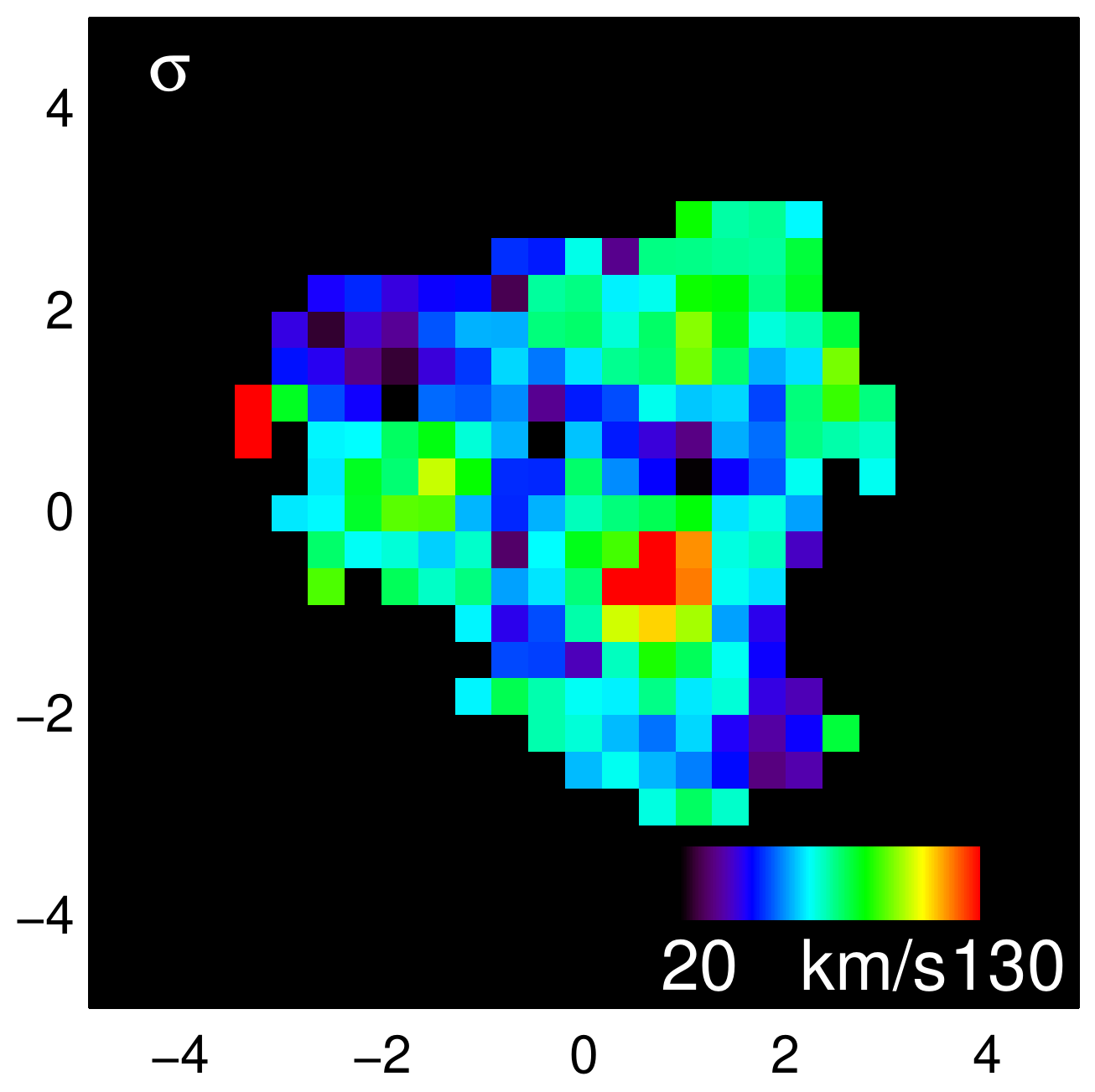}
\includegraphics[width=0.32\columnwidth]{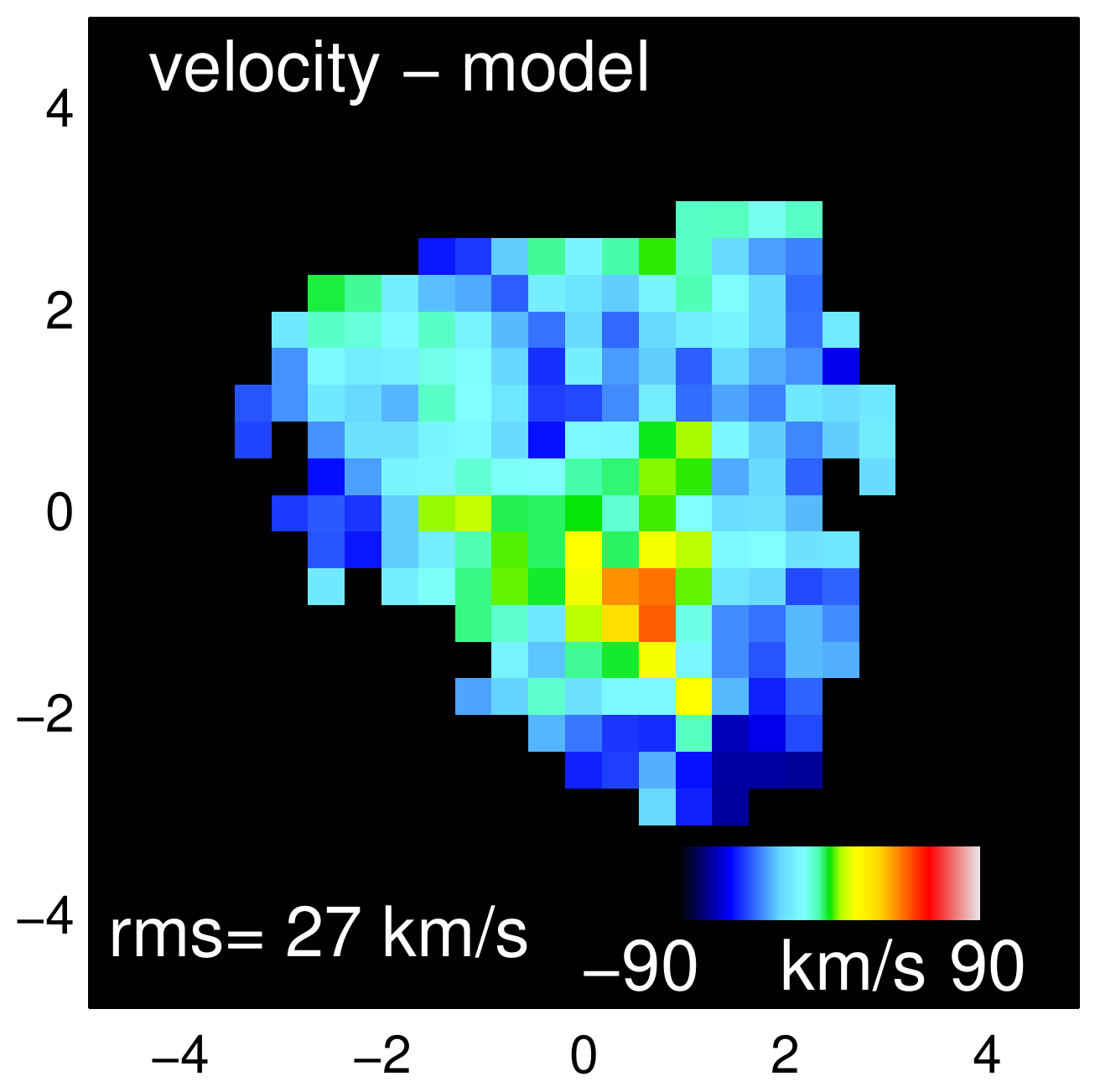}
\includegraphics[width=0.345\columnwidth]{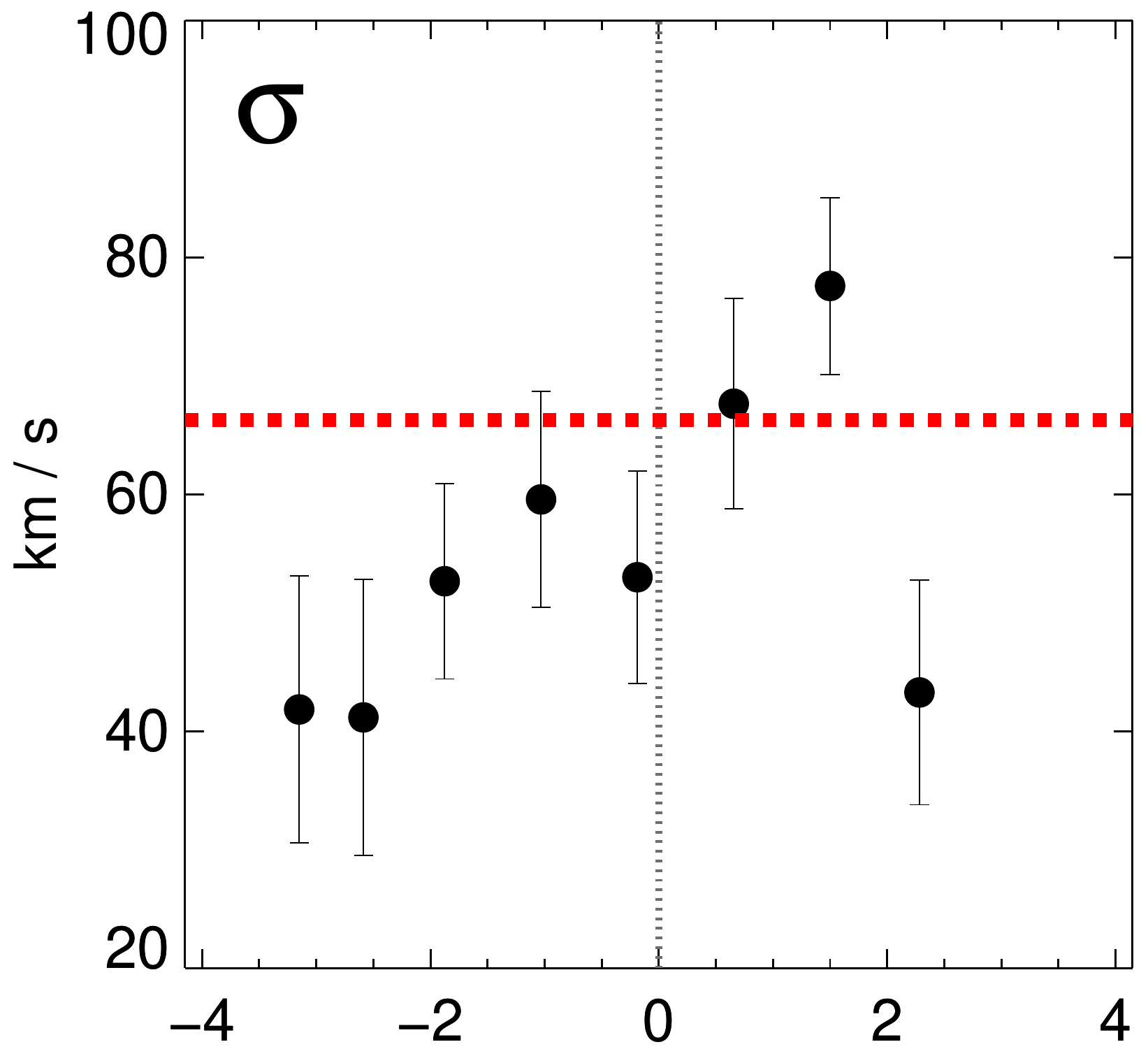}
\includegraphics[width=0.373\columnwidth]{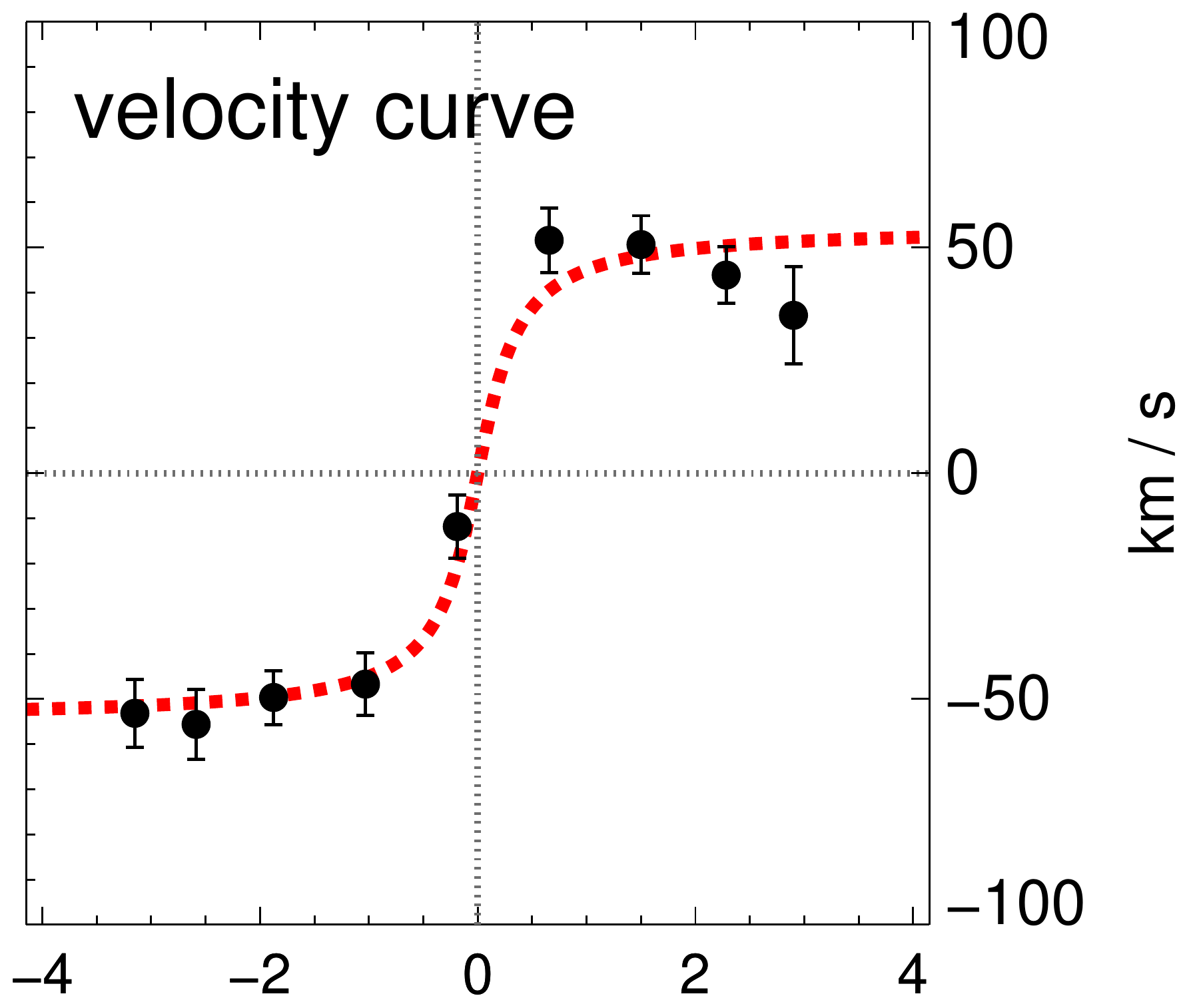}\\
\vspace{1mm}
\includegraphics[width=0.343\columnwidth]{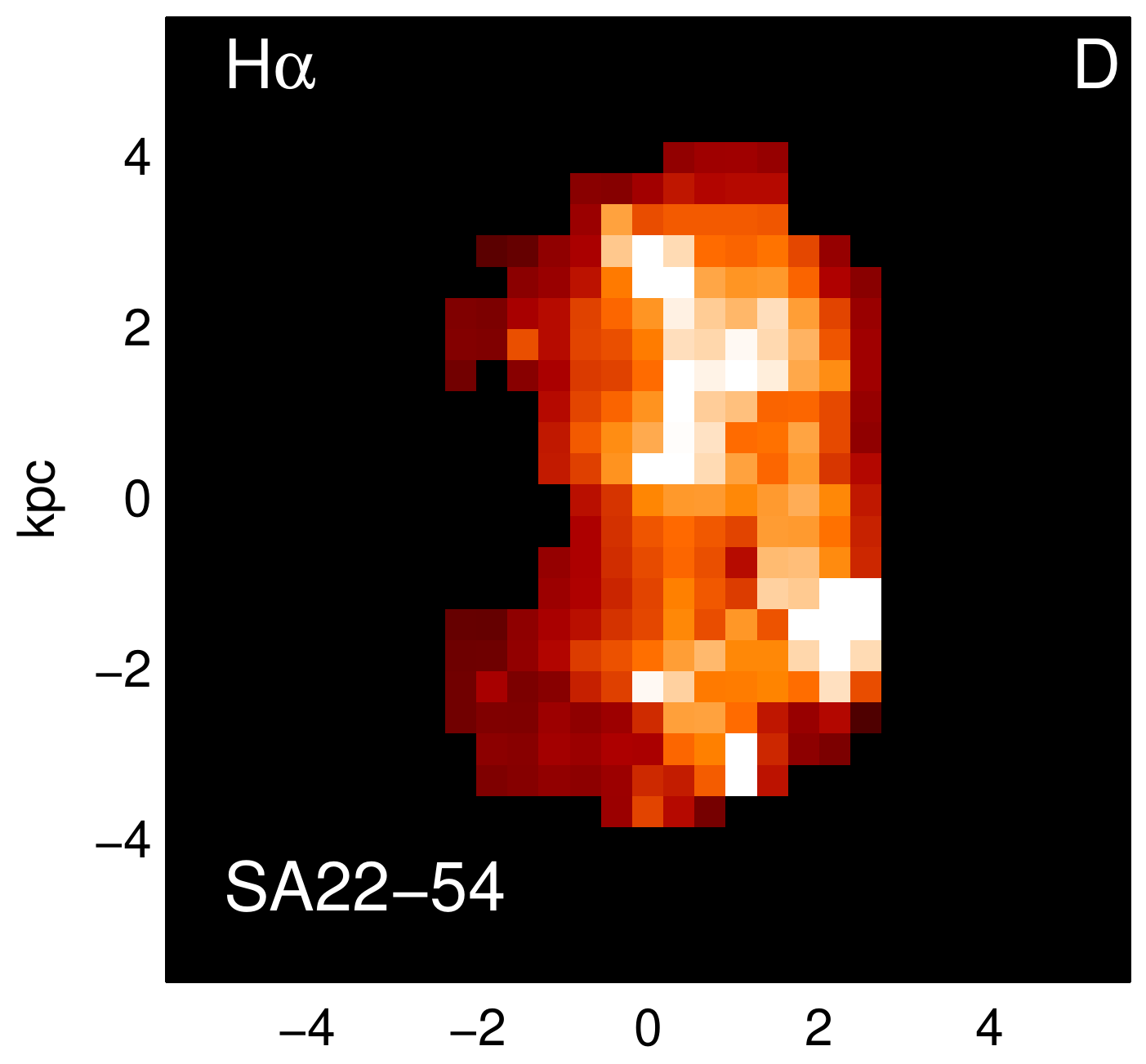}
\includegraphics[width=0.32\columnwidth]{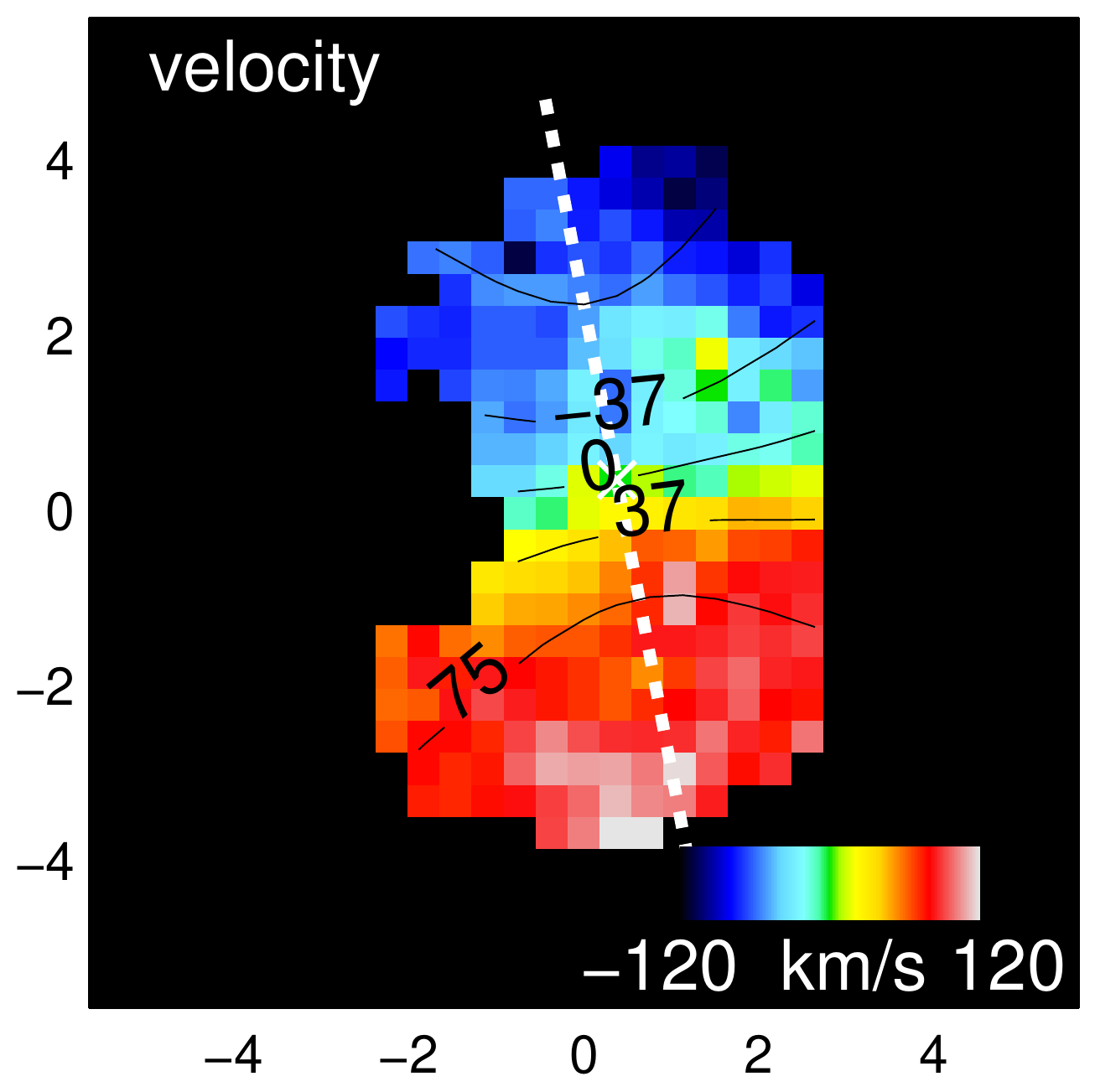}
\includegraphics[width=0.32\columnwidth]{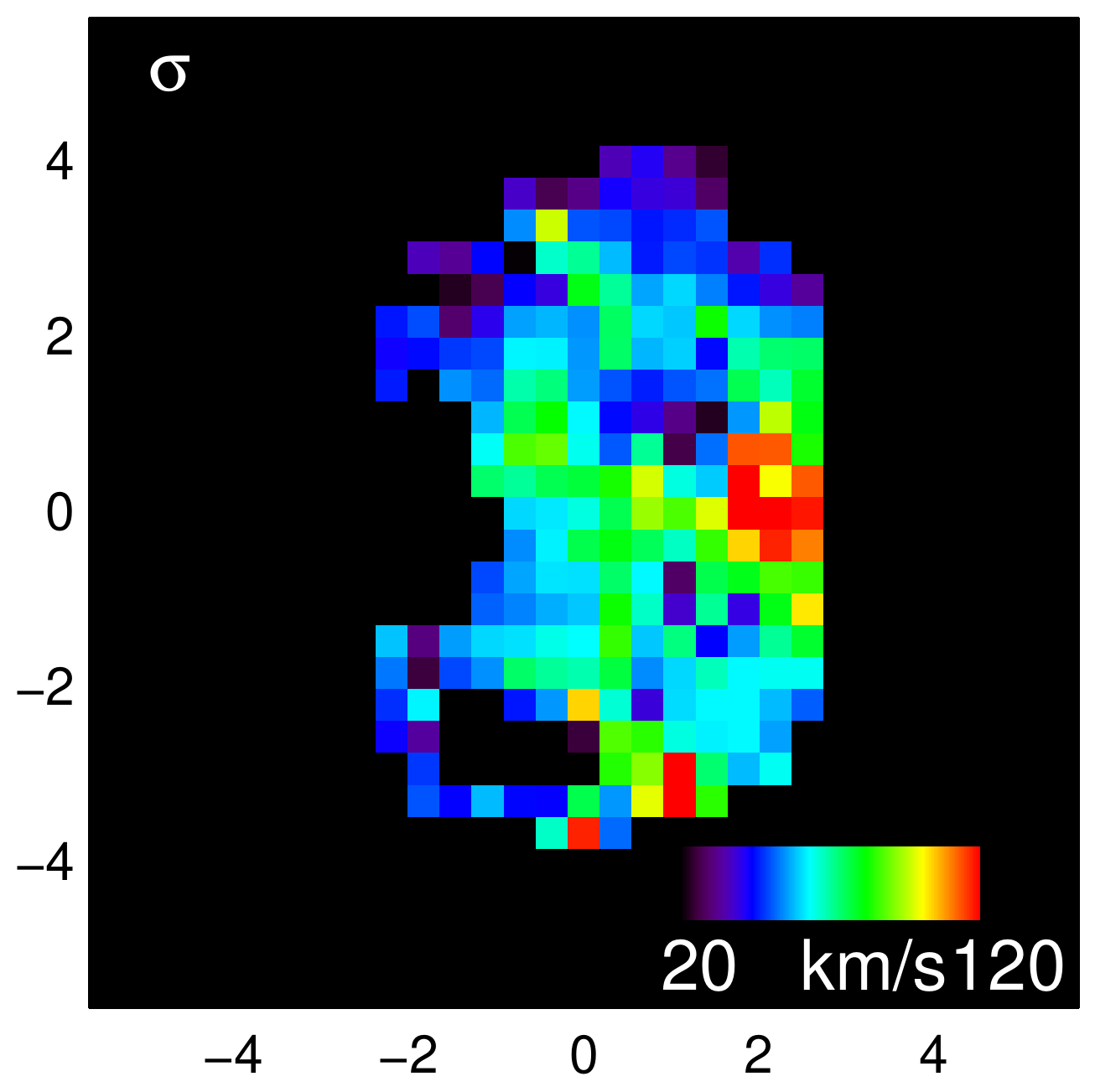}
\includegraphics[width=0.32\columnwidth]{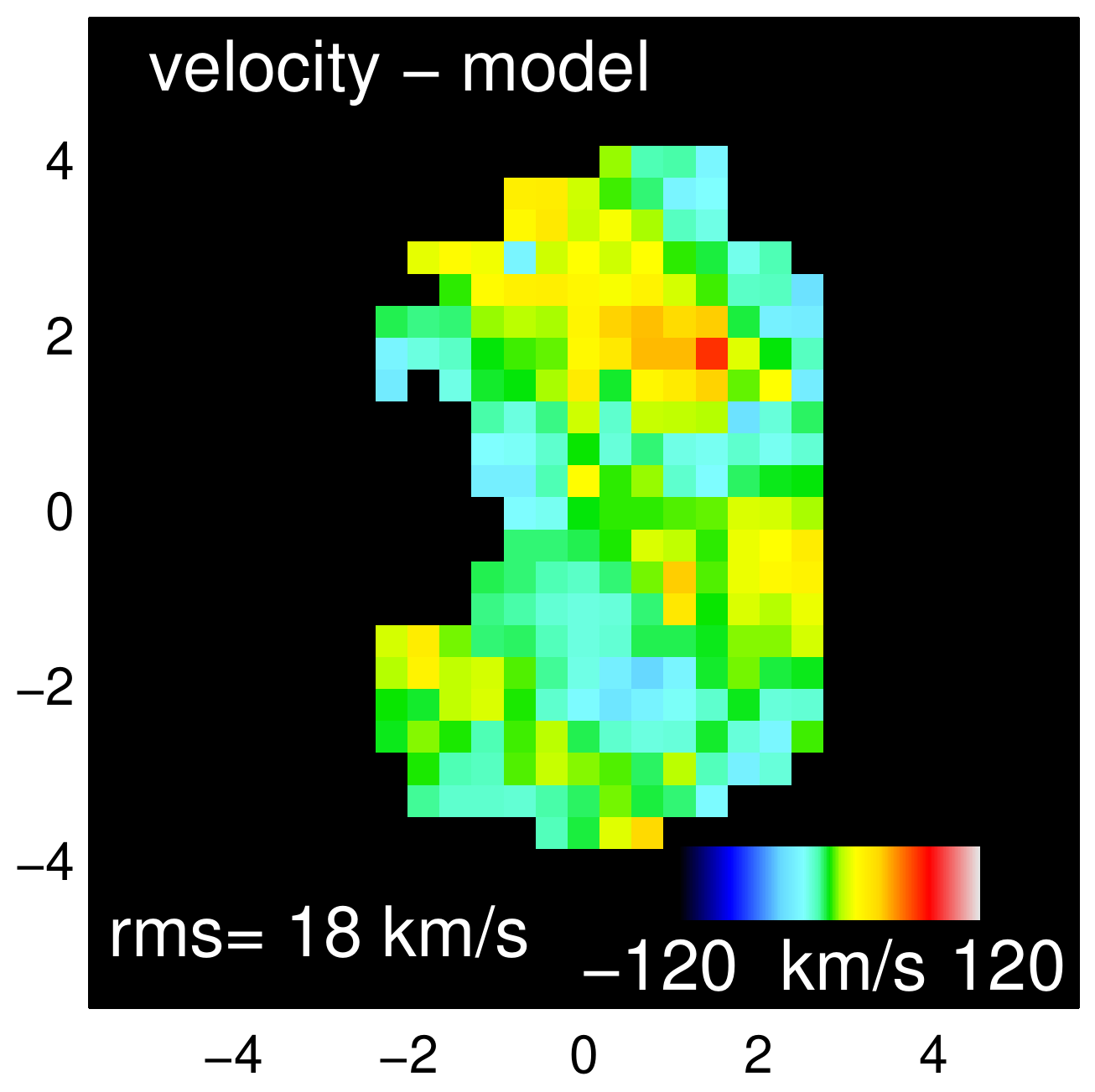}
\includegraphics[width=0.345\columnwidth]{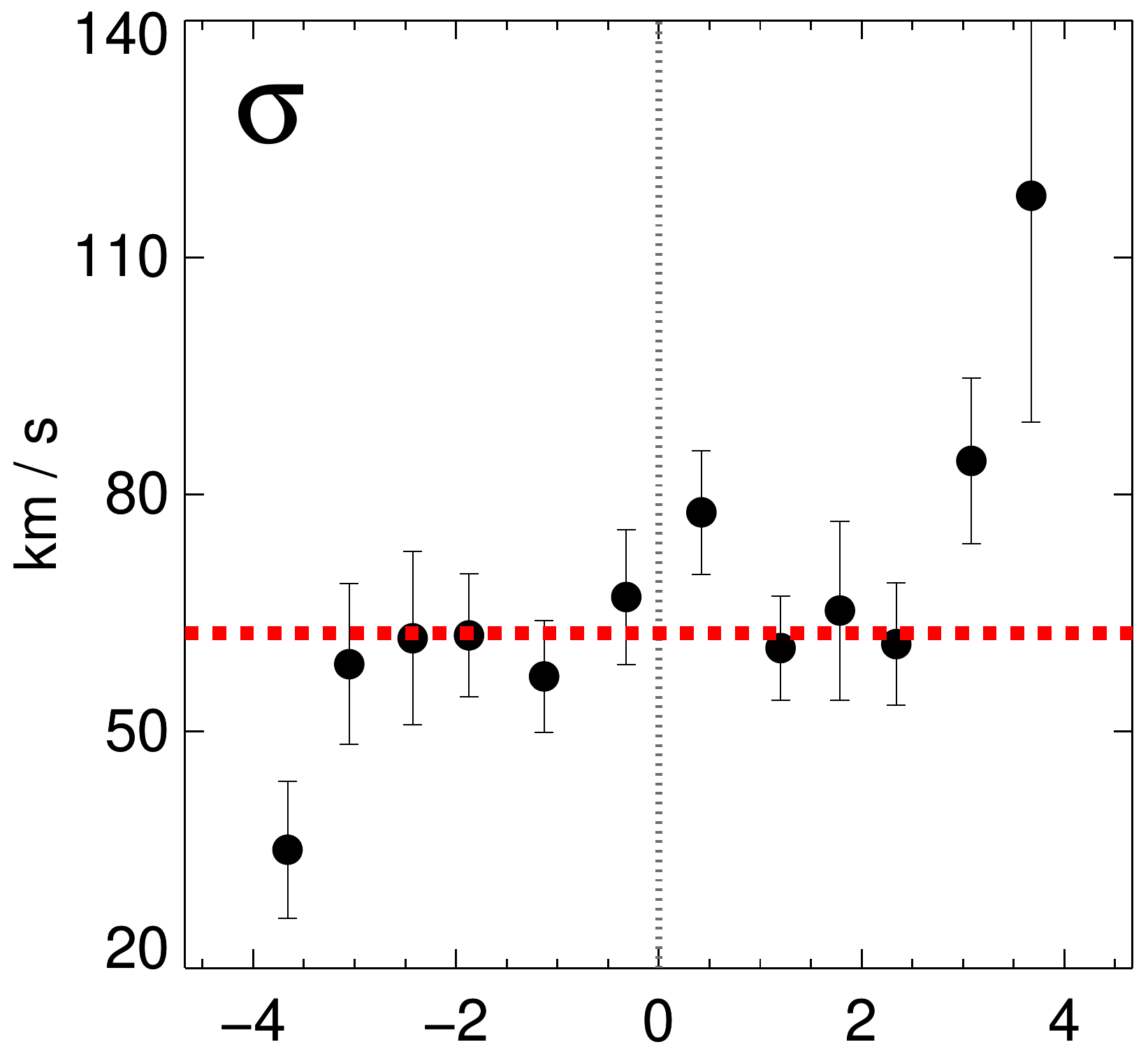}
\includegraphics[width=0.373\columnwidth]{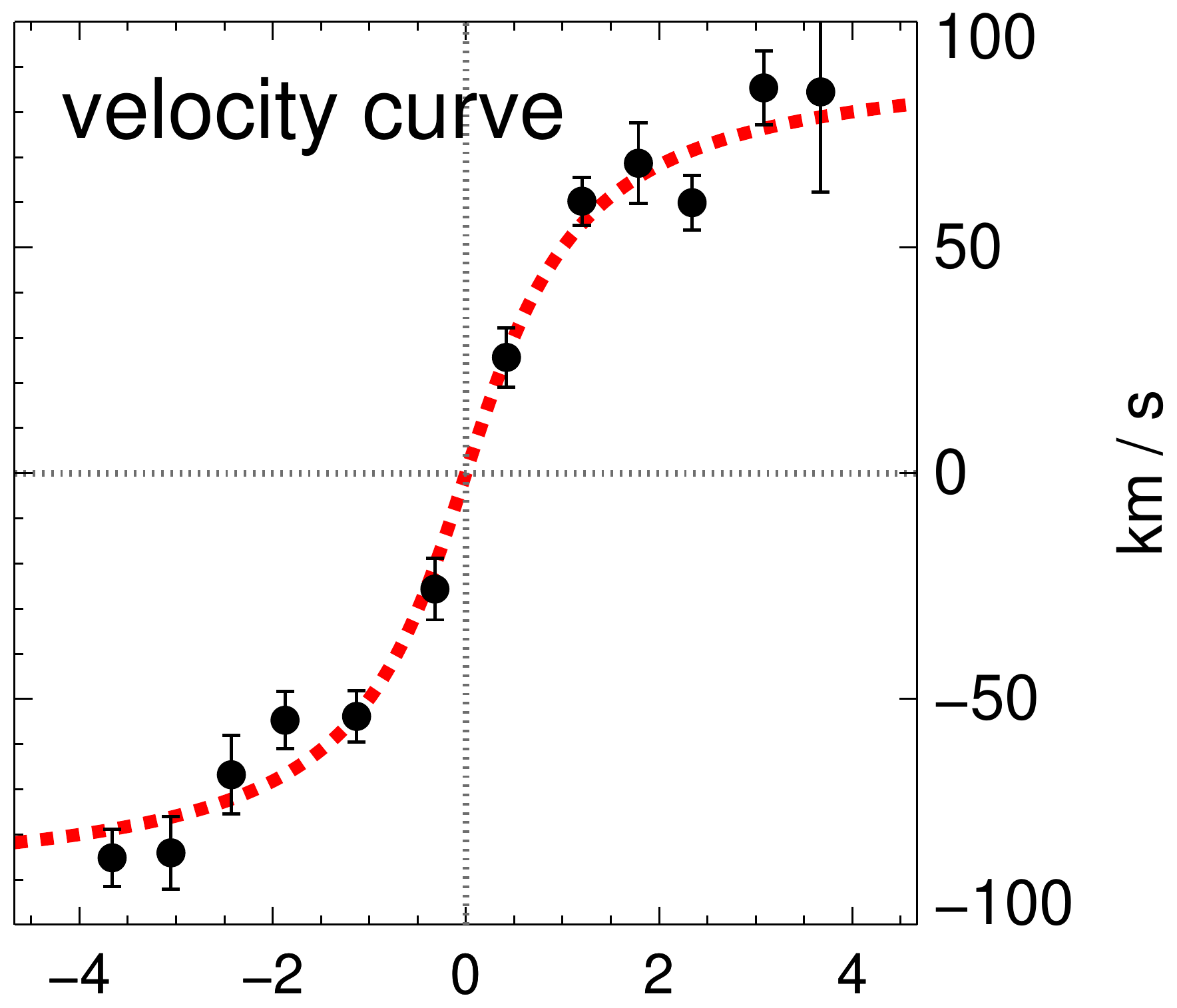}\\
\vspace{5mm}
\includegraphics[width=0.343\columnwidth]{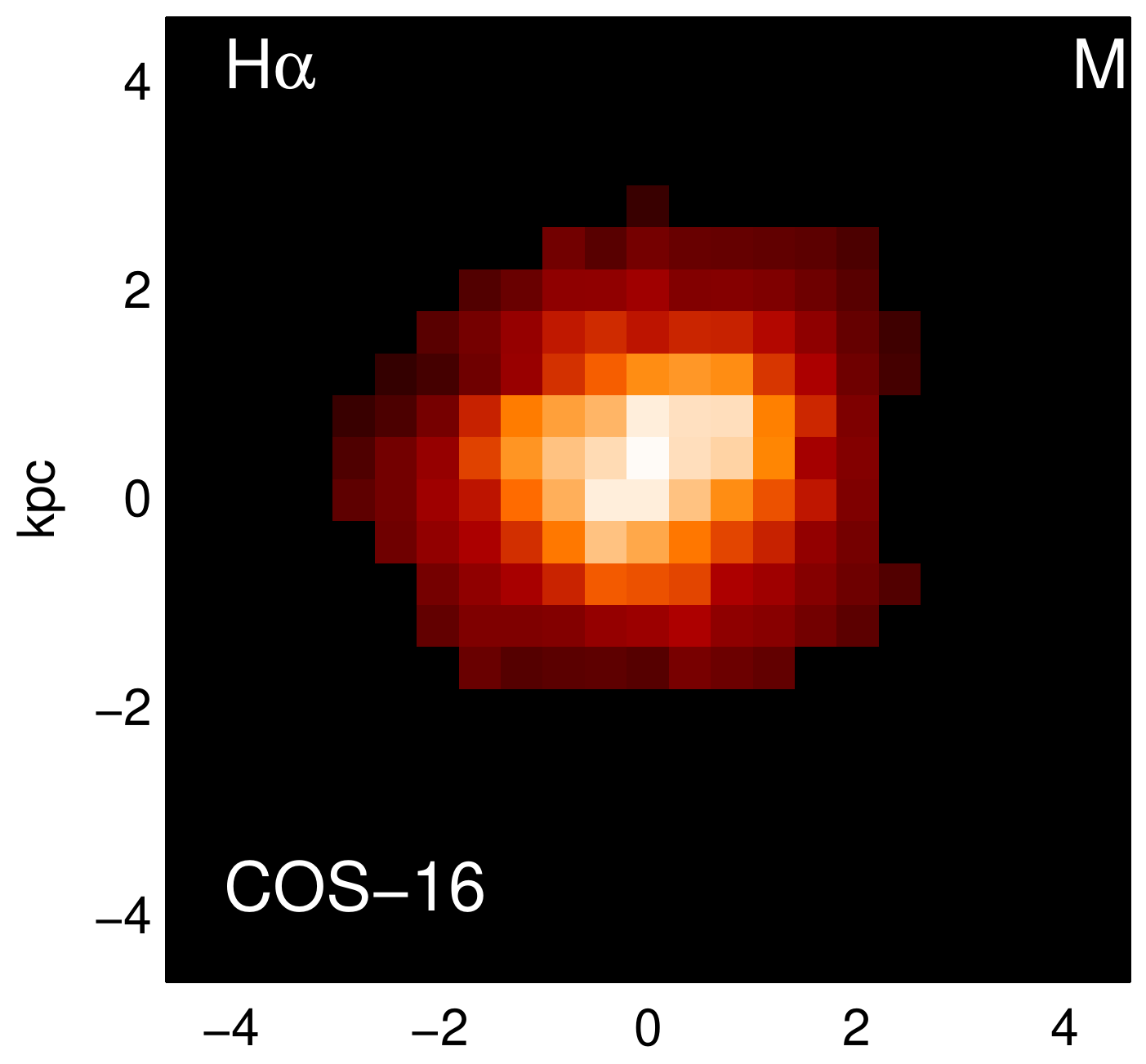}
\includegraphics[width=0.32\columnwidth]{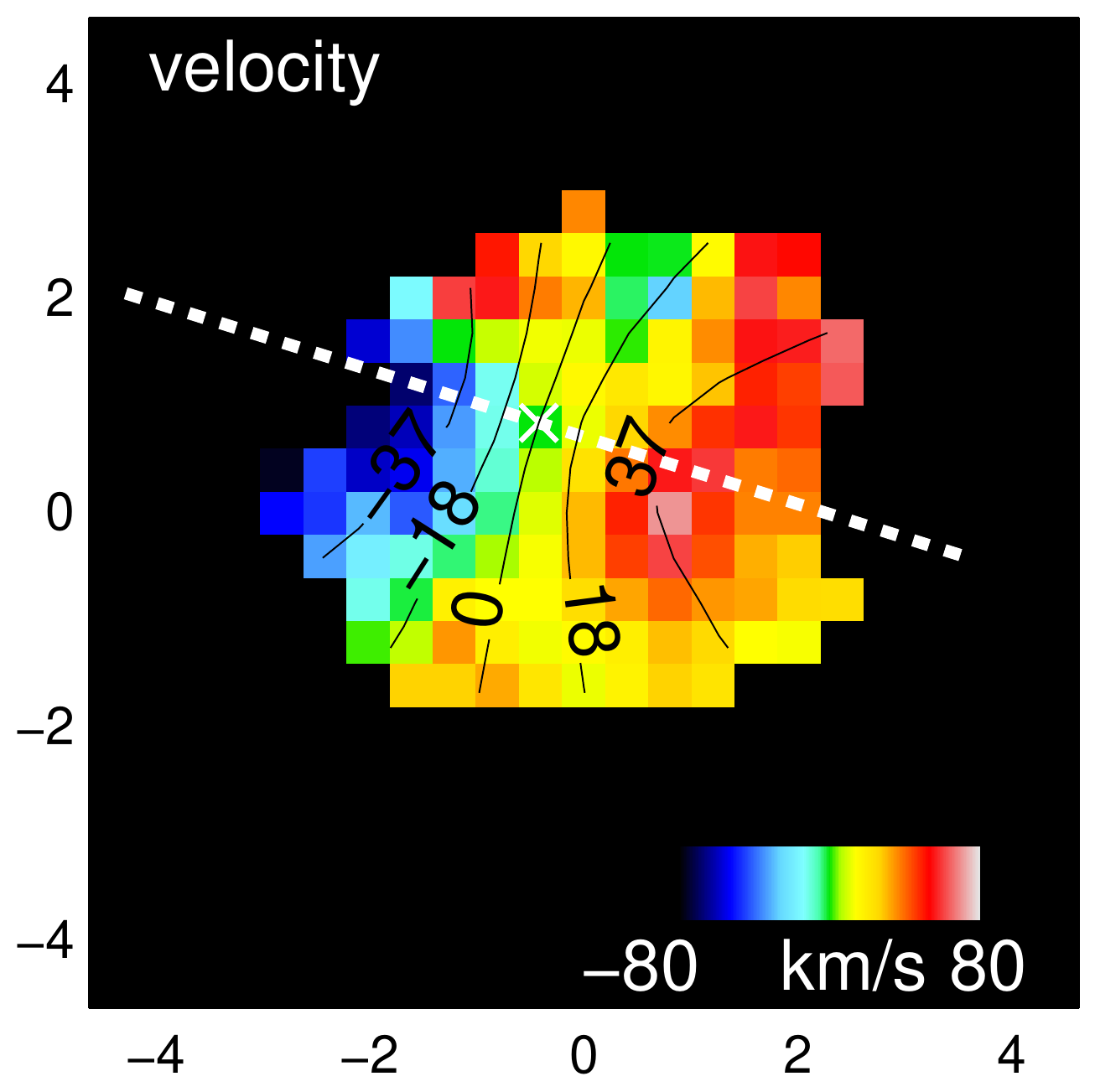}
\includegraphics[width=0.32\columnwidth]{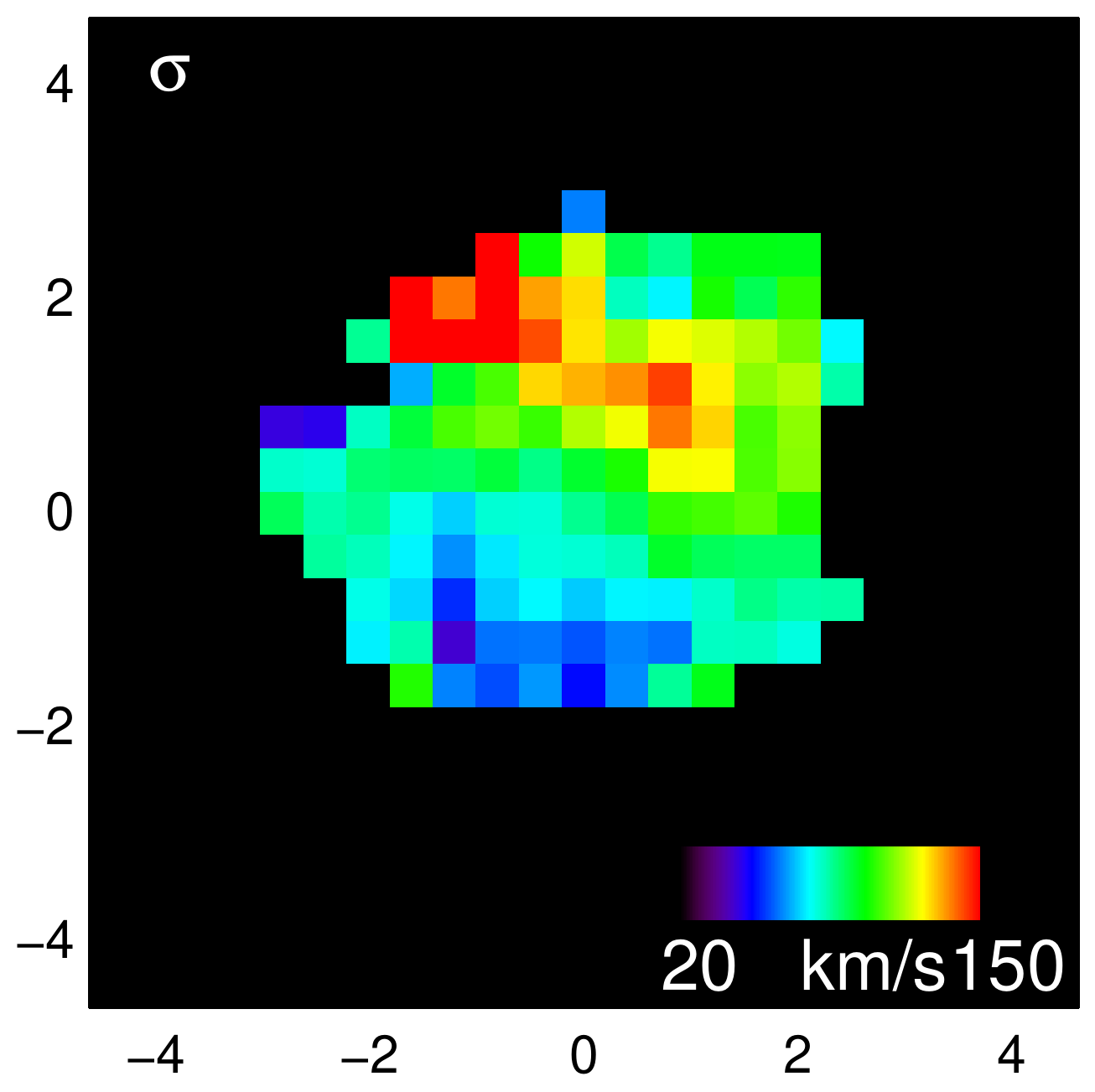}
\includegraphics[width=0.32\columnwidth]{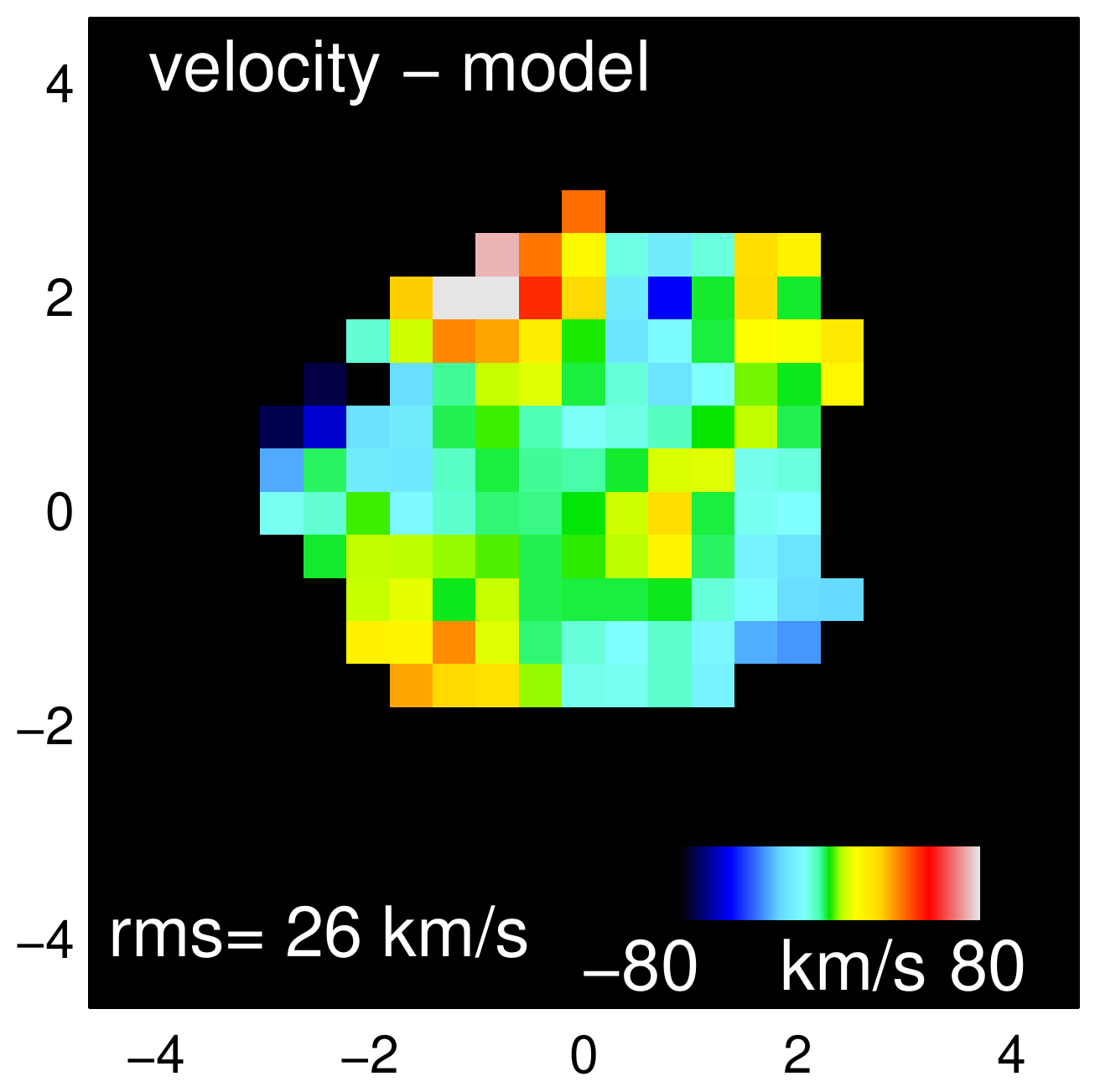}
\includegraphics[width=0.345\columnwidth]{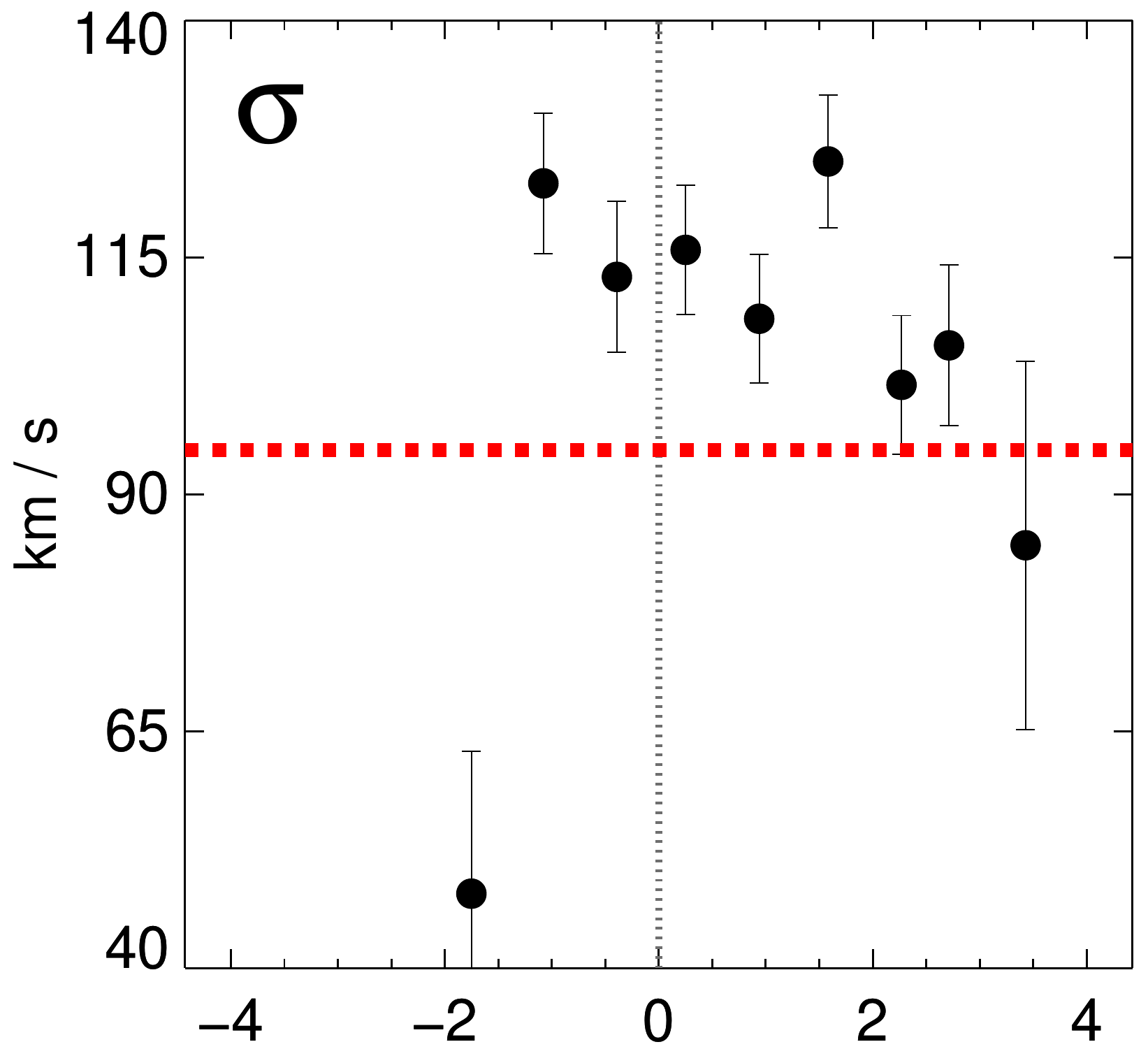}
\includegraphics[width=0.373\columnwidth]{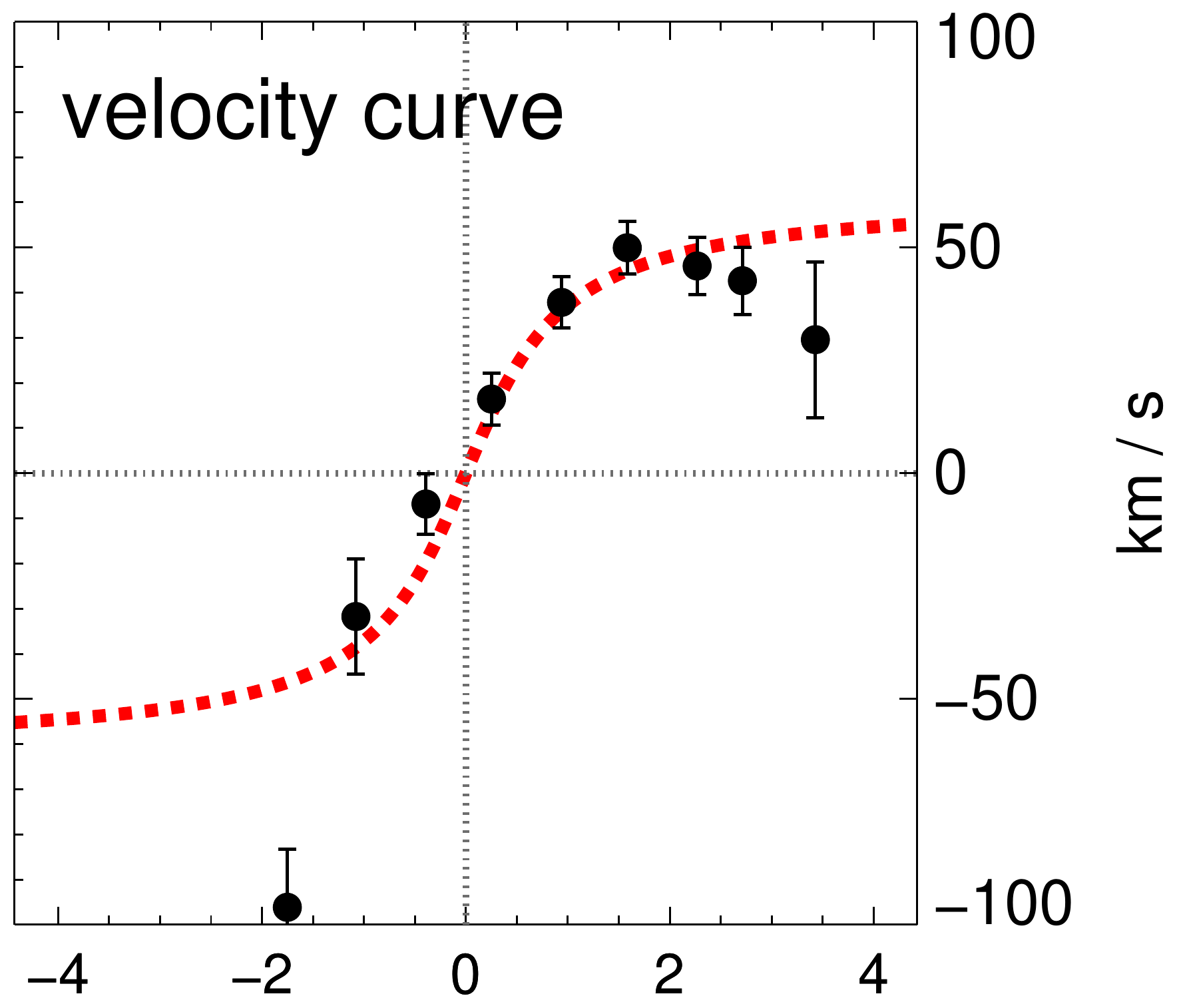}\\
\vspace{1mm}
\includegraphics[width=0.343\columnwidth]{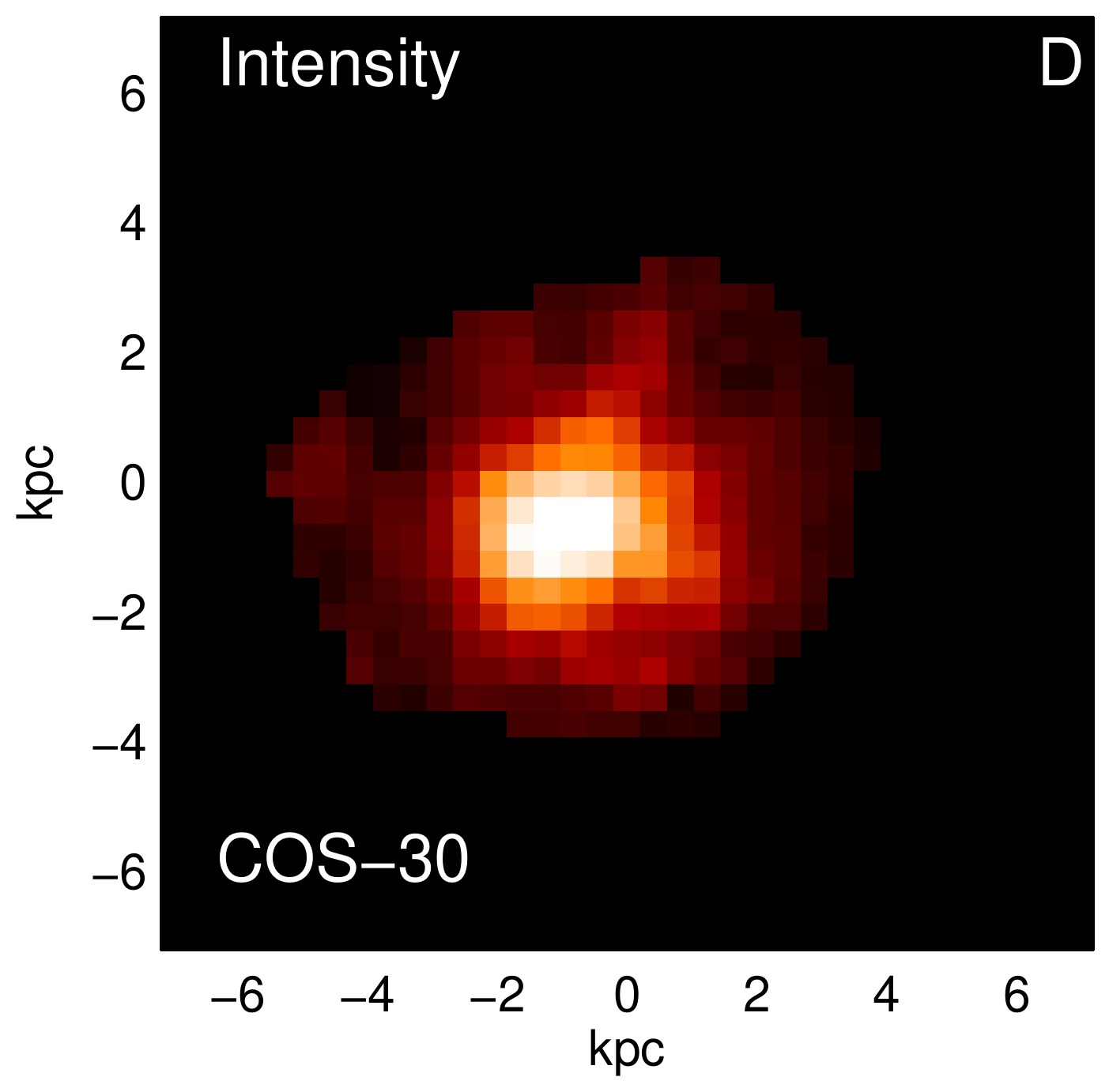}
\includegraphics[width=0.32\columnwidth]{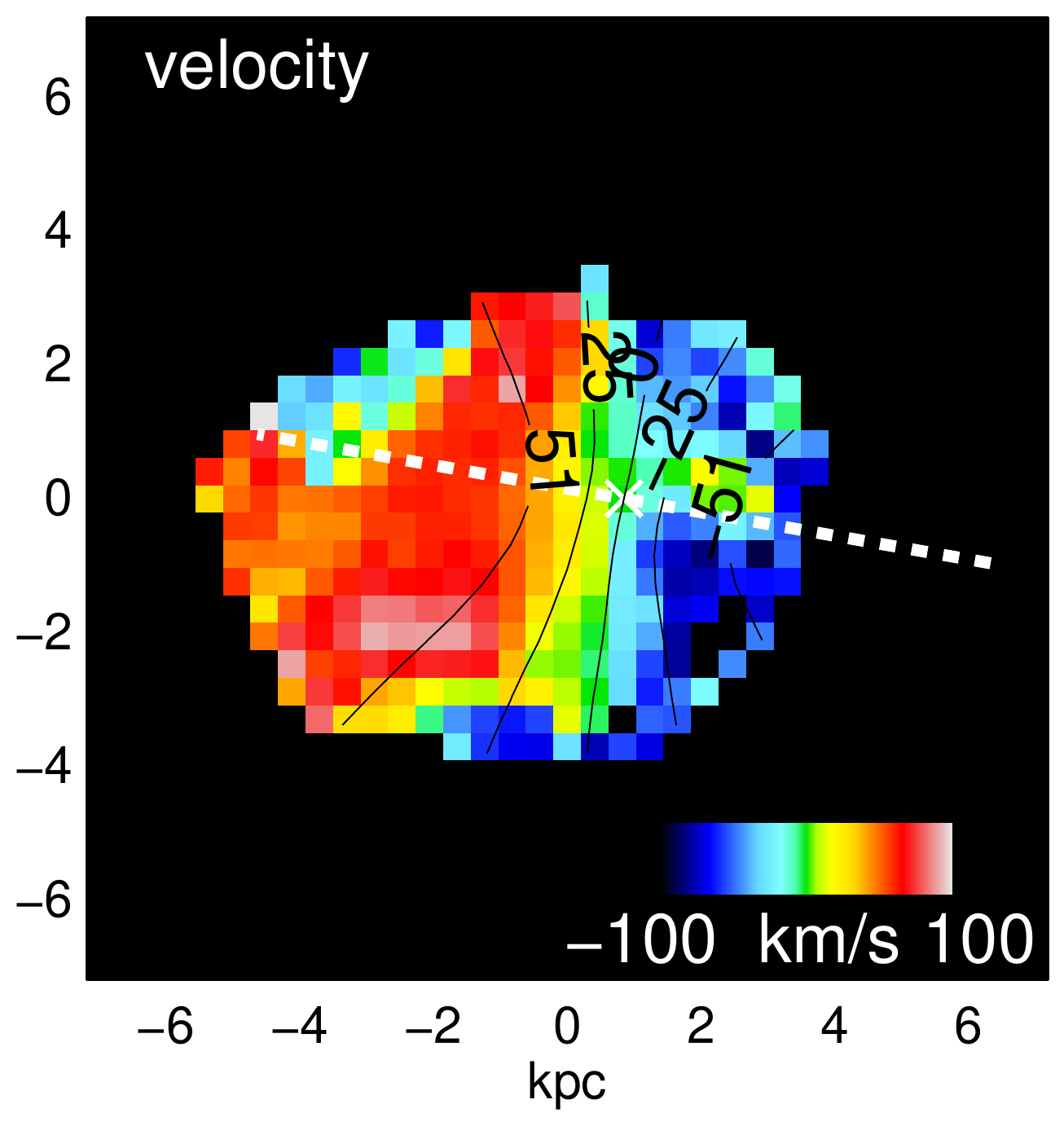}
\includegraphics[width=0.32\columnwidth]{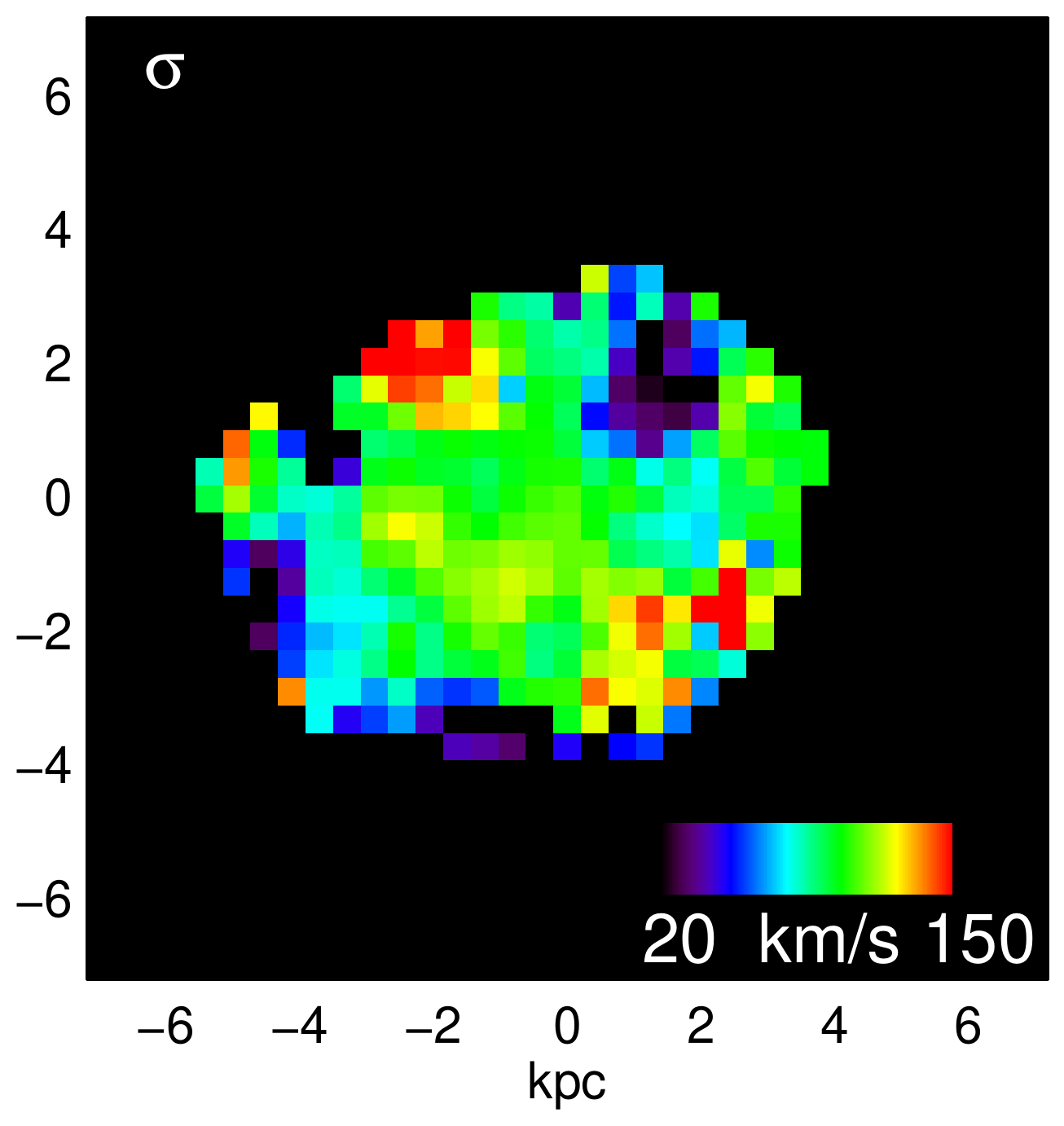}
\includegraphics[width=0.32\columnwidth]{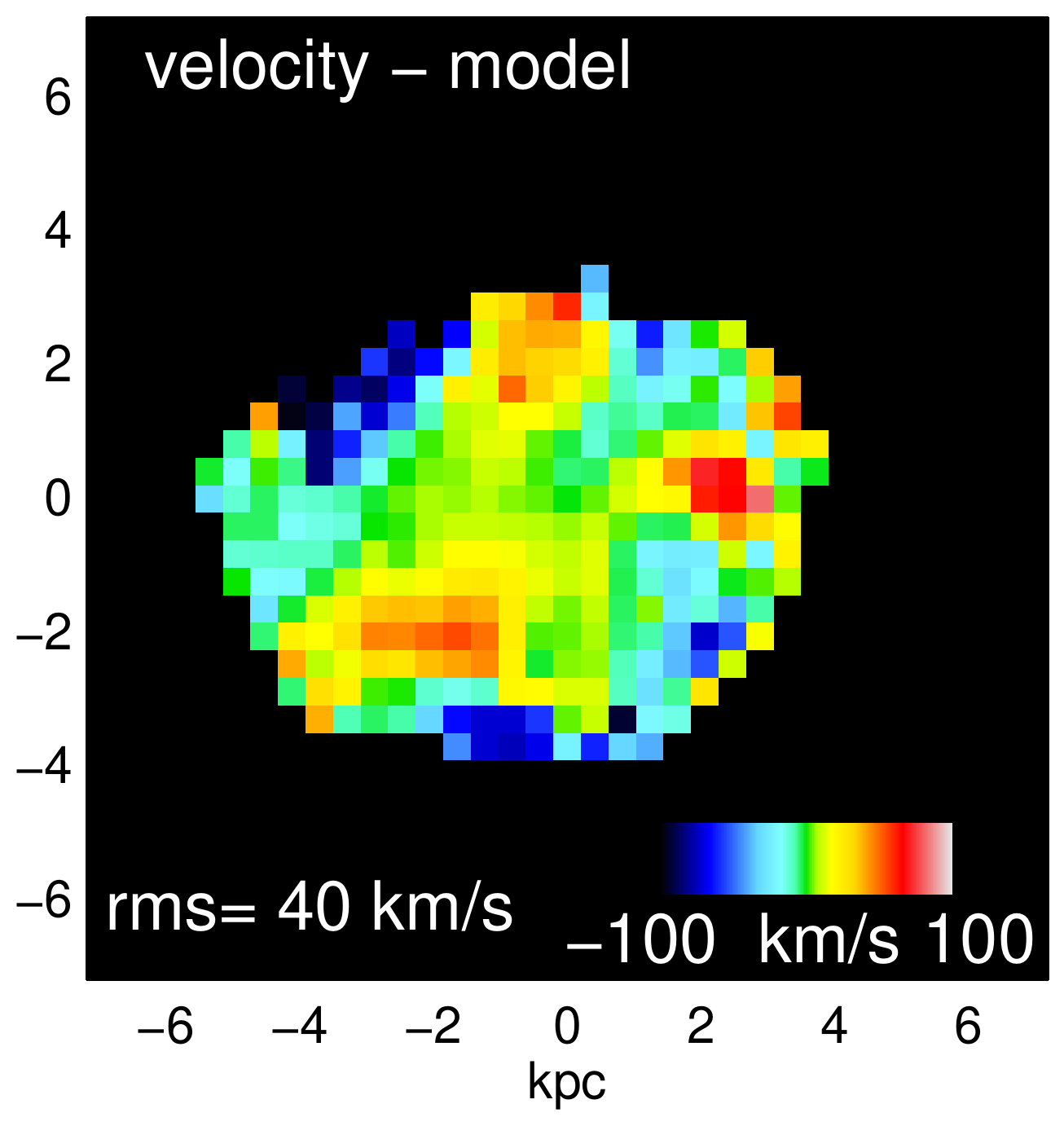}
\includegraphics[width=0.345\columnwidth]{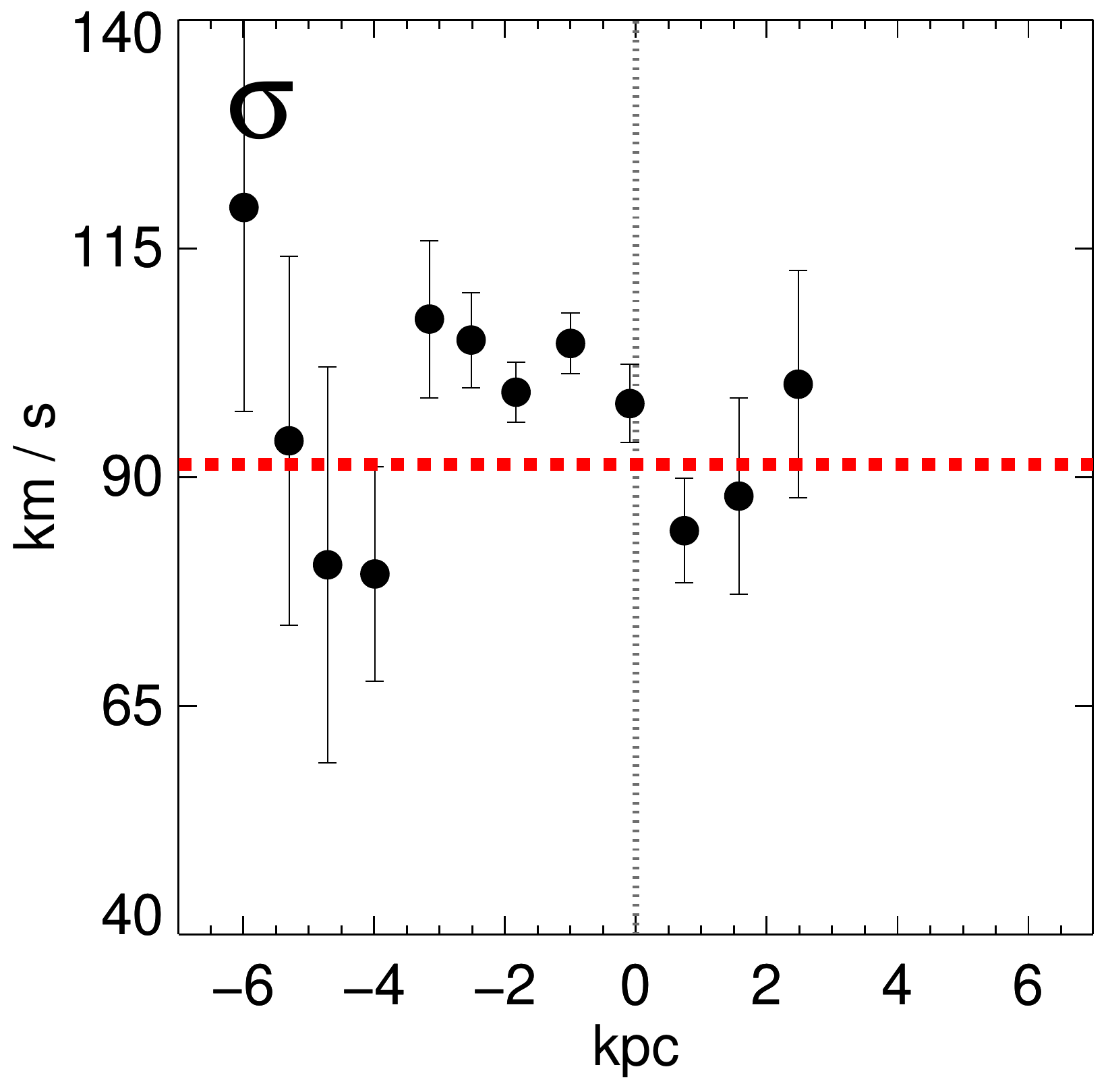}
\includegraphics[width=0.373\columnwidth]{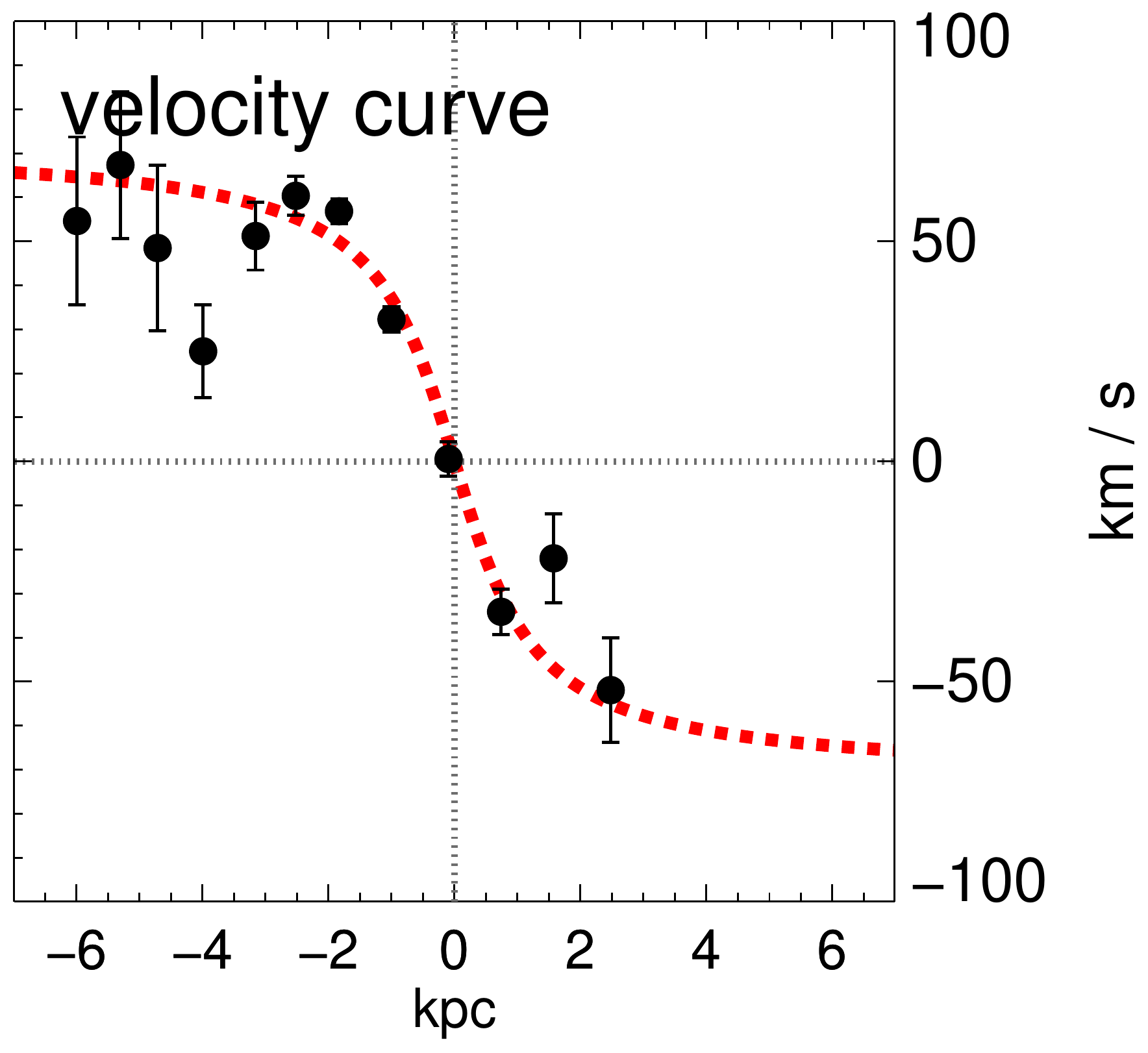}\\
\caption{\label{fig:maps}
H$\alpha$ intensity, velocity, line-of-sight velocity dispersion ($\sigma$), residual fields, one
dimensional velocity dispersion profile and one dimensional velocity profile (columns) 
for eleven galaxies from our observed sample (rows).
The H$\alpha$ intensity map also shows the classification done by kinemetry analysis 
(see \S\ref{sec:analysis}): six galaxies were classified as disks (D), four as mergers (M) and one as 
unresolved/compact (C). The unresolved/compact source (SA22-01) have no modelling. The velocity field has overplotted the kinematical 
centre, the mayor kinematic axis and velocity contours of the best-fit two dimensional kinematical disk model. 
The line-of-sight velocity dispersion ($\sigma$) field is corrected for the local 
velocity gradient ($\Delta$V/$\Delta$R) across the PSF. The residual map is constructed by subtracting 
the best-fit kinematic model from the velocity map: the root-mean-square (r.m.s) of these residuals are given in each
panel. The one dimensional velocity profiles are derived from the two dimensional velocity field using 
the best-fit kinematical parameters and a slit width of $\sim$1 kpc across the major kinematic axis. 
The error bars shows the 1$\sigma$ uncertainty. In the velocity dispersion profile plots, the red-dashed line shows 
the mean galactic velocity dispersion value. The dotted grey line represents the best-fit dynamical centre (Table~\ref{tab:table2}). 
In the last column, the red-dashed line show the best one-dimensional fit using an arctan model for each source. 
The dotted vertical and horizontal grey lines represent the best-fit dynamical centre and the zero velocity point respectively.}
\end{figure*}

\begin{figure*}
\flushleft
\includegraphics[width=0.343\columnwidth]{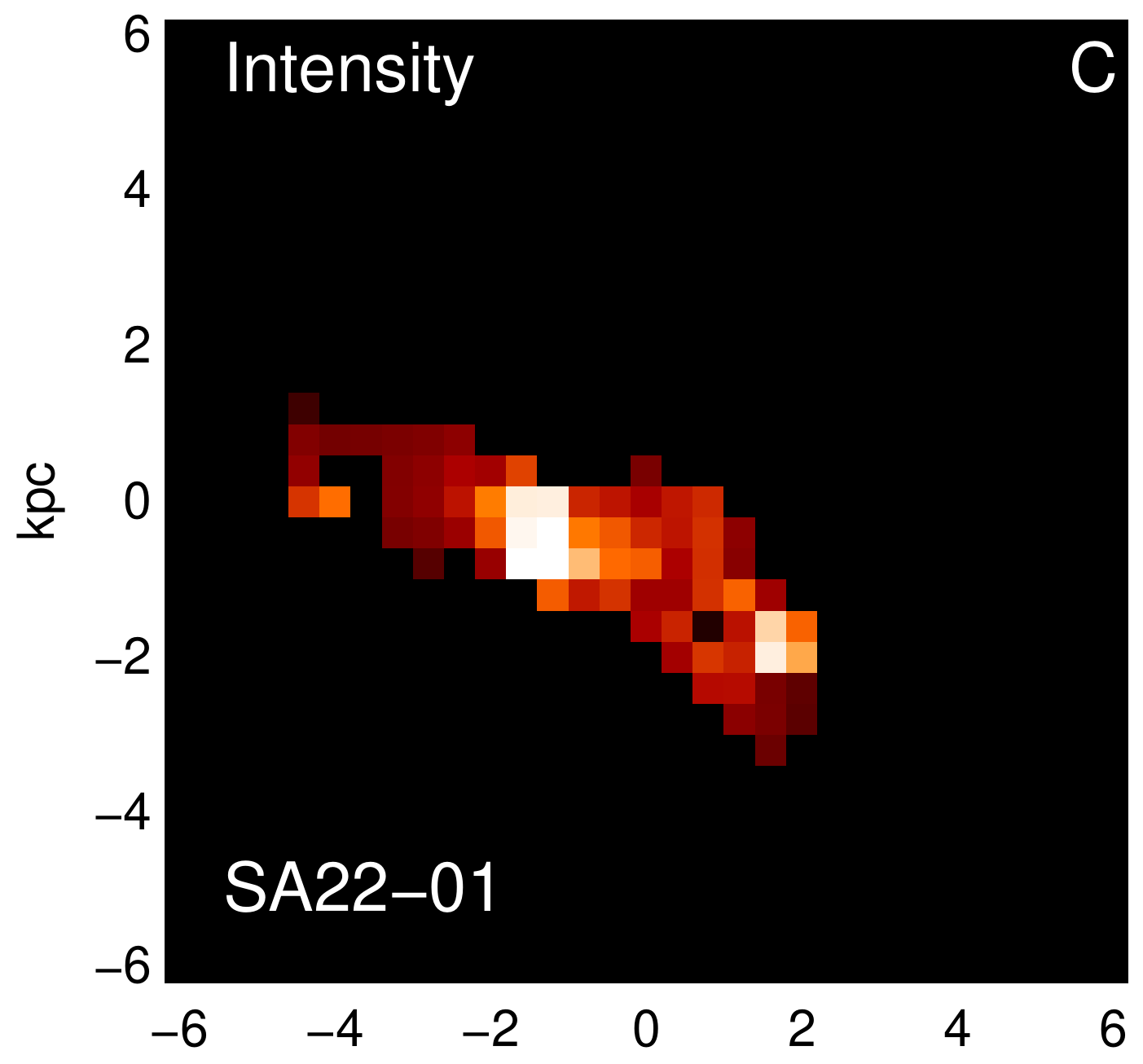}
\includegraphics[width=0.32\columnwidth]{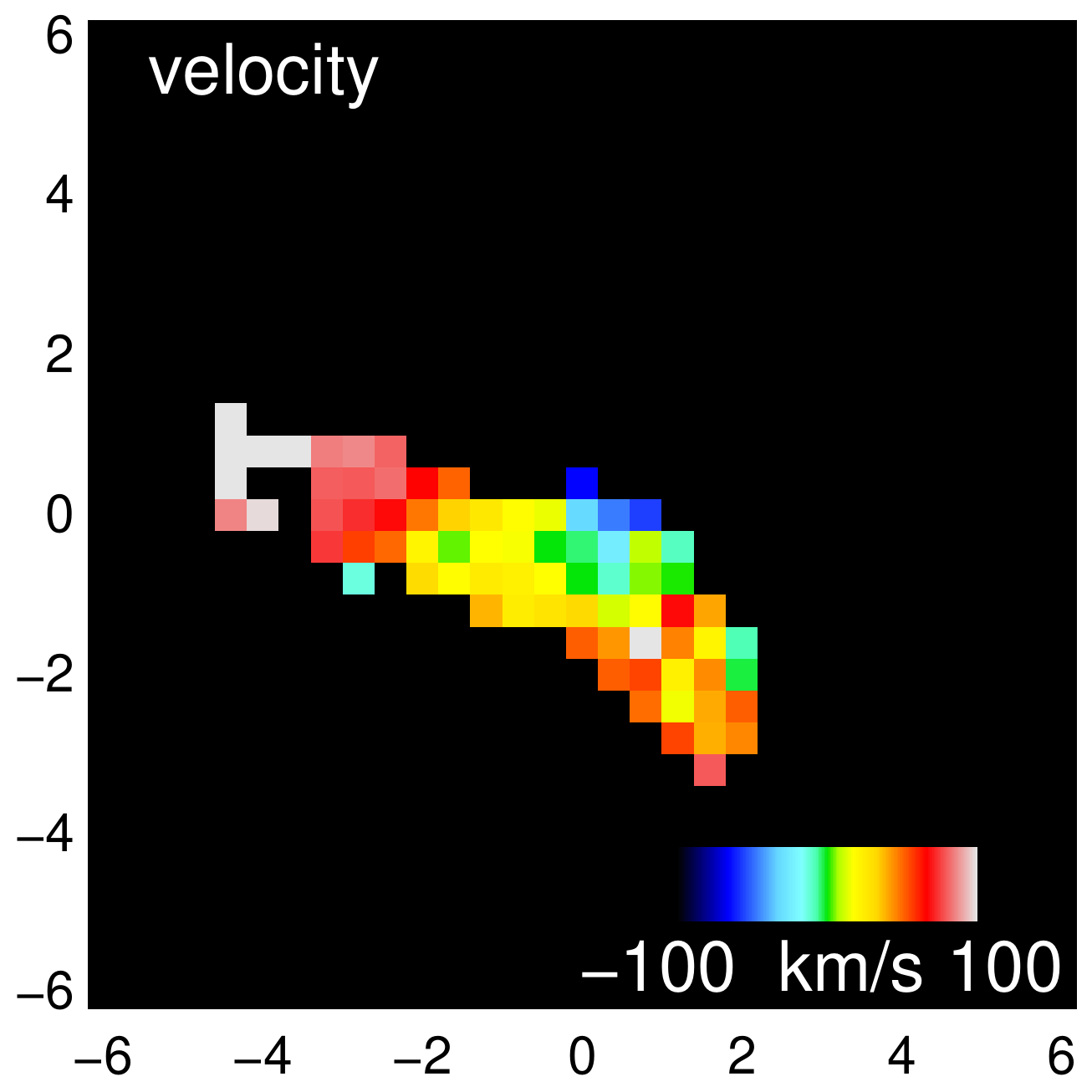}
\includegraphics[width=0.32\columnwidth]{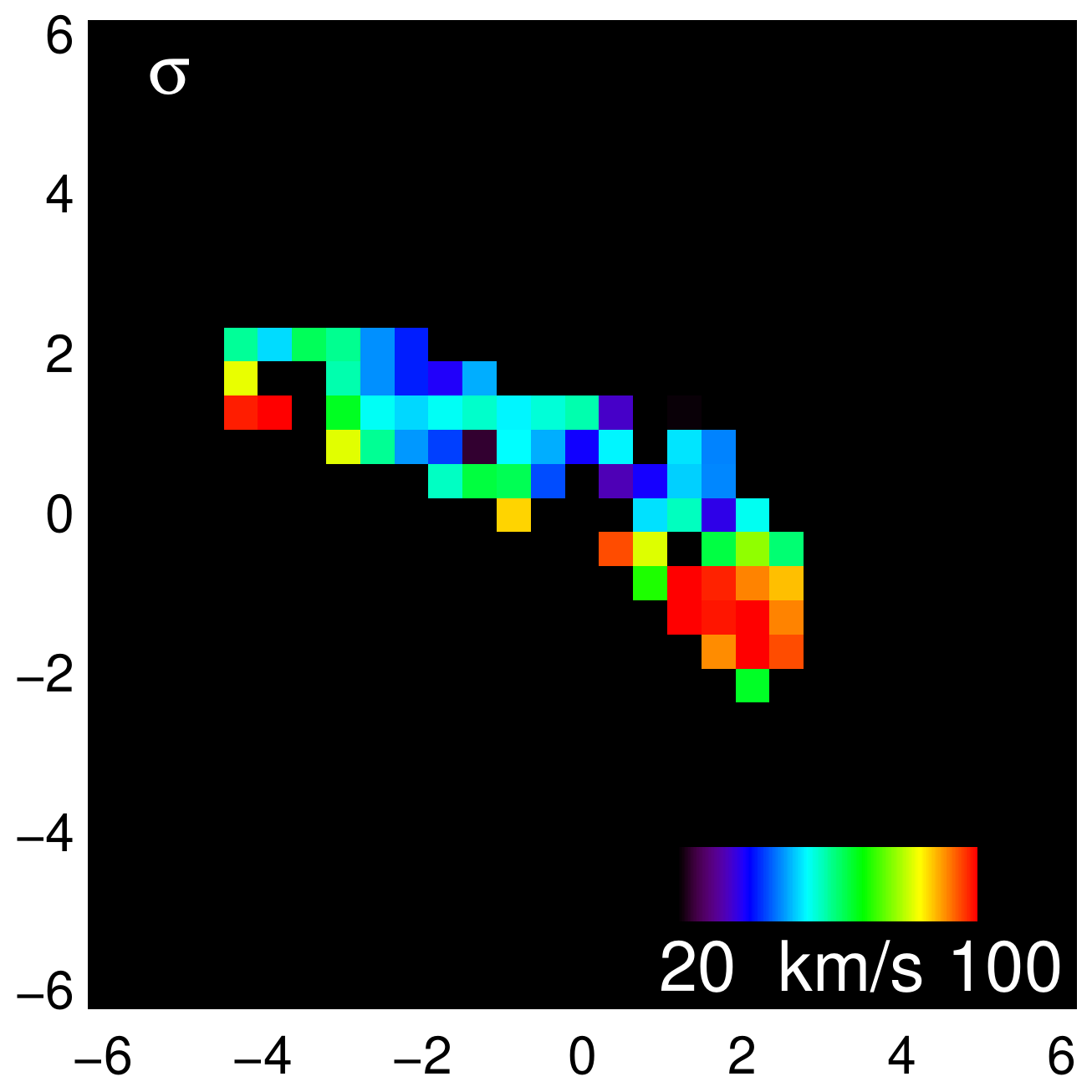}\\
\includegraphics[width=0.343\columnwidth]{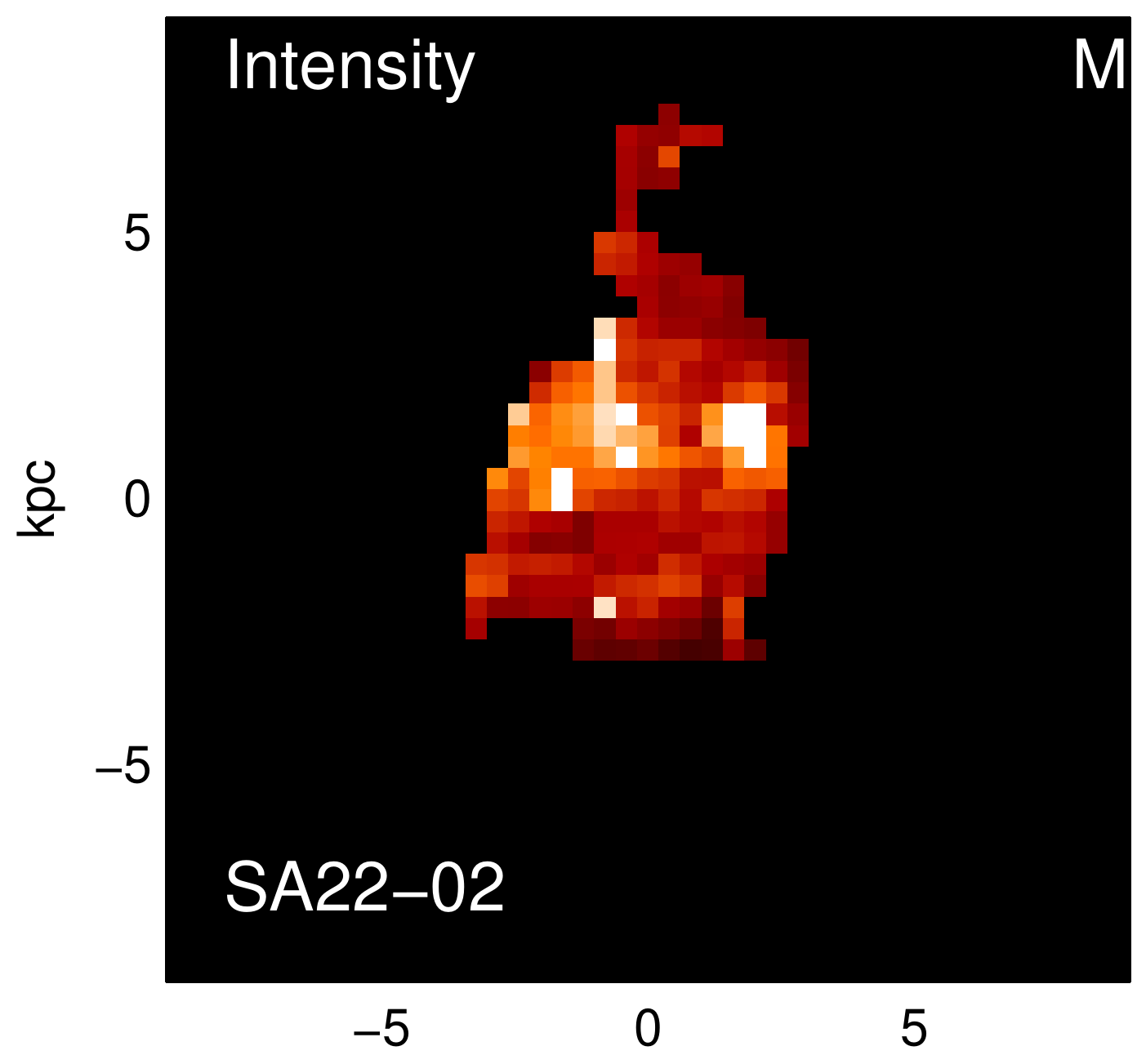}
\includegraphics[width=0.32\columnwidth]{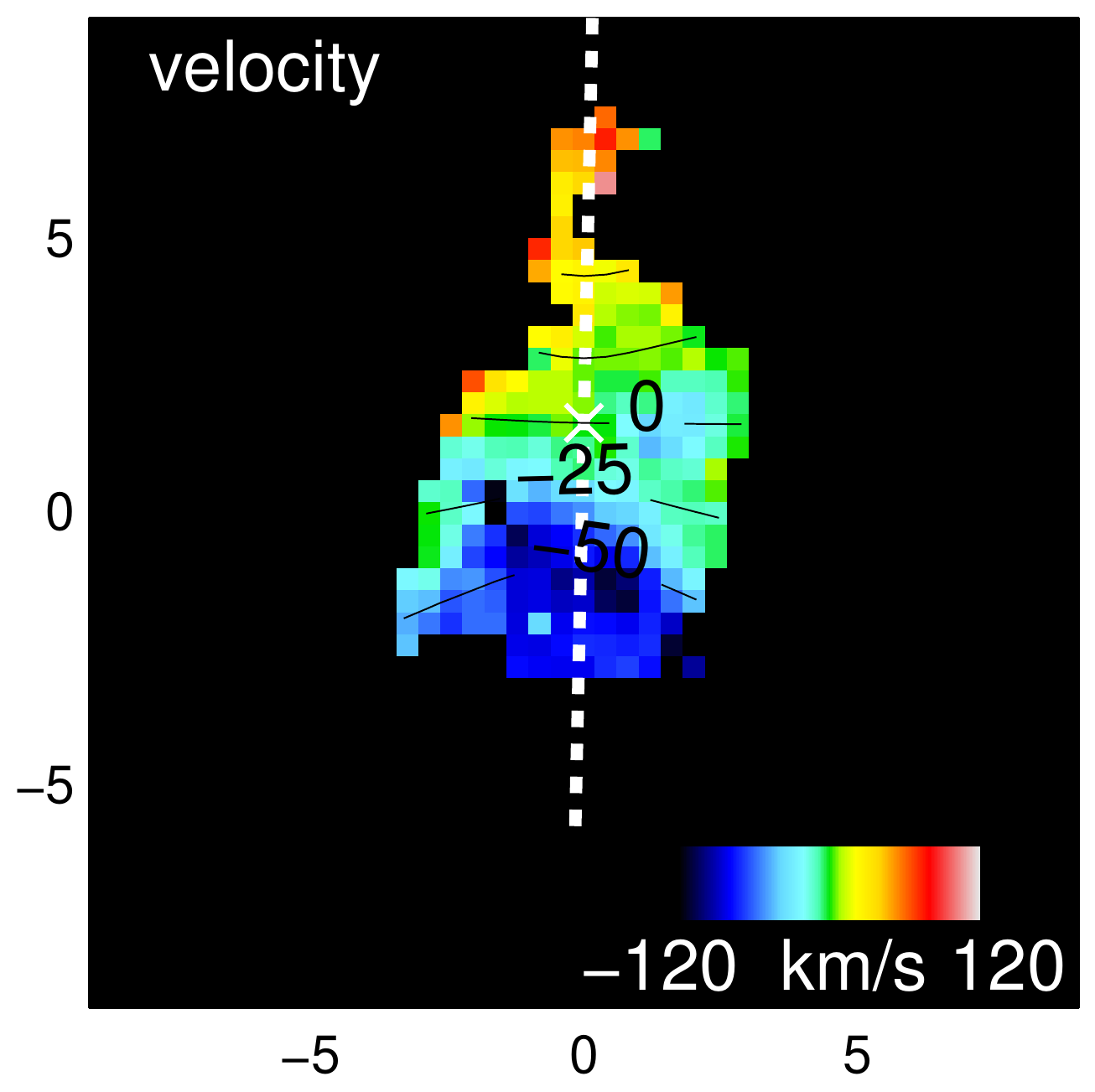}
\includegraphics[width=0.32\columnwidth]{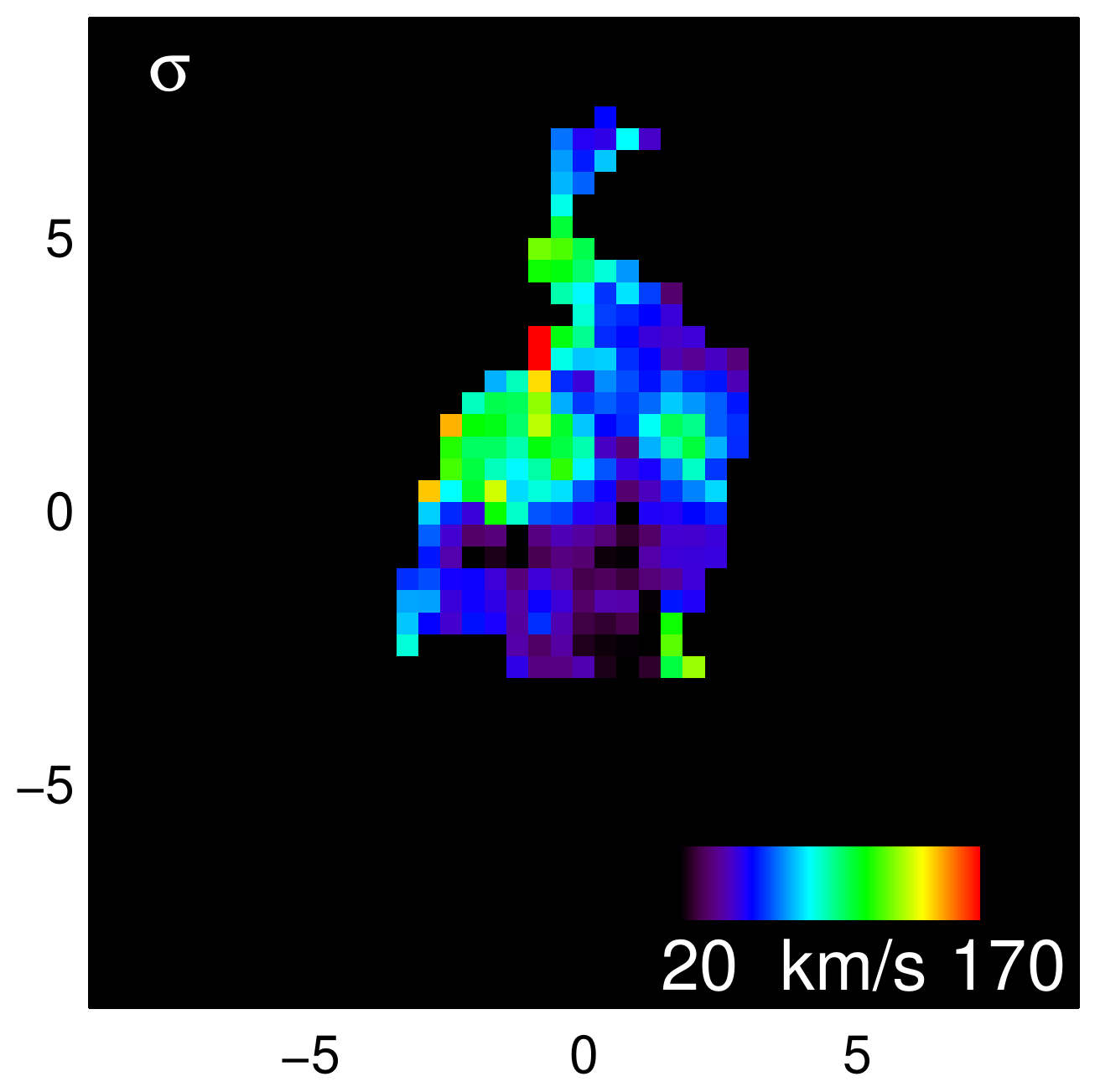}
\includegraphics[width=0.32\columnwidth]{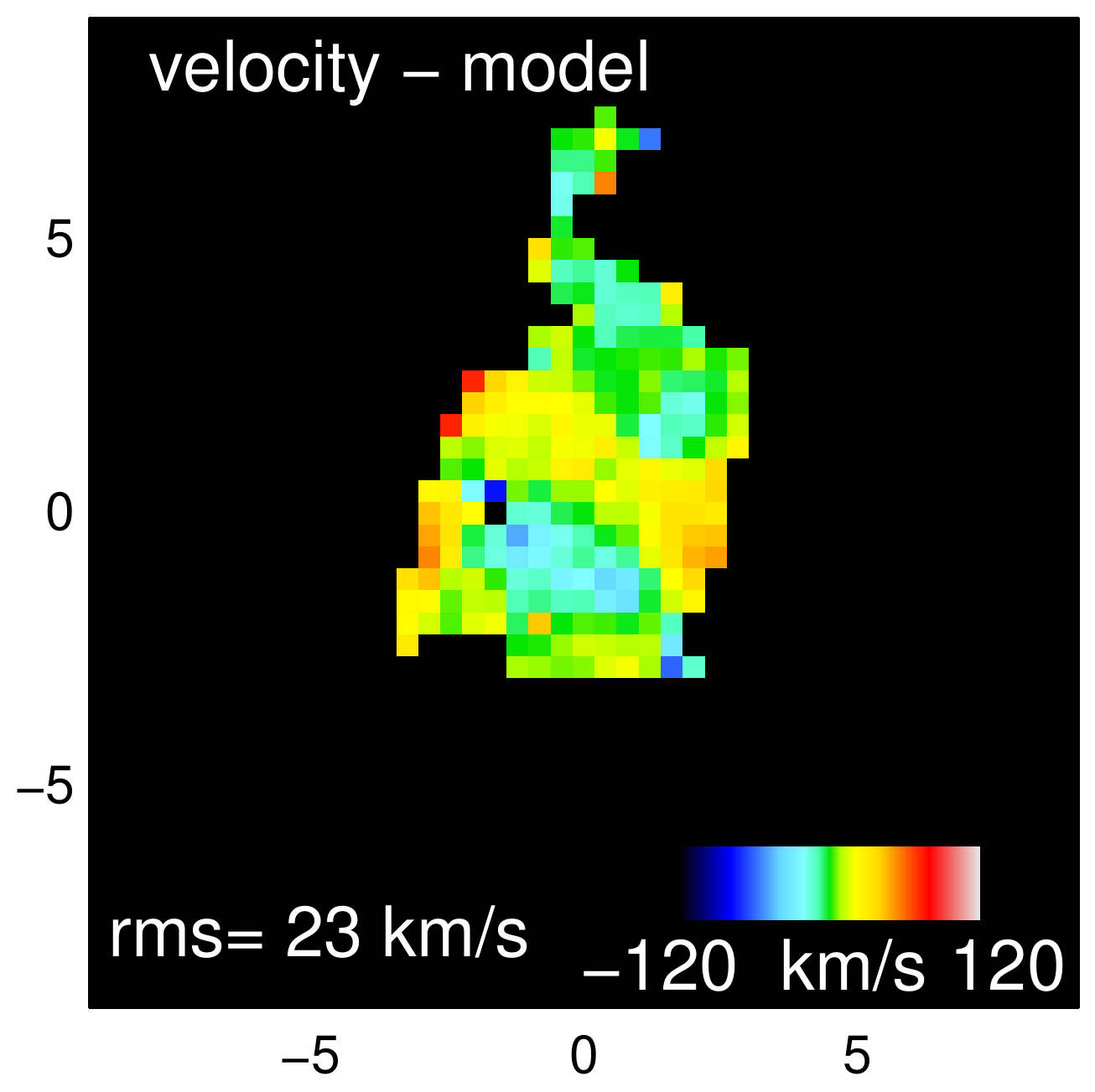}
\includegraphics[width=0.345\columnwidth]{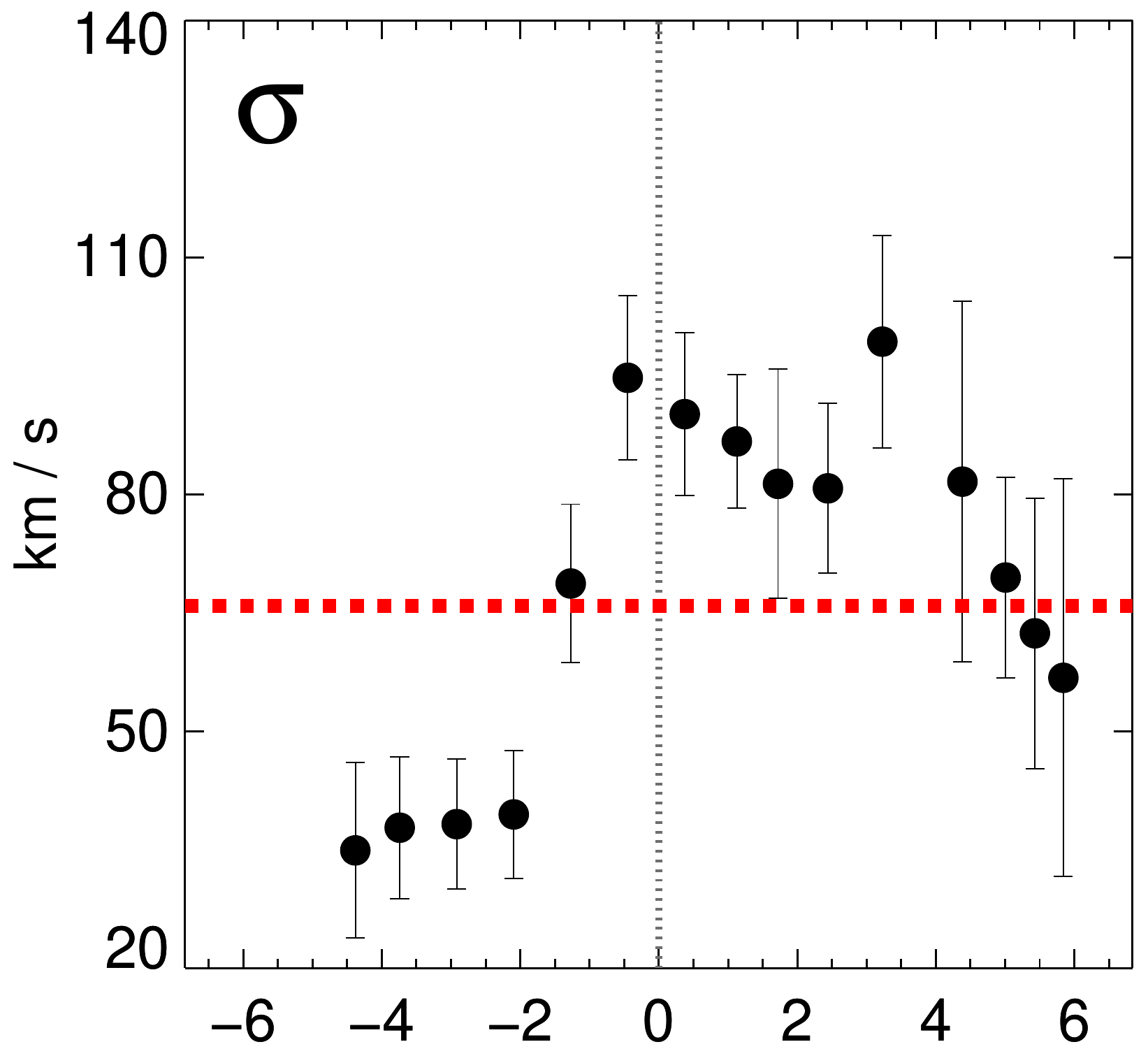}
\includegraphics[width=0.373\columnwidth]{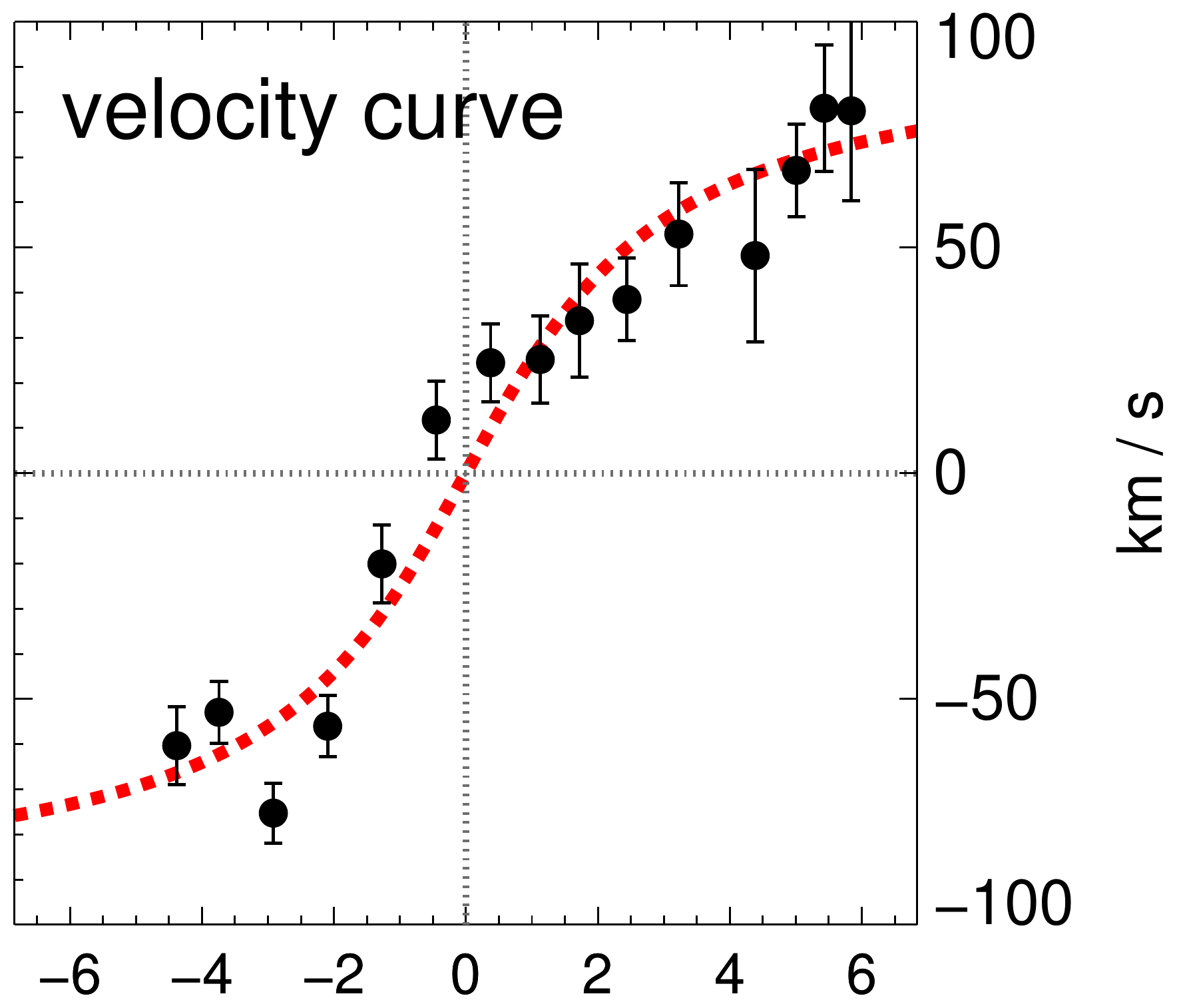}\\
\includegraphics[width=0.343\columnwidth]{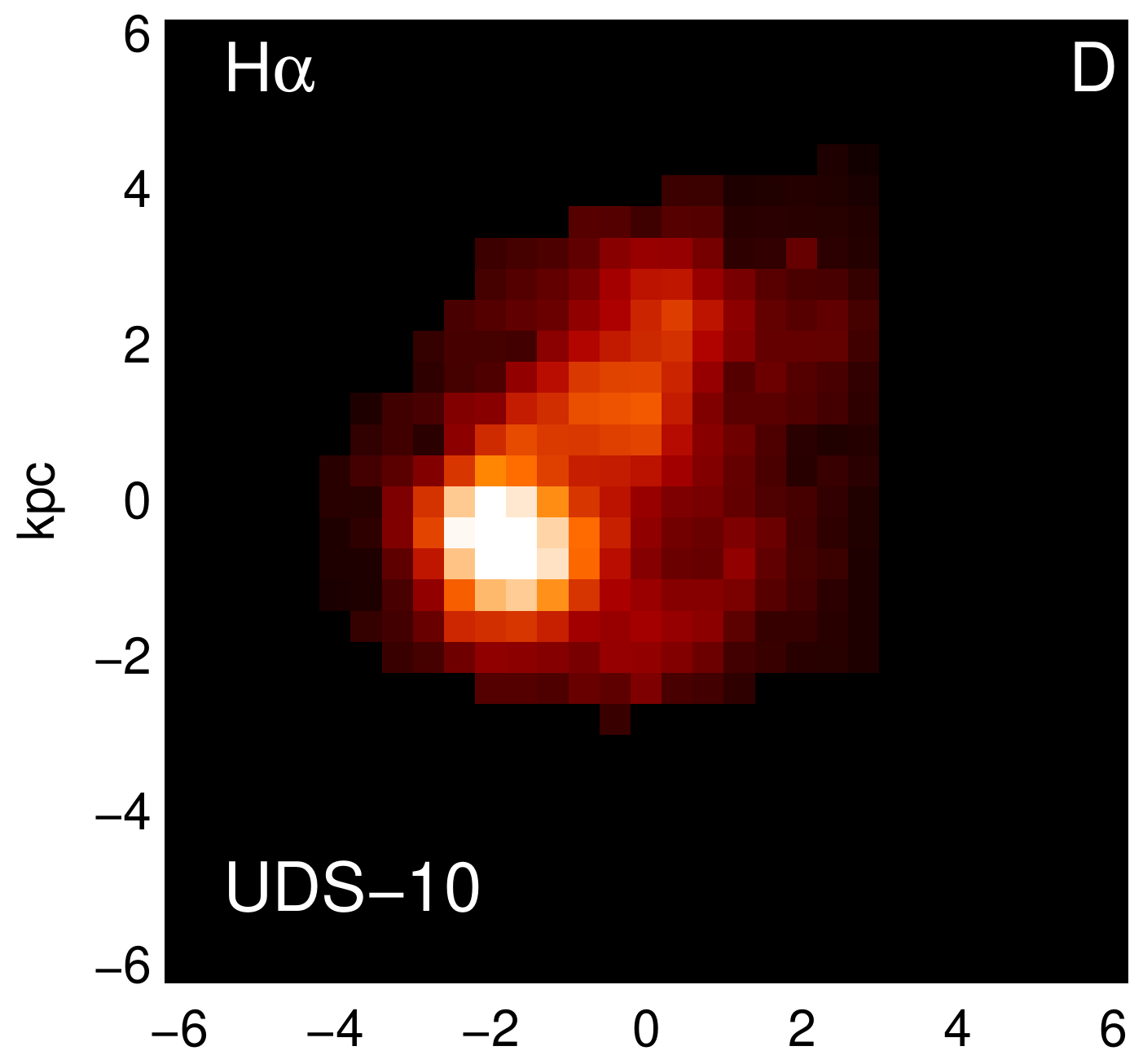}
\includegraphics[width=0.32\columnwidth]{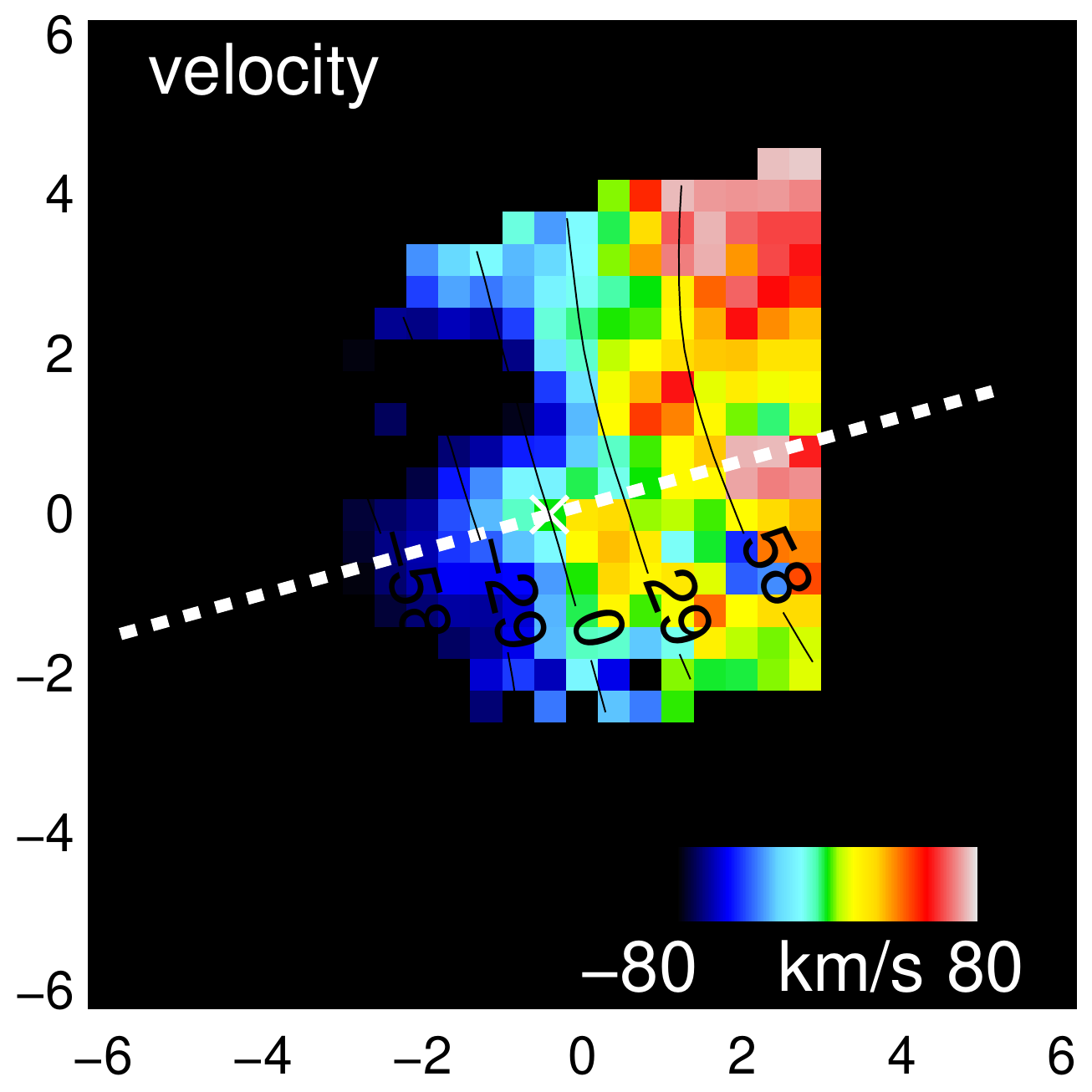}
\includegraphics[width=0.32\columnwidth]{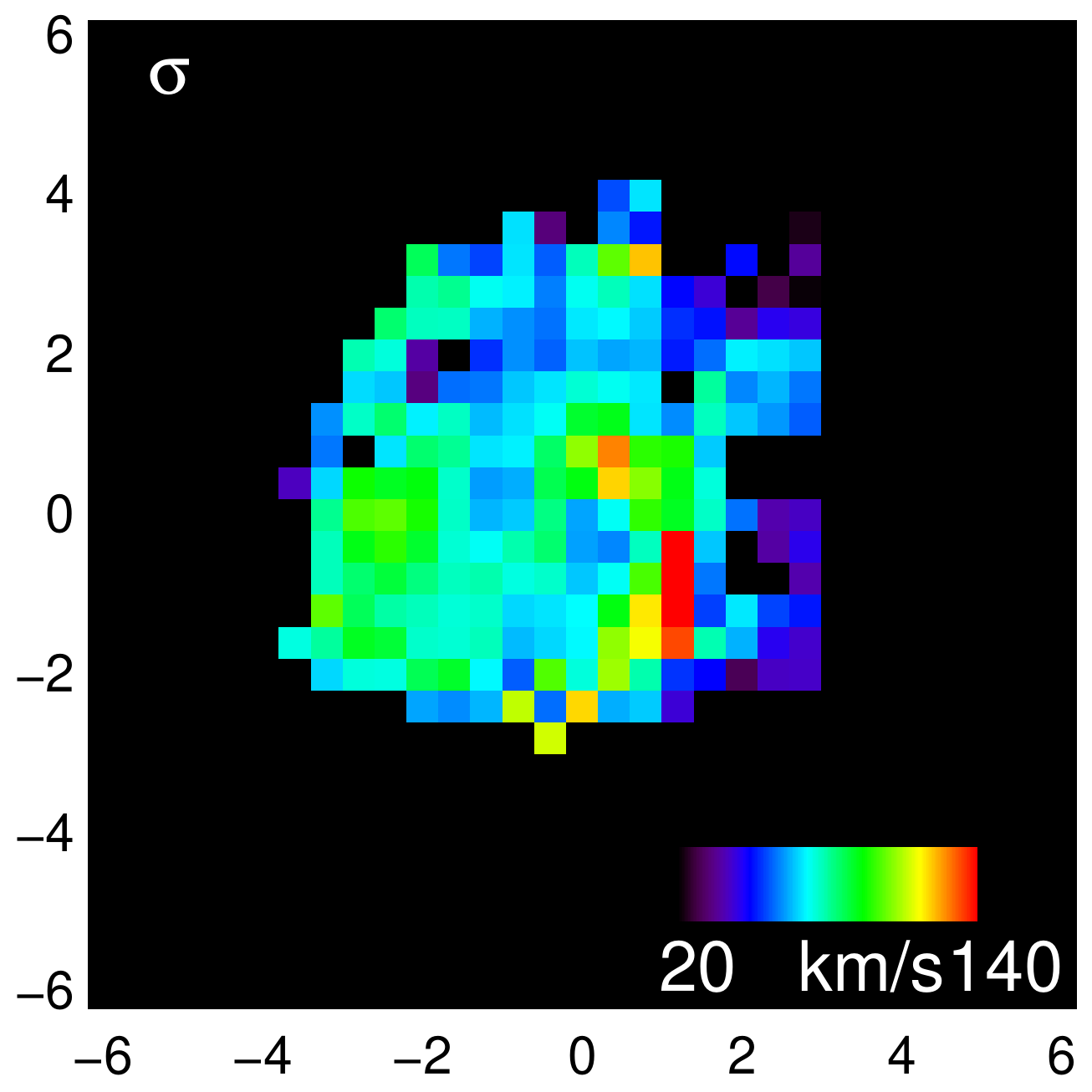}
\includegraphics[width=0.32\columnwidth]{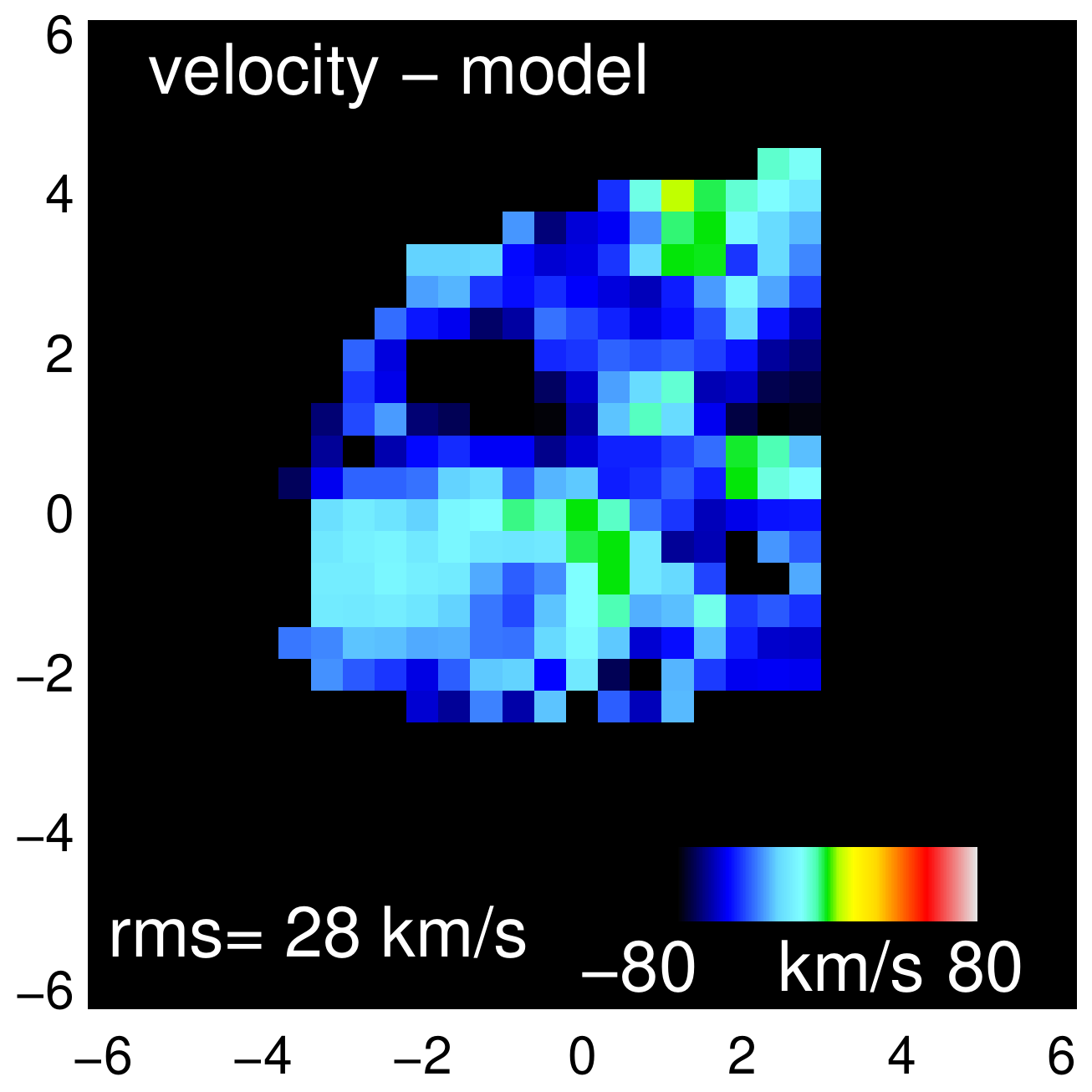}
\includegraphics[width=0.345\columnwidth]{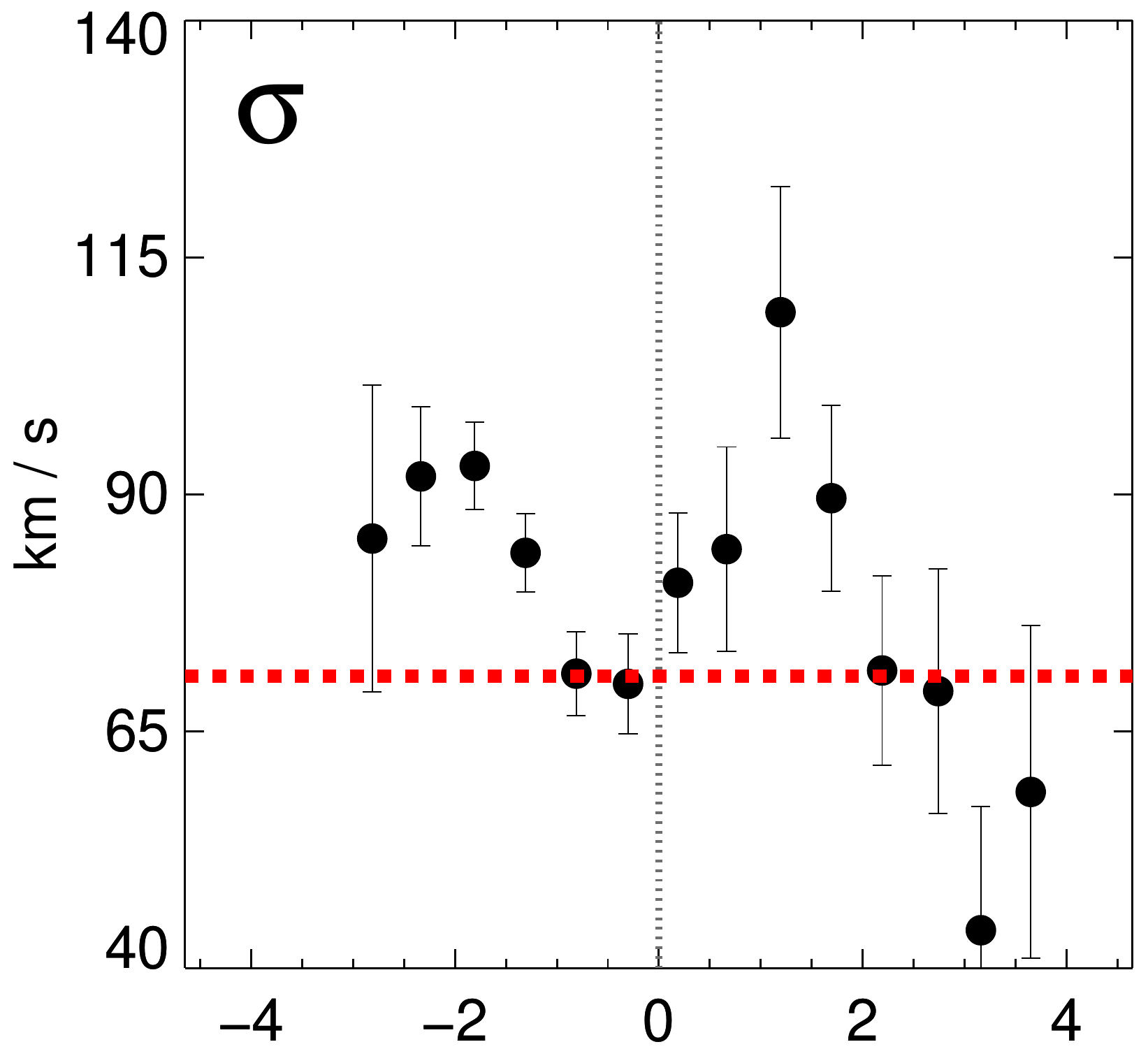}
\includegraphics[width=0.373\columnwidth]{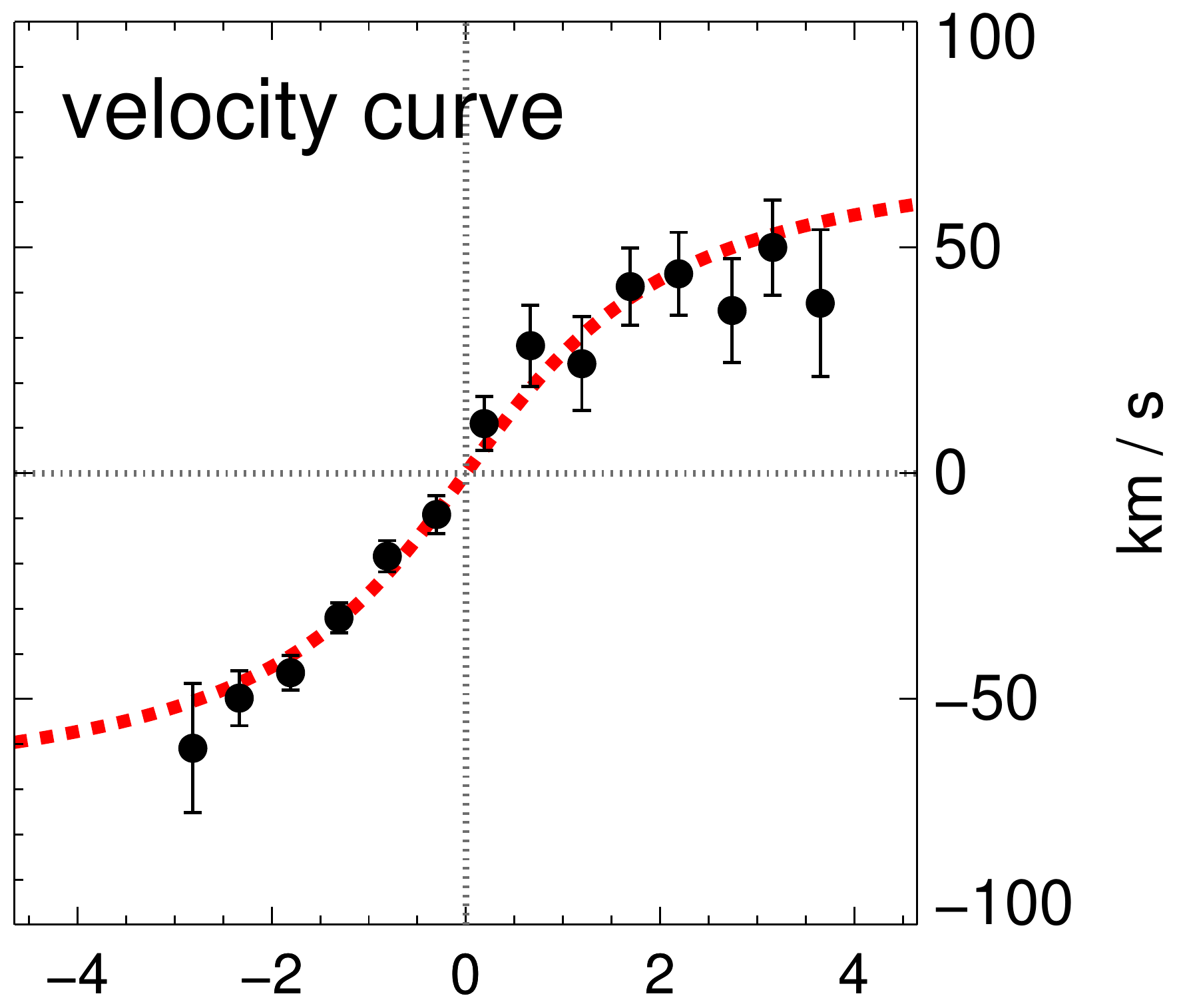}\\
\includegraphics[width=0.343\columnwidth]{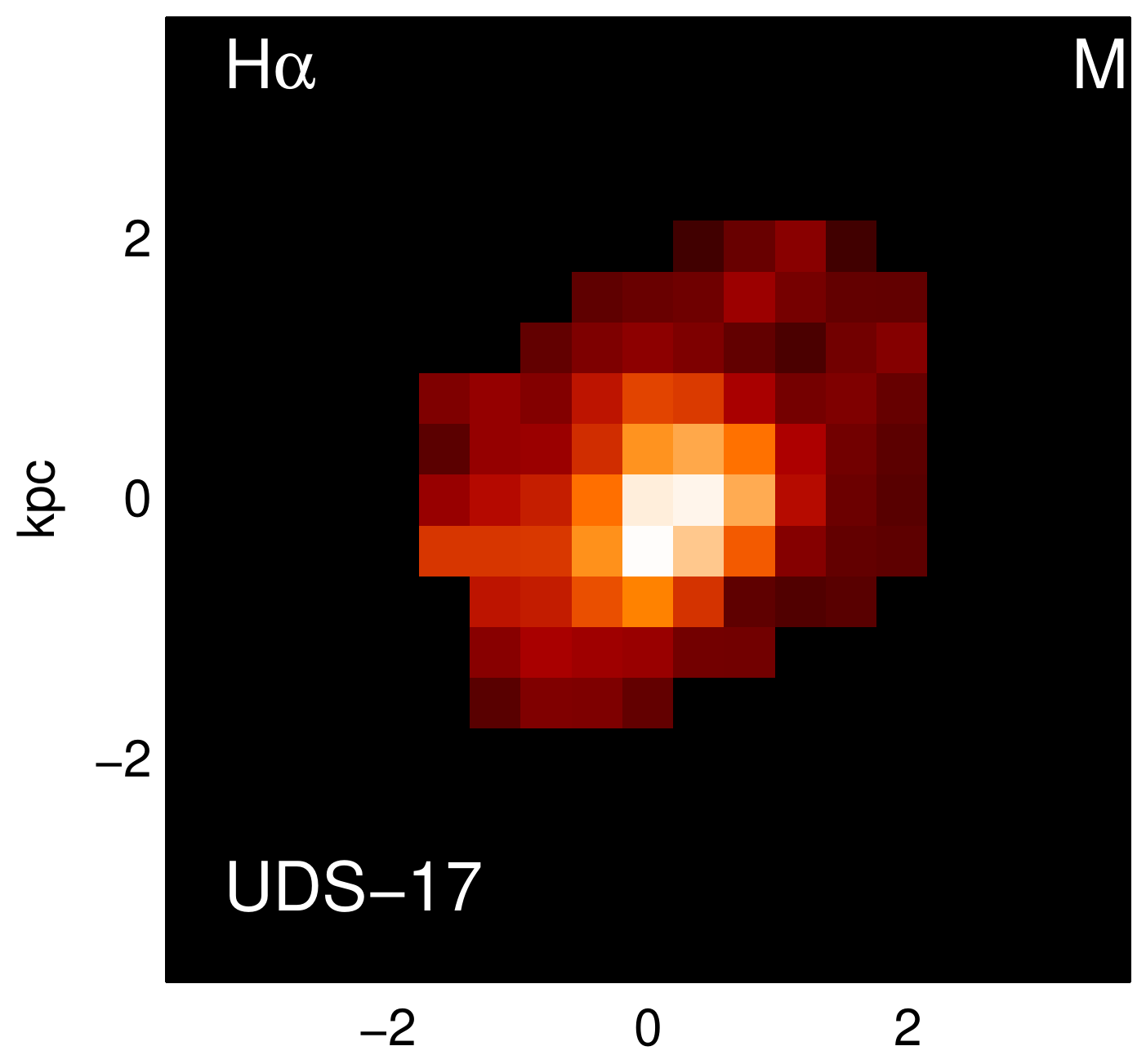}
\includegraphics[width=0.32\columnwidth]{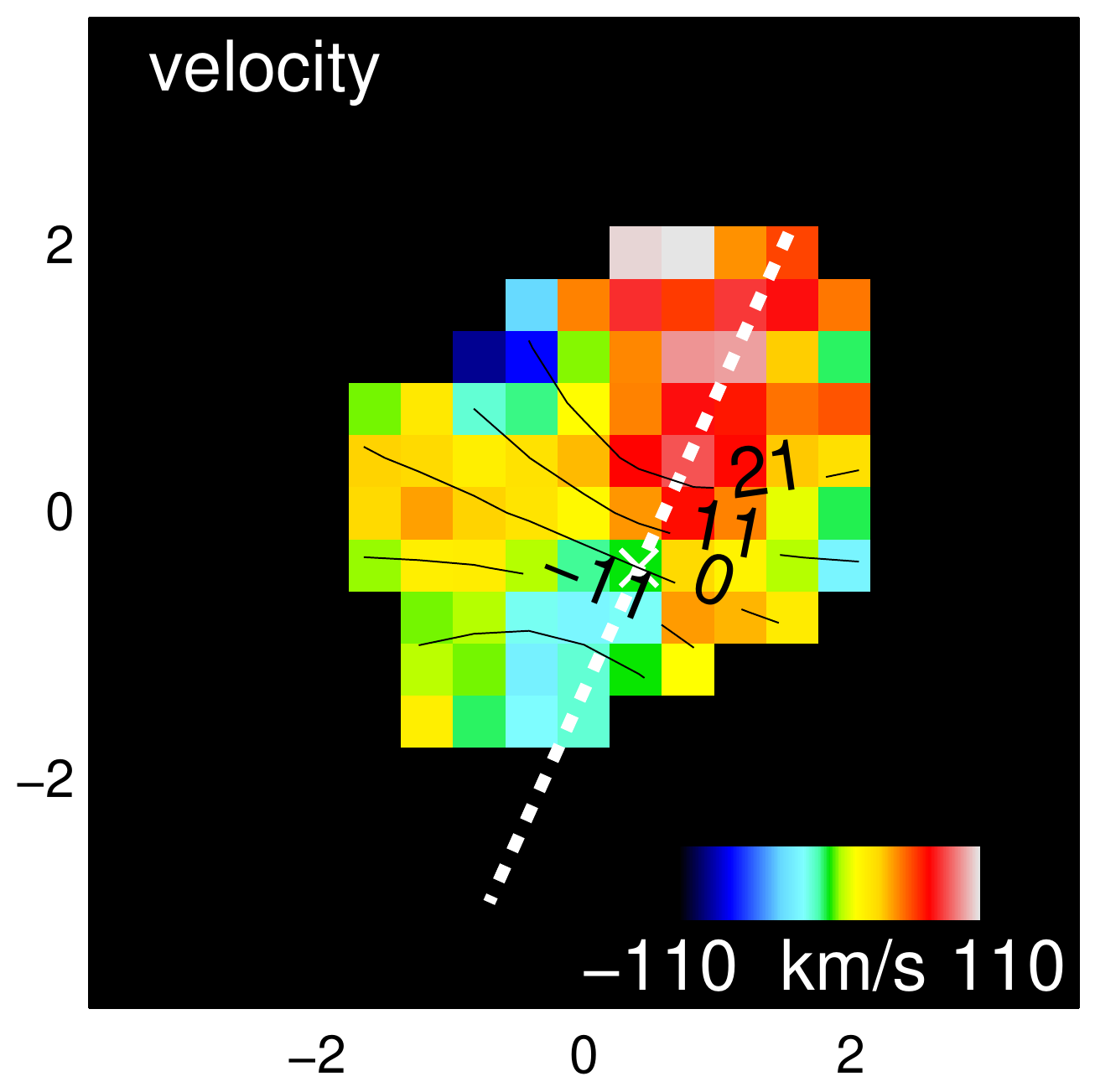}
\includegraphics[width=0.32\columnwidth]{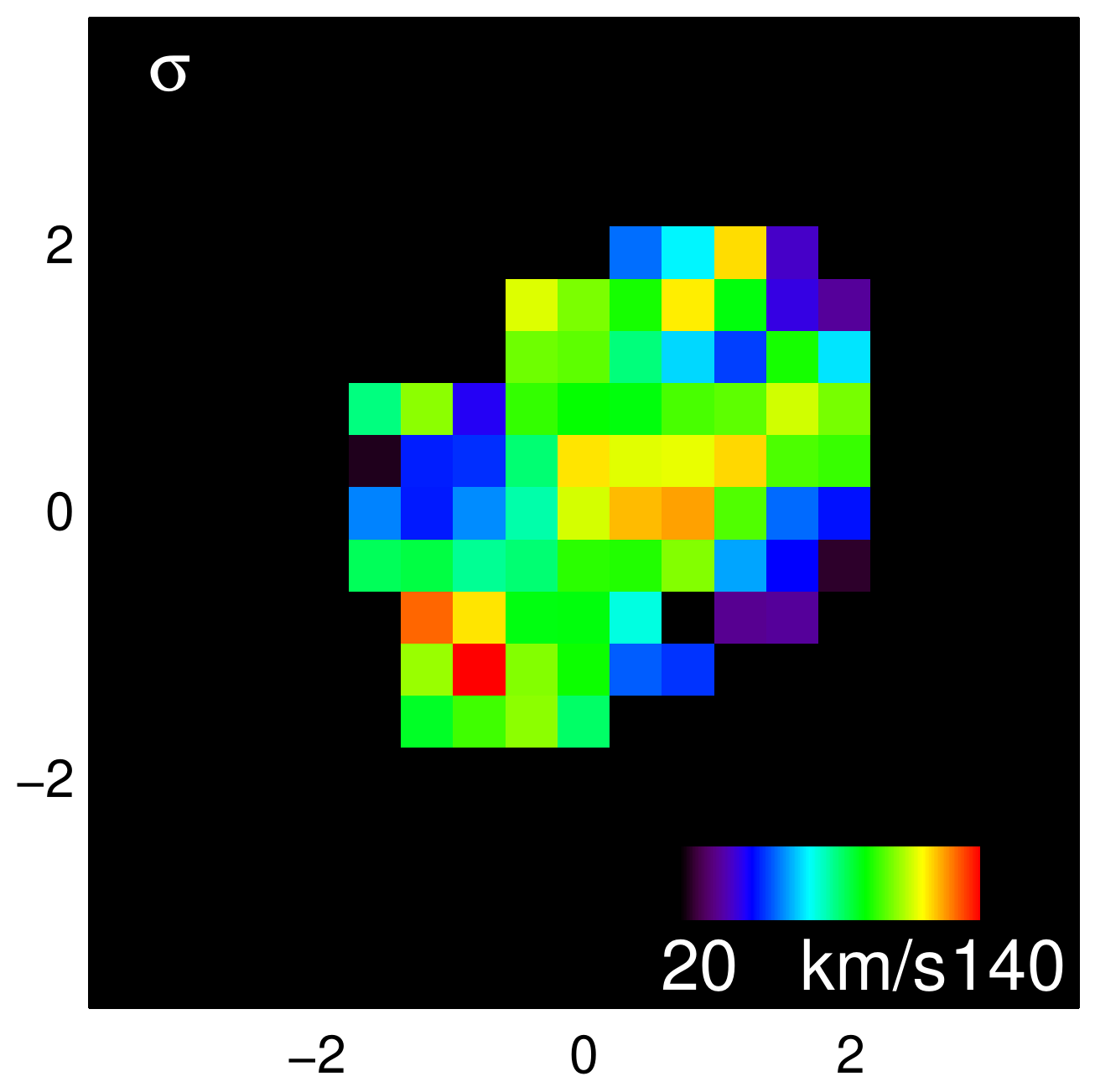}
\includegraphics[width=0.32\columnwidth]{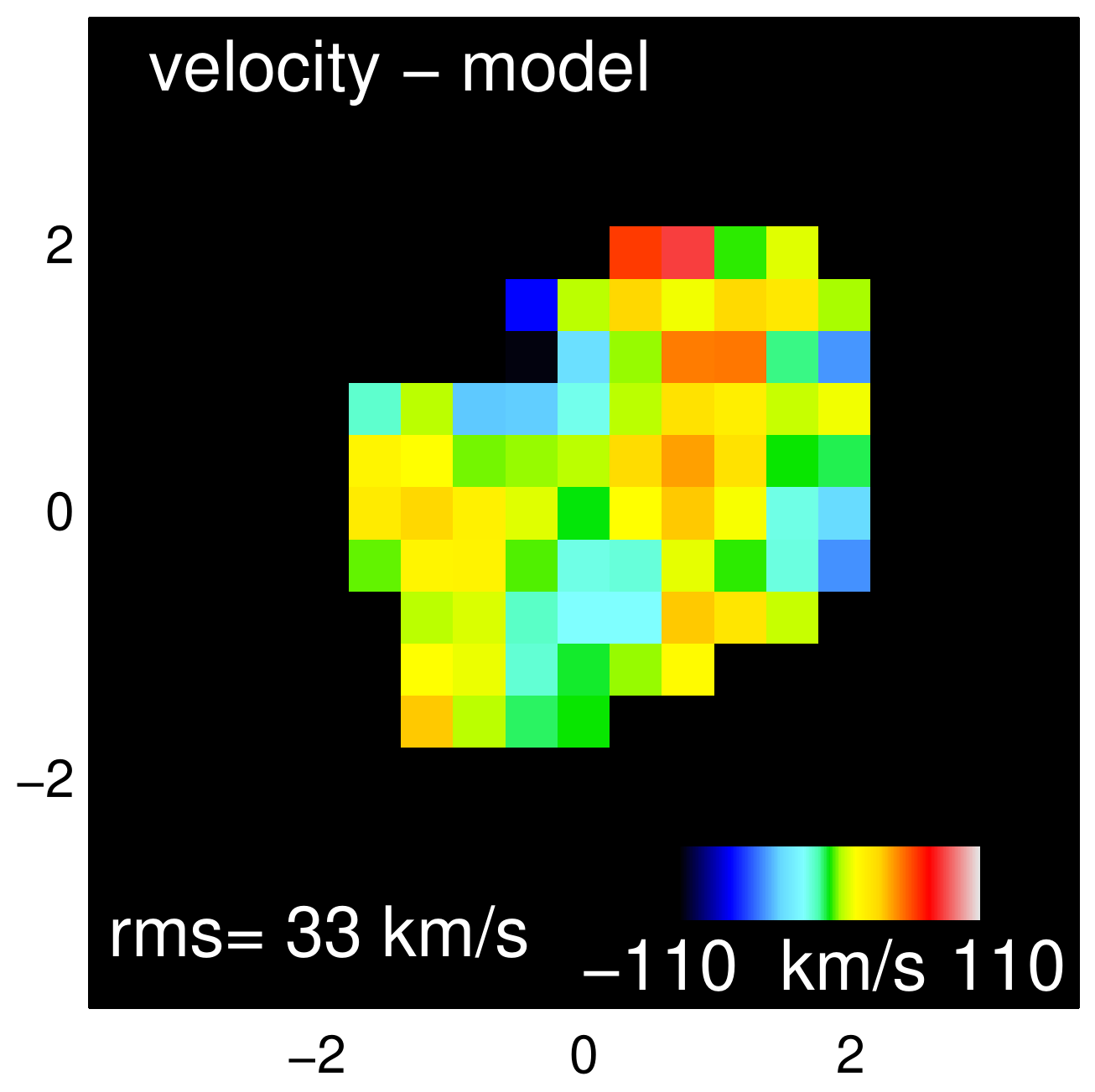}
\includegraphics[width=0.345\columnwidth]{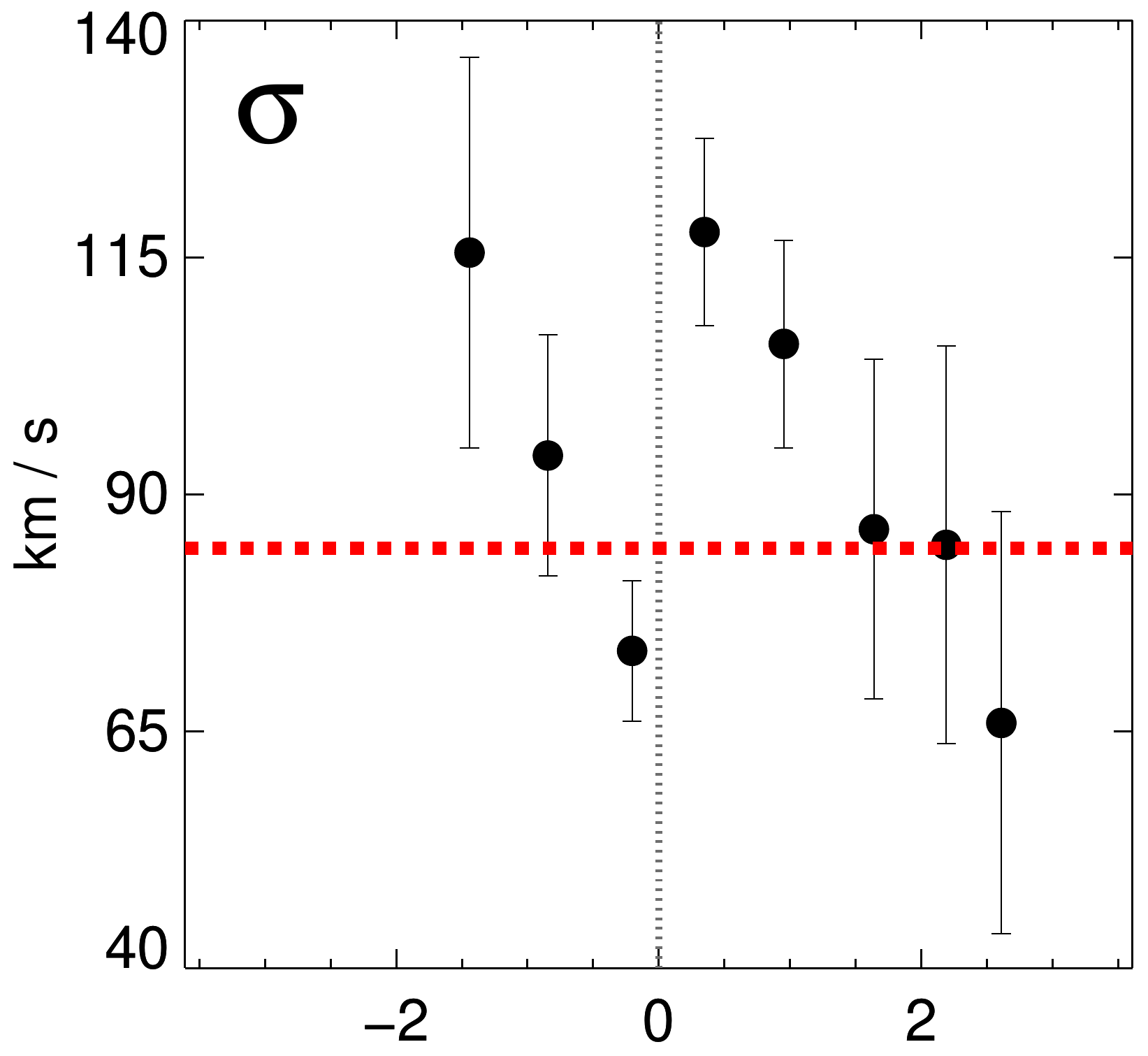}
\includegraphics[width=0.373\columnwidth]{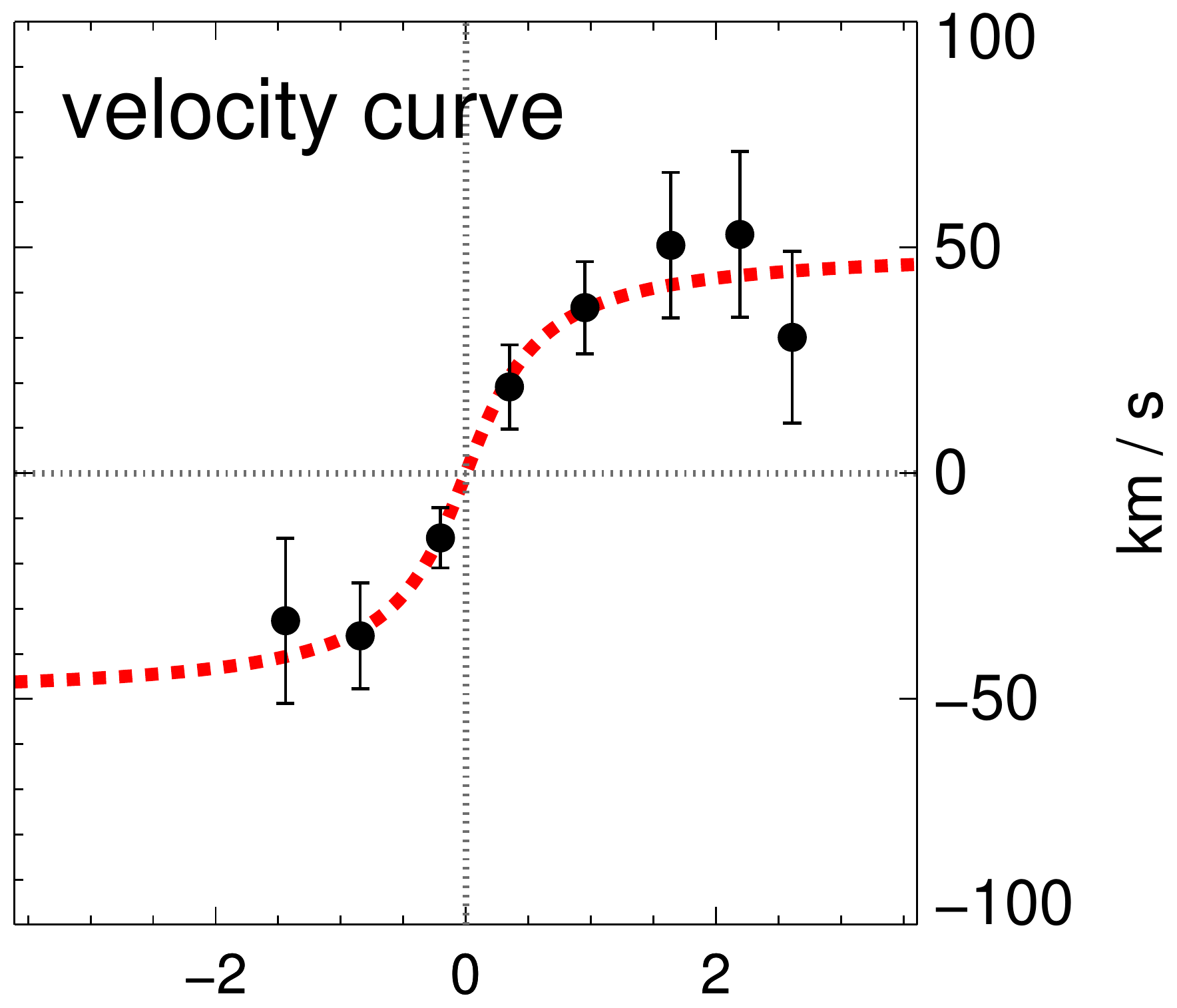}\\
\includegraphics[width=0.343\columnwidth]{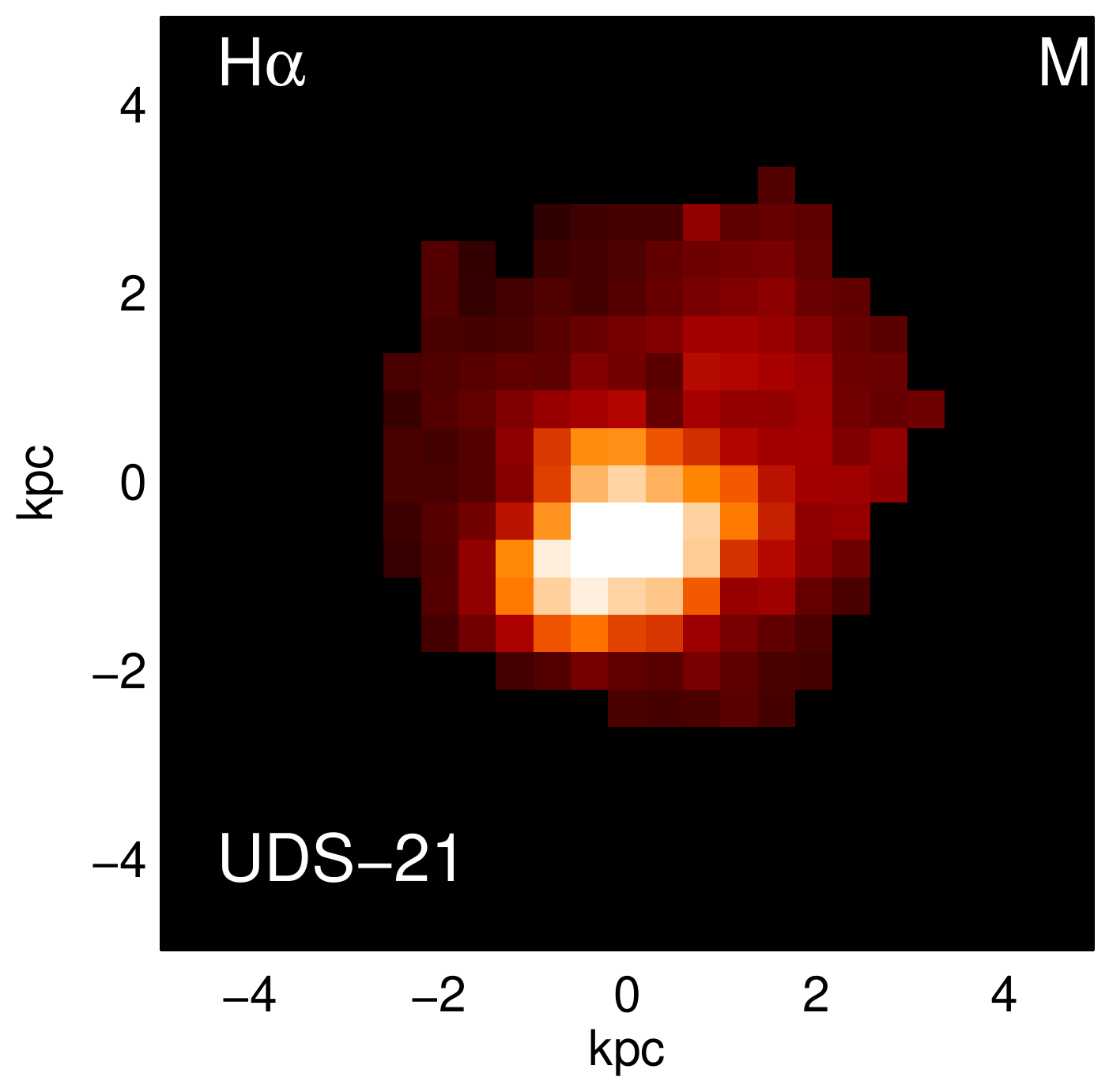}
\includegraphics[width=0.32\columnwidth]{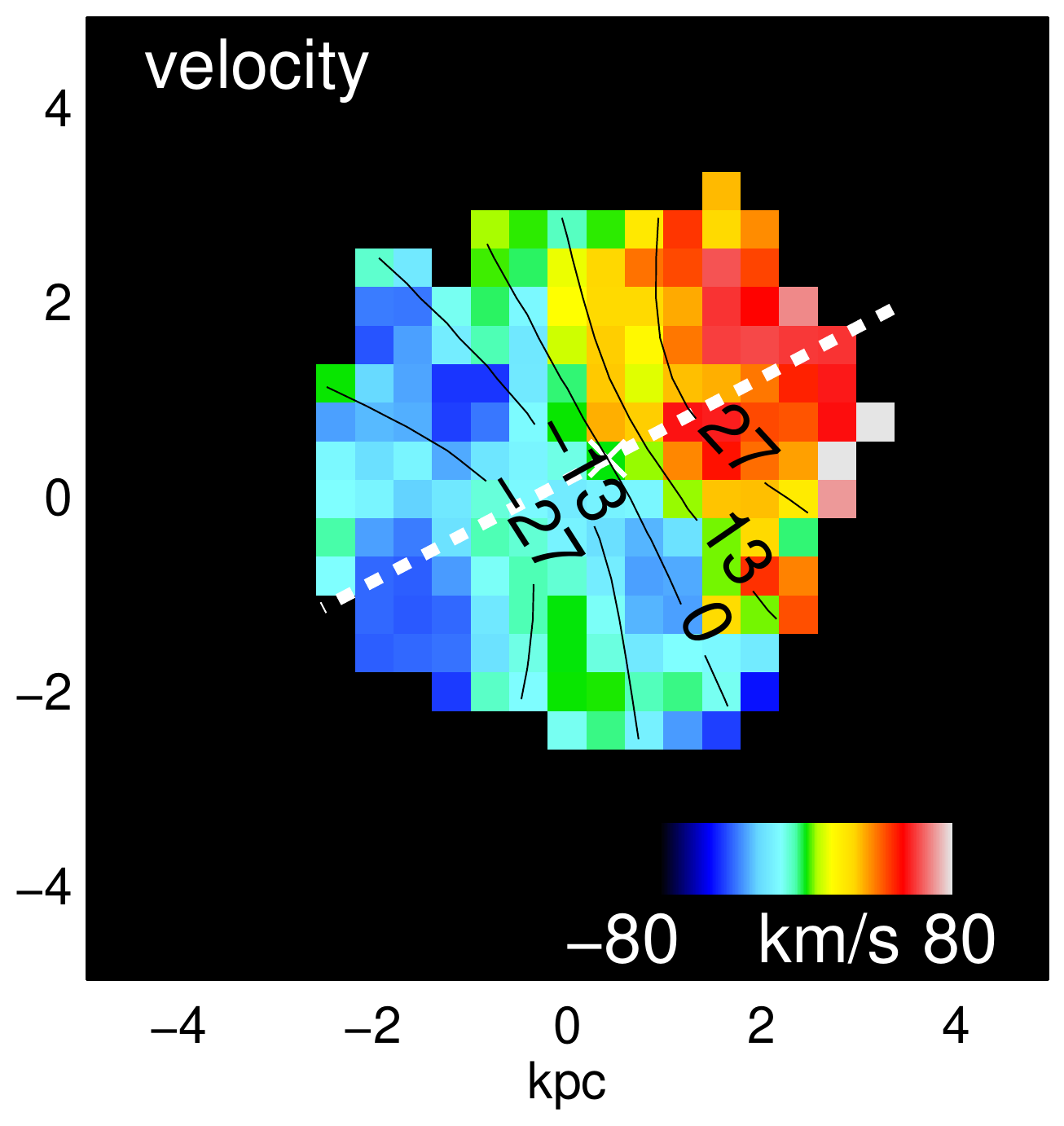}
\includegraphics[width=0.32\columnwidth]{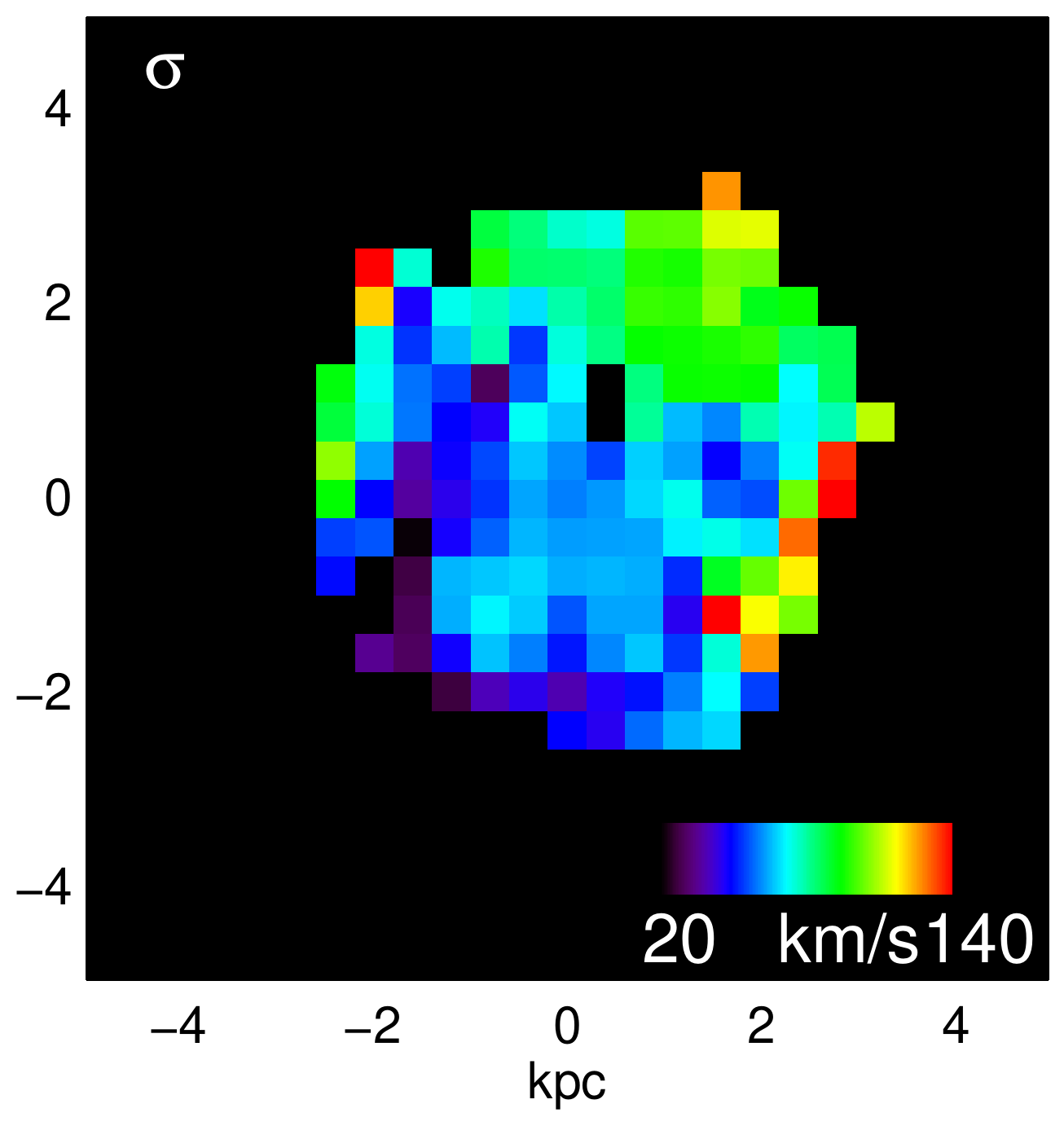}
\includegraphics[width=0.32\columnwidth]{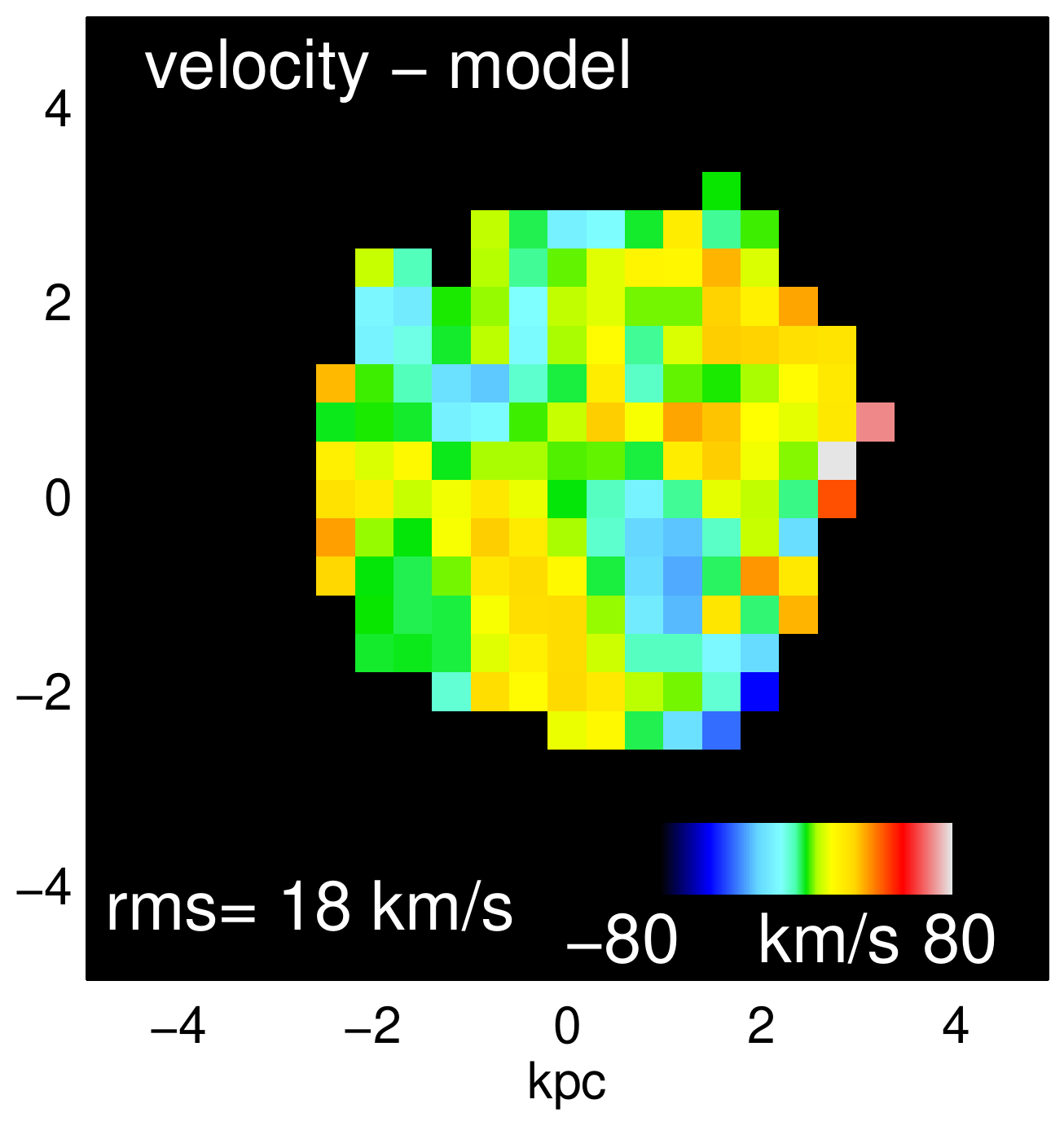}
\includegraphics[width=0.345\columnwidth]{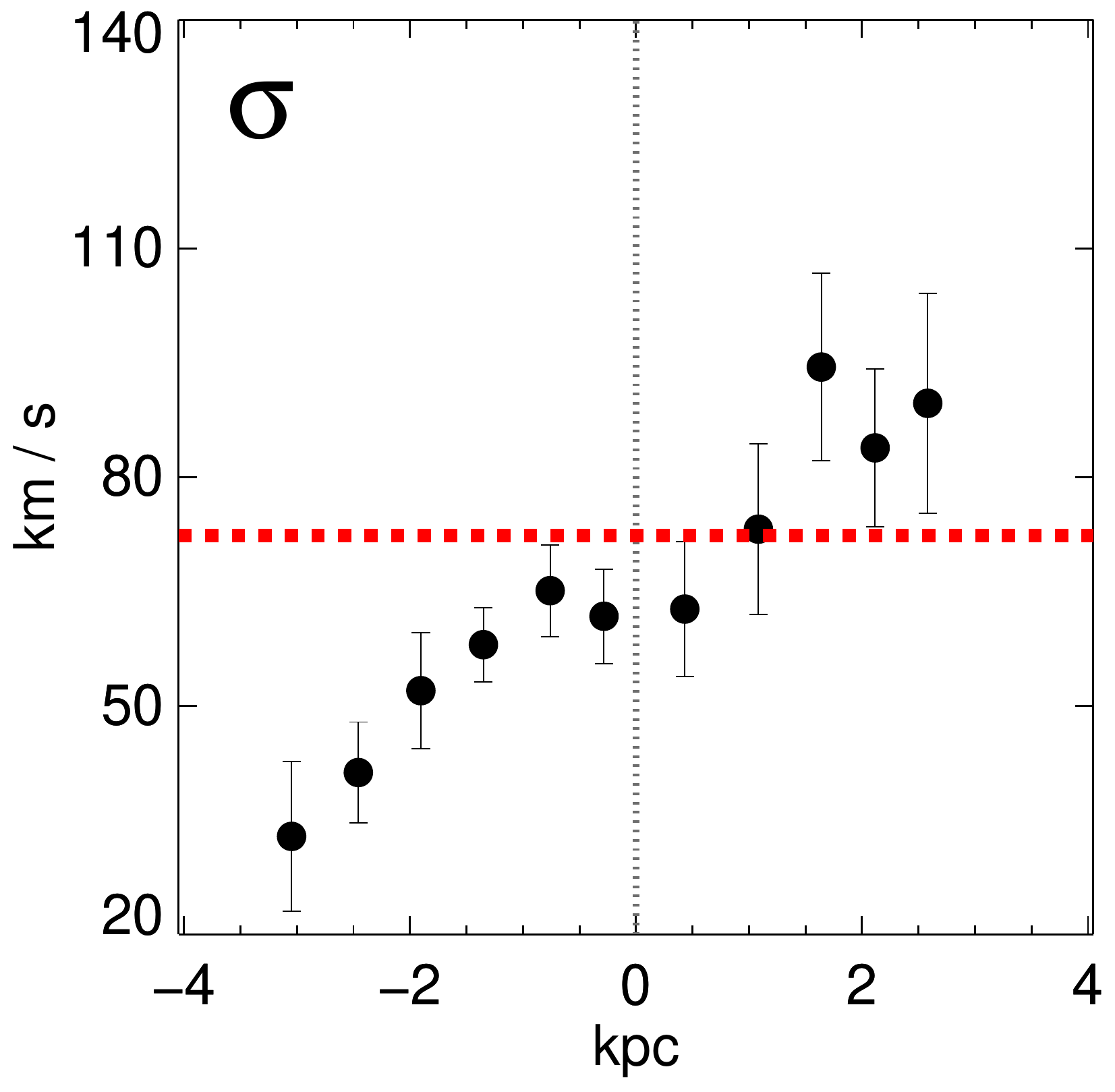}
\includegraphics[width=0.373\columnwidth]{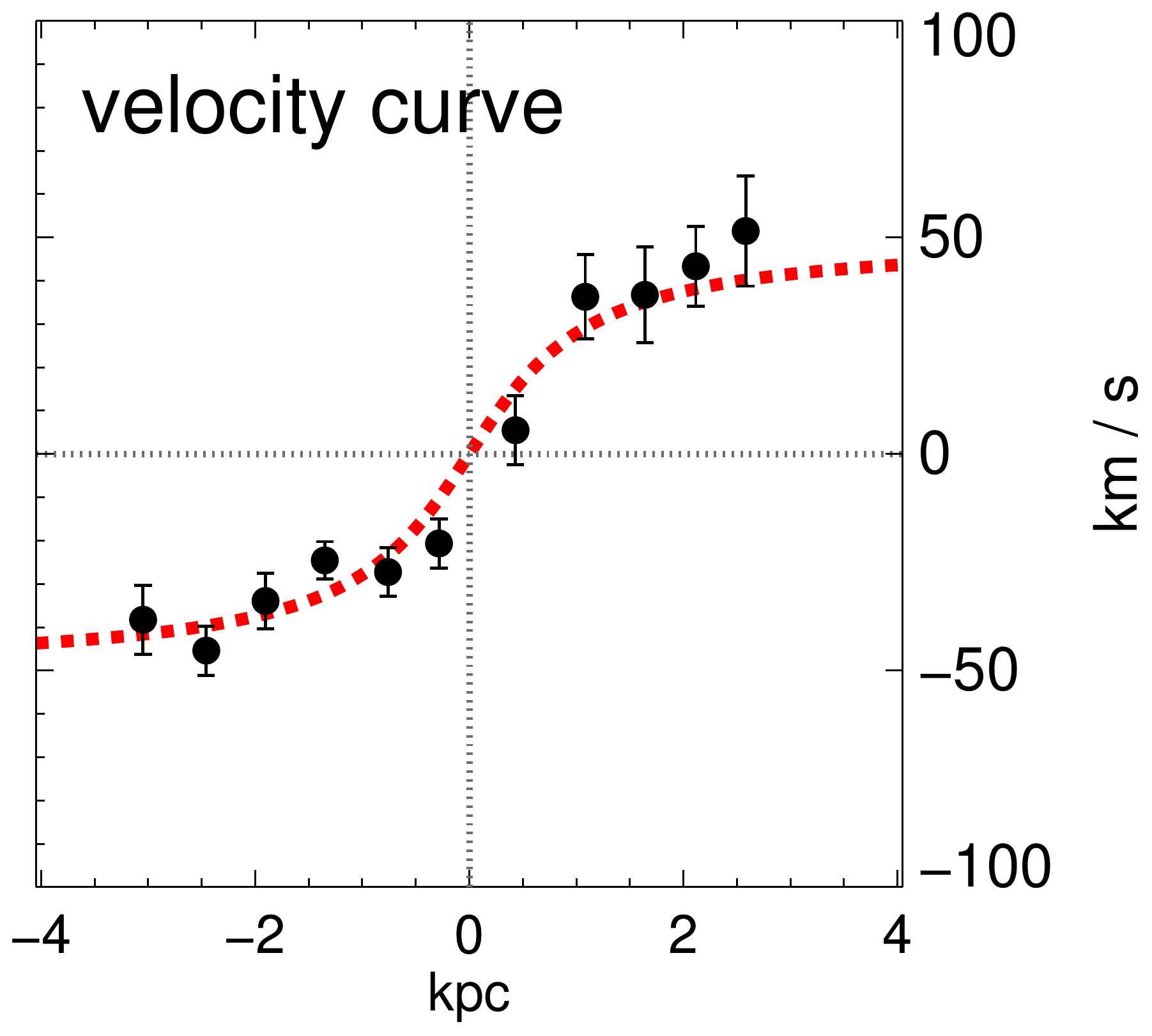}\\
\centering{\textbf{Figure 3. Continued.}}
\end{figure*}

\subsection{Average ISM properties}
\label{sec:ISM_properties}

To analyse the H$\alpha$ and [N\,{\sc ii}] line fluxes for our targets we first collapse 
each data-cube into a one-dimensional spectrum (see Fig.~\ref{fig:integrated_spectra}). 
In eight cases we detect the [N\,{\sc ii}]$\lambda$6583
emission line, deriving a median ratio of [N\,{\sc
    ii}]/H${\alpha}=0.27\pm0.02$, with a range between 0.10<[N\,{\sc
    ii}]/H${\alpha}$<0.43 (Table~\ref{tab:table1}).  None of the
galaxies display strong active galactic nucleus (AGN) signatures in
their rest-frame optical spectra (Fig.~\ref{fig:integrated_spectra}).

To search for fainter lines and obtain the mean properties of our observed sample 
we de-redshift each spectrum to rest-frame and co-add them (weighted by flux),
yielding the composite spectrum shown in Fig.~\ref{fig:integrated_spectra}. 
Weighting by flux instead of signal-to-noise helps to smooth residual features seen in 
low S/N spectra (e.g. SA22-54, UDS-17 in Fig.~\ref{fig:integrated_spectra}).
In this stacked spectrum, we measure a [N\,{\sc ii}]/H${\alpha}$ ratio
of 0.25$\pm$0.04 which is consistent with the median ratio derived for our sample. 
We also make a weak detection of the [S\,{\sc ii}]$\lambda\lambda$6716,6731 
doublet and derive a flux ratio of I$_{6716}$/I$_{6731}$=1.04$\pm$0.31. If we 
assume a typical H\,{\sc ii} region temperature of 10$^4$\,K, then the measured 
I$_{6716}$/I$_{6731}$ ratio corresponds to an electron density in the range 
of 100--1000\,cm$^{-3}$ \citep{Osterbrock1989}, and an upper limit to the ionised 
gas mass in the ISM of 4--40\,$\times$\,10$^{10}$\,M$_{\odot}$ for a disk galaxy 
with half-light radius of $\sim$2.4\,kpc (Table~\ref{tab:table1}). 
For an isobaric density distribution of the ionized gas, the density is defined in terms of the mean 
ISM pressure $P$ and mean electron temperature ($T_{\rm e}\sim10^4$\,K), through $P/k_{\rm B}\sim$\,$T_{\rm e}$\,$n_{\rm e}$.
Therefore we estimate a median ISM pressure of $P/k_{\rm B}\sim10^{6-7}$\,K\,cm$^{-3}$
which is $\sim$100-1000 times higher than the typical ISM pressure in the Milky Way ($\sim10^4$\,K\,cm$^{-3}$)
and consistent with other high-$z$ galaxy ISM pressure estimates \citep{Swinbank2015}. 
Although this value has considerable uncertainty, the derived pressure is compatible 
with hydrodynamic models which suggest that typical pressure in the ISM of star-forming
galaxies should increases from $\sim10^4$\,K\,cm$^{-3}$ at $z=0.1$ to 
$\sim10^{6-7}$\,K\,cm$^{-3}$ at $z=2$ \citep{Crain2015}. 
The I$_{6716}$/H$\alpha$ flux ratio reflects the ionisation strength of the ISM.
We measure I$_{6716}$/H$\alpha$=0.12$\pm$0.03. Considering also the derived [N\,{\sc ii}]/H${\alpha}$ flux ratio,
then we suggest an ionisation parameter of $\log_{10}$(U/cm$^3$)=$-$3.6$\pm$0.3 \citep{Osterbrock1989,Collins2001}.
Those median values are in agreement with \citet{Swinbank2012a}, \citet{Stott2013a} 
and \citet{Sobral2013a,Sobral2015}.

\begin{table*}
	\centering
	TABLE 2: DYNAMICAL PROPERTIES\\
	\begin{tabular}{lccccccccc} 
		\\
		\hline
		ID & inc. & $\sigma$ &V$_{\rm asym}$ & V$_{2.2}$ & $\chi^2_\nu$ & K$_{\rm Tot}$& Class\\ 
		   & (deg) & (km/s) & (km/s) & (km/s) & & & \\
		\hline
		SA22-17 & 72  & 57$\pm$13  & 75$\pm$2   & 62$\pm$4   & 1.1 & 0.36$\pm$0.04 & D \\
		SA22-26 & 53  & 46$\pm$11  & 142$\pm$3  & 120$\pm$12 & 1.5 & 0.24$\pm$0.03 & D \\
		SA22-28 & 65  & 66$\pm$8   & 60$\pm$3   & 52$\pm$7   & 1.7 & 0.22$\pm$0.03 & D \\
		SA22-54 & 63  & 62$\pm$10  & 104$\pm$2  & 95$\pm$5   & 1.3 & 0.14$\pm$0.02 & D \\
		\noalign{\smallskip}
		COS-16  & 53  & 95$\pm$8   & 77$\pm$11  & 59$\pm$10  & 1.9 & 0.99$\pm$0.09 & M \\
		COS-30  & 63  & 91$\pm$13  & 81$\pm$3   & 61$\pm$3   & 2.9 & 0.16$\pm$0.02 & D \\
		\noalign{\smallskip}
		SA22-01 & ... & ...        & ...        & ...        & ... & ...           & C \\
		SA22-02 & 71  & 66$\pm$9   & 100$\pm$3  & 85$\pm$12  & 2.0 & 0.81$\pm$0.09 & M \\
		UDS-10  & 32  & 71$\pm$10  & 143$\pm$10 & 85$\pm$7   & 3.2 & 0.24$\pm$0.04 & D \\
	 	UDS-17  & 71  & 84$\pm$14  & 53$\pm$6   & 40$\pm$7   & 9.0 & 0.90$\pm$0.08 & M \\
		UDS-21  & 40  & 72$\pm$11  & 78$\pm$14  & 58$\pm$12  & 1.6 & 0.75$\pm$0.07 & M \\
		\hline
		Mean    & 58  & 71$\pm$3   & 91$\pm$2   & 72$\pm$3   & 2.6 & 0.48$\pm$0.02 & ... \\ 
		\hline
	\end{tabular}
	\caption{\label{tab:table2}
	  Dynamical properties of the galaxies in our sample. `inc.'\ is the inclination angle defined by the angle between the line of sight and the plane of the
          galaxy disk (for a face-on galaxy, inc = 0 deg.). $\sigma$ is the average velocity dispersion across the galaxy image corrected for ``beam smearing'' effects due to PSF;
			see \S\ref{sec:galaxy_dyn}. V$_{\rm asym}$ and V$_{2.2}$ are inclination corrected. The $\chi^2_\nu$ of the best two-dimensional fit for each source
			is given in column six. K$_{\rm Tot}$ is the kinemetry coefficient. The classes in the final column denote Disk(D), Merger(M) 
			and Compact(C) (see \S\ref{sec:analysis} for more details of these parameters).}
\end{table*}

\subsection{Galaxy Dynamics} 
\label{sec:galaxy_dyn}
To measure the dynamics of each galaxy, we fit the H$\alpha$ and
[N\,{\sc ii}]$\lambda\lambda$6548,6583 emission lines
pixel-by-pixel. Following \citet{Swinbank2012a} we use a $\chi^2$ minimisation procedure, 
estimating the noise per spectral channel from an area that does not contain source emission. 
We first attempt to identify a H$\alpha$ line in each $0\farcs1\times0\farcs1$ pixel
($\sim$1$\times$1 kpc, which corresponds to the approximate PSF),
although if the fit fails to detect the emission line, the area is
increased by considering the neighbouring pixels, for example using
the averaged signal from an area of 3$\times$3 pixels. We use the criterion 
that the fit requires a S/N\,>\,5 to detect the emission line in each pixel,
and when this criterion is met then we simultaneously fit the H$\alpha$ and [N\,{\sc ii}]
$\lambda\lambda$6548,6583 emission allowing the centroid,
intensity and width of the Gaussian profile to vary (the FWHM of the
H$\alpha$ and [N\,{\sc ii}] lines are coupled in the fit). 

Even at $\sim$\,kpc-scale resolution, there is a contribution to the
line widths of each pixel from the large-scale velocity motions across
the galaxy, which must be corrected for \citep{Davies2011}. This is calculated
for each pixel where the H$\alpha$ emission is detected. We calculate
the local luminosity-weighted velocity gradient ($\Delta$V) across the
PSF ($\Delta$R) and subtract this from the measured velocity dispersion
\citep[see][for more details]{Stott2016}. We show the H$\alpha$ intensity, 
velocity and line of sight velocity dispersion maps in Fig.~\ref{fig:maps} 
for our sample.

\section{ANALYSIS, RESULTS \& DISCUSSION}
\label{sec:analysis}

In Fig.~\ref{fig:maps} we can see a variety of H$\alpha$ structures,
including various levels of clumpiness of the emission within our
sample. However we note that resolution effects tend to smooth kinematic
deviations making galaxy velocity fields appear more disky than they actually are \citep{Bellocchi2012}.
Fig.~\ref{fig:maps} also shows that there are strong velocity
gradients in many cases (e.g.\ SA22-28, SA22-54) with peak-to-peak
differences (V$_{\rm max}$\,sin(i)) ranging from 90--180\,km\,s$^{-1}$
and ratio of peak-to-peak difference to line-of-sight velocity
dispersion ($\sigma$) of V$_{\rm max}$\,sin(i)/$\sigma$ = 1.1--3.8.
This is in concordance with previous observations of galaxies at $z\sim$2
\citep{Starkenburg2008,Law2009,Forster2009,Gnerucci2011b,Genzel2011}.
Assuming that the dynamics of the underlying mass distribution are coupled
to the measured kinematics of the ionized gas, then these observed high-$z$ 
galaxies are consistent with highly turbulent systems. 

Although a ratio of V$_{\rm max}$/$\sigma$ = 0.4 has been used to 
crudely differentiate rotating systems from mergers \citep{Forster2009},
more detailed kinematic modelling is essential to reliably distinguish these two populations. 
We therefore attempt to model the two-dimensional velocity field by first
identifying the dynamical centre and the kinematic major axis. We
follow \citet{Swinbank2012a} to construct two-dimensional models with
an input rotation curve following an arctan function
(V(r)=$\frac{2}{\pi}$V$_{\rm asym}$arctan(r/r$_{\rm t}$)), where
V$_{\rm asym}$ is the asymptotic rotational velocity and r$_{\rm t}$
is the effective radius at which the rotation curve turns over
\citep{Courteau1997}. This model has six free parameters
(V$_{\rm asym}$, r$_{\rm t}$, [x/y] centre, position angle (PA) and disk
inclination) and a genetic algorithm \citep{Charbonneau1995} is used to
find the best fit (see \citealt{Swinbank2012a} for more details). The best-fit
kinematics maps and velocity residuals are shown in
Fig.~\ref{fig:maps}, the best-fit inclination and disk rotation speeds
are given in Table~\ref{tab:table2}. The mean deviation from the
best-fit models within the sample (indicated by the typical RMS) is
<data\,$-$\,model>\,=\,27$\pm$2\,km\,s$^{-1}$ with a range of
<data\,$-$\,model>\,=\,18--40\,km\,s$^{-1}$. These offsets are probably
the product of an un-relaxed dynamical component indicated by the high
mean velocity dispersion $\sigma$\,=\,71$\pm$1\,km\,s$^{-1}$ of our sample 
(Table~\ref{tab:table2}), dynamical substructures, or effects of gravitational 
instability within the disk.

We use the dynamical centre and position angle derived from the
best-fit dynamical model to extract the one-dimensional rotation curve
across the major kinematic axis of each galaxy (see
Fig.~\ref{fig:maps}).
Three targets (SA22-26, SA22-54 and SA22-02) do not show a flattening of the velocity 
curve at large radii, so V$_{\rm asym}$ can only be estimated using an extrapolation
of the true rotational velocity for these targets.

In order to distinguish between rotation and motion from disturbed 
kinematics we use `kinemetry' which measures the asymmetry of the 
velocity field and line-of-sight velocity dispersion maps for each galaxy
\citep{Shapiro2008}. This technique has been well calibrated and
tested at low redshift \citep[e.g.][]{Krajnovic2006}, whilst at high
redshift it has been used to determine the strength of deviations of
the observed velocity and dispersion maps from an ideal rotating disk
\citep{Alaghband2012, Shapiro2008, Sobral2013a,
  Swinbank2012a}. Briefly, kinemetry proceeds to analyse the two dimensional
velocity and velocity dispersion maps using azimuthal kinematic
profiles in an outward series of best fitting elliptical rings. The
kinematic profile as a function of angle is then expanded
harmonically, which is equivalent to a Fourier transformation
which has coefficients k$_{\rm n}$ at each tilted ring
(see \citealt{Krajnovic2006} for more details).

Defining the velocity asymmetry (K$_{\rm V}$) and the velocity
dispersion asymmetry (K$_{\sigma}$) using the eigen k$_{\rm n}$ coefficients from 
the velocity and velocity dispersion maps respectively, we measure the level
of asymmetries from an ideal disk in our galaxies (we omit SA22-01 from
this analysis as it is not well-resolved). For an ideal disk, the values
K$_{\rm V}$ and K$_{\sigma}$ will be zero. In a merging system, strong
deviations from the idealised case produces large K$_{\rm V}$ and
K$_{\sigma}$ values. The total asymmetry is defined as K$^2_{\rm Tot}$
= K$^2_{\rm V}$+ K$^2_{\sigma}$ and we use this quantity to
differentiate disks (K$_{\rm Tot}$<0.5) from mergers (K$_{\rm
  Tot}$>0.5) following \citet{Shapiro2008}.
The K$_{\rm Tot}$  errors are derived by bootstrapping 
via Monte-Carlo simulations the errors in measured velocities, velocity 
dispersions, and estimated dynamical parameters of each galaxy.

\begin{figure}
 \centering
 \includegraphics[width=0.95\columnwidth]{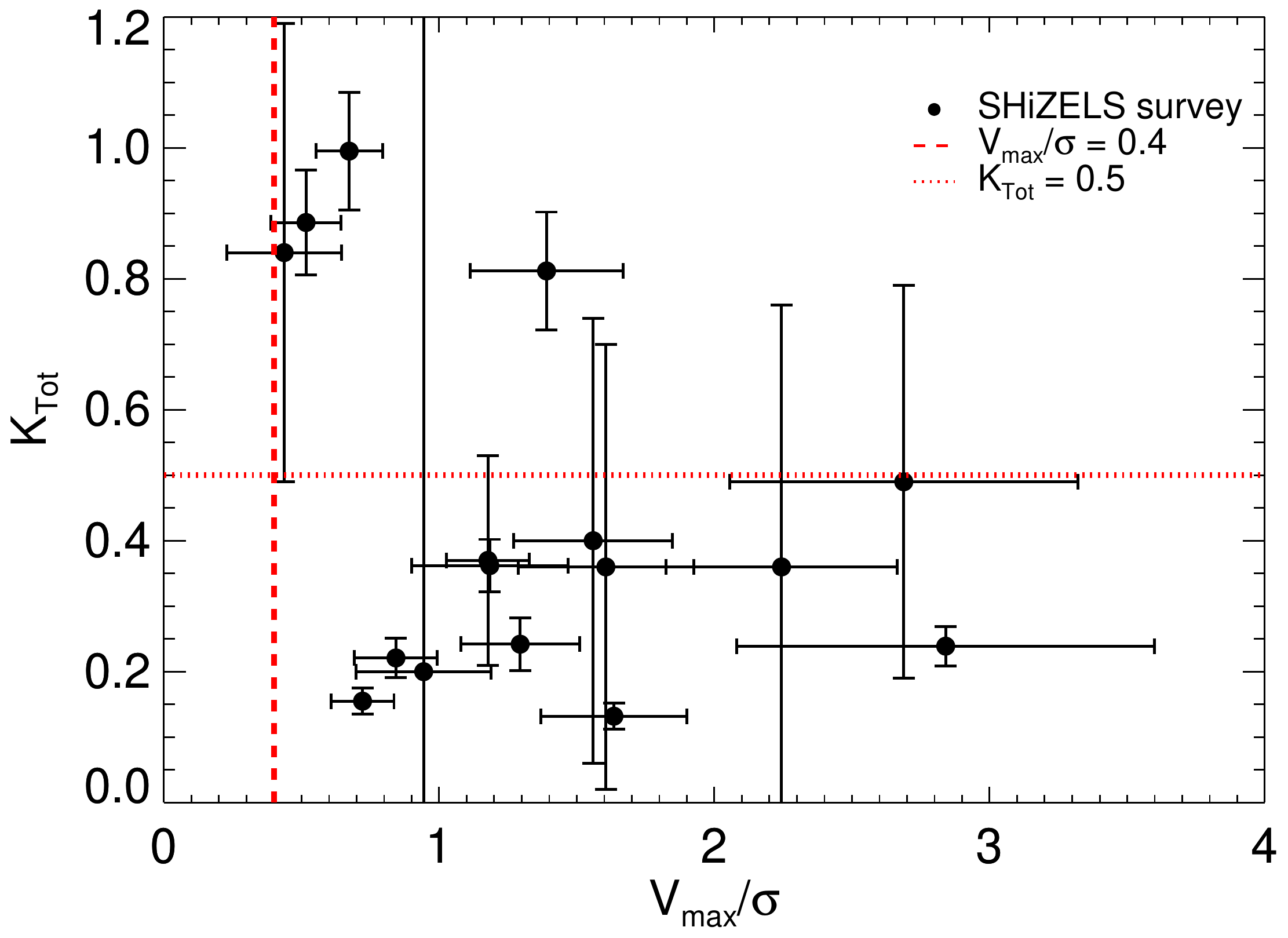}
 \caption{\label{fig:MC}
	  The kinematic measure K$_{\rm Tot}$ (see \S\ref{sec:analysis}) against the V$_{\rm max}$/$\sigma$ ratio for the SHiZELS Survey. 
	  The red-dashed line shows the V$_{\rm max}$/$\sigma=0.4$ ratio which has been used to crudely differentiate rotating systems from 
	  mergers \citep{Forster2009}. The red-dotted line shows the K$_{\rm Tot}=0.5$ value which is used to distinguish between galaxy disks 
	  from mergers \citep{Shapiro2008}. Although there is no strong correlation between both quantities, it is notable that galaxies classified 
	  as mergers by kinemetry criterion tend to show lower V$_{\rm max}$/$\sigma$ ratio, however not as low as 0.4. This suggests that the 
	  V$_{\rm max}$/$\sigma=0.4$ criterion tend to under-estimate the total number of mergers in a given galaxy sample.}
\end{figure}

\citet{Bellocchi2012} proposed a modified kinemetry criterion 
(K$_{\rm Tot,B12}$), which try to distinguish between post-coalescence mergers 
and disks. As the major merger evolves, the central region tends to
relax rapidly into a disk meanwhile the outer parts remain out of equilibrium. 
Therefore the outer regions retain better the memory of a merger event \citep{Kronberger2007}.
In order to consider this effect, \citet{Bellocchi2012} weights more highly the outskirts 
of each galaxy when combining the asymmetries measured from the velocity and velocity dispersion maps. 

These two kinemetry criteria have been compared with a visual classification 
scheme done at higher spatial resolution. \citet{Hung2015} observed 
eighteen (U)LIRGs at $z<0.088$ with the {\it Hubble Space Telescope} ({\it HST}\,) 
Advanced Camera for Surveys (ACS) and considered another six sources from the Digitized 
Sky Survey (DSS). They classified galaxies by inspecting their optical morphologies  
\citep{KLarson2016} and then they obtained IFS data for this sample 
from the Wide Field Spectrograph (WiFeS). They artificially redshifted their 
local IFS observations to $z=1.5$ to make a comparison with IFU seeing-limited 
observations (0$\farcs$5) at high-$z$. \citet{Hung2015} concluded that \citet{Shapiro2008}'s
kinemetry criterion (K$_{\rm Tot}$) tend to underestimate the merger fraction 
whereas \citet{Bellocchi2012}'s kinemetry criterion (K$_{\rm Tot,B12}$) overestimated 
the number of mergers within the same sample. 
Hereafter, we will use the kinemetry criterion defined by \citet{Shapiro2008} 
to classify our targets, considering that our merger fraction values are likely to be lower 
limits at each redshift.

From the kinemetry criterion, we classify four targets as merger
systems and six targets as rotating systems (see
Table~\ref{tab:table2}). In addition, from the kinemetry criterion
error rate \citep[see][for more details]{Shapiro2008}, we expect that
$\sim$1 merger is being misclassified as a disk and $\sim$1 disk is
being misclassified as merger. The fraction of rotating systems
within our sample is $\sim$60\%, which is consistent within 1$\sigma$ with that 
found from other H$\alpha$ IFU surveys at similar high redshift
\citep[e.g.][]{Forster2009, Jones2010b, Wisnioski2011, Swinbank2012a}.
We note that most of our mergers are identified in galaxies at $z\sim$2.23 and the large
error estimates are inherent of the low statistics of our sample.

In Fig.~\ref{fig:MC} we plot the K$_{\rm Tot}$ parameter against 
the V$_{\rm max}$/$\sigma$ ratio for our sample and that presented by \citet{Swinbank2012a}. 
All of these galaxies were observed at $\sim$kpc-scale resolution using AO. 
We find no correlation between both quantities. Although galaxies classified as mergers by 
kinemetry tend to lie in the region with lower V$_{\rm max}$/$\sigma$ ratio,
we find that the V$_{\rm max}$/$\sigma$ = 0.4 merger criterion is not consistent 
with the more sophisticated kinematic estimate K$_{\rm Tot}$, suggesting that the former
criterion under-estimates the total number of mergers within a given galaxy sample. This also 
suggests that a detailed kinematic analysis is needed in order to classify mergers from galaxy disks.

Hereafter, we will refer to the `SHiZELS' survey as the compilation of the observations presented in this work
with the previous observations by \citet{Swinbank2012a}. In this previous campaign they observed nine 
H$\alpha$-selected star-forming galaxies between $z=0.84-2.23$ with SINFONI. This sample was also drawn 
from the HiZELS survey. The median M${_\star}$ and SFR are $\sim2\times$10$^{10}$\,M$_\odot$ and 
$\sim$7\,M$_\odot$\,yr$^{-1}$ respectively \citep[see][for more details]{Swinbank2012a}.

\subsection{The Stellar-Mass Tully-Fisher and M$_\star$--S$_{0.5}$ Relations}
\label{sec:tf}

\begin{figure}
 \centering
 \includegraphics[width=0.95\columnwidth]{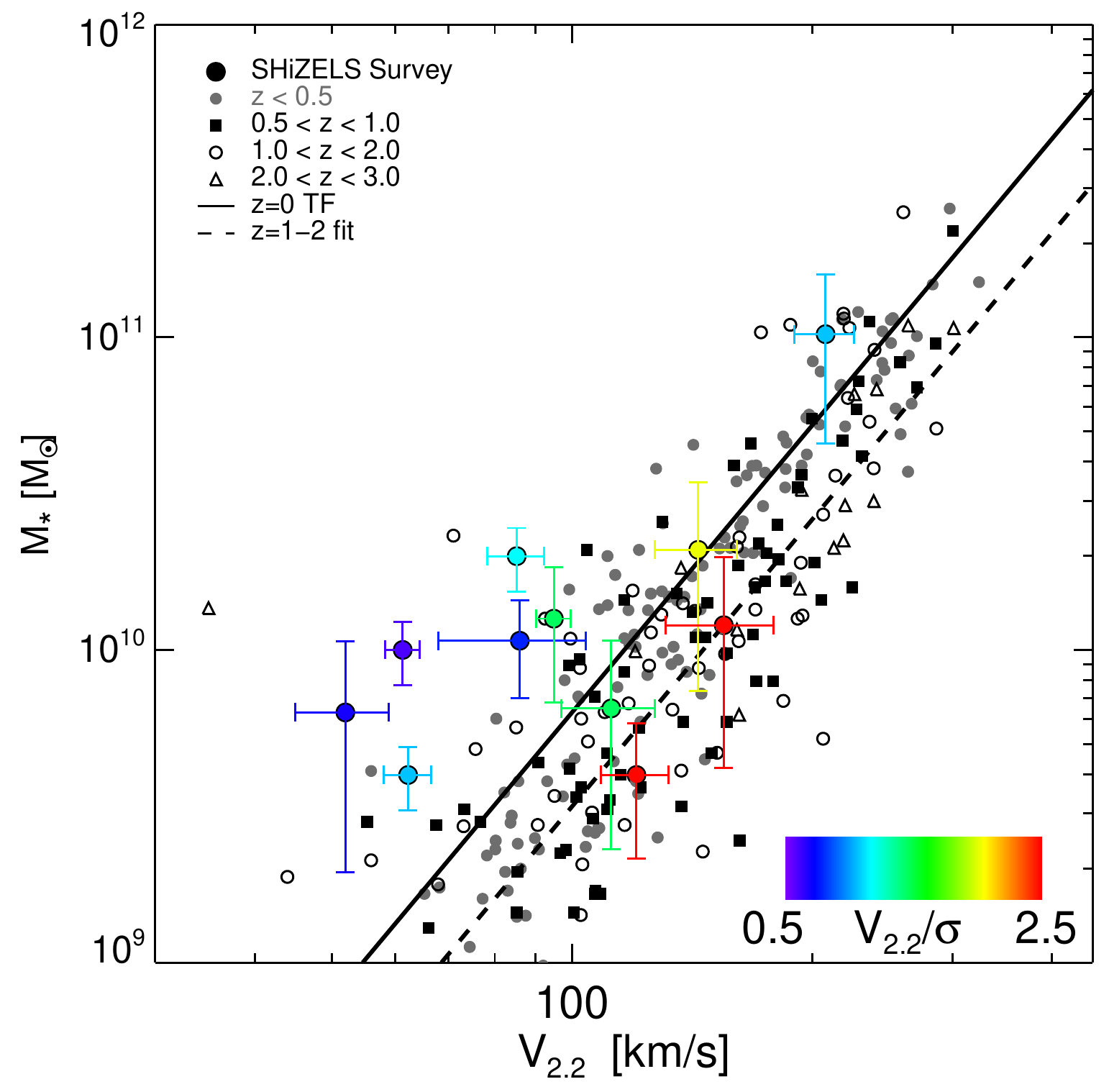}
 \includegraphics[width=0.95\columnwidth]{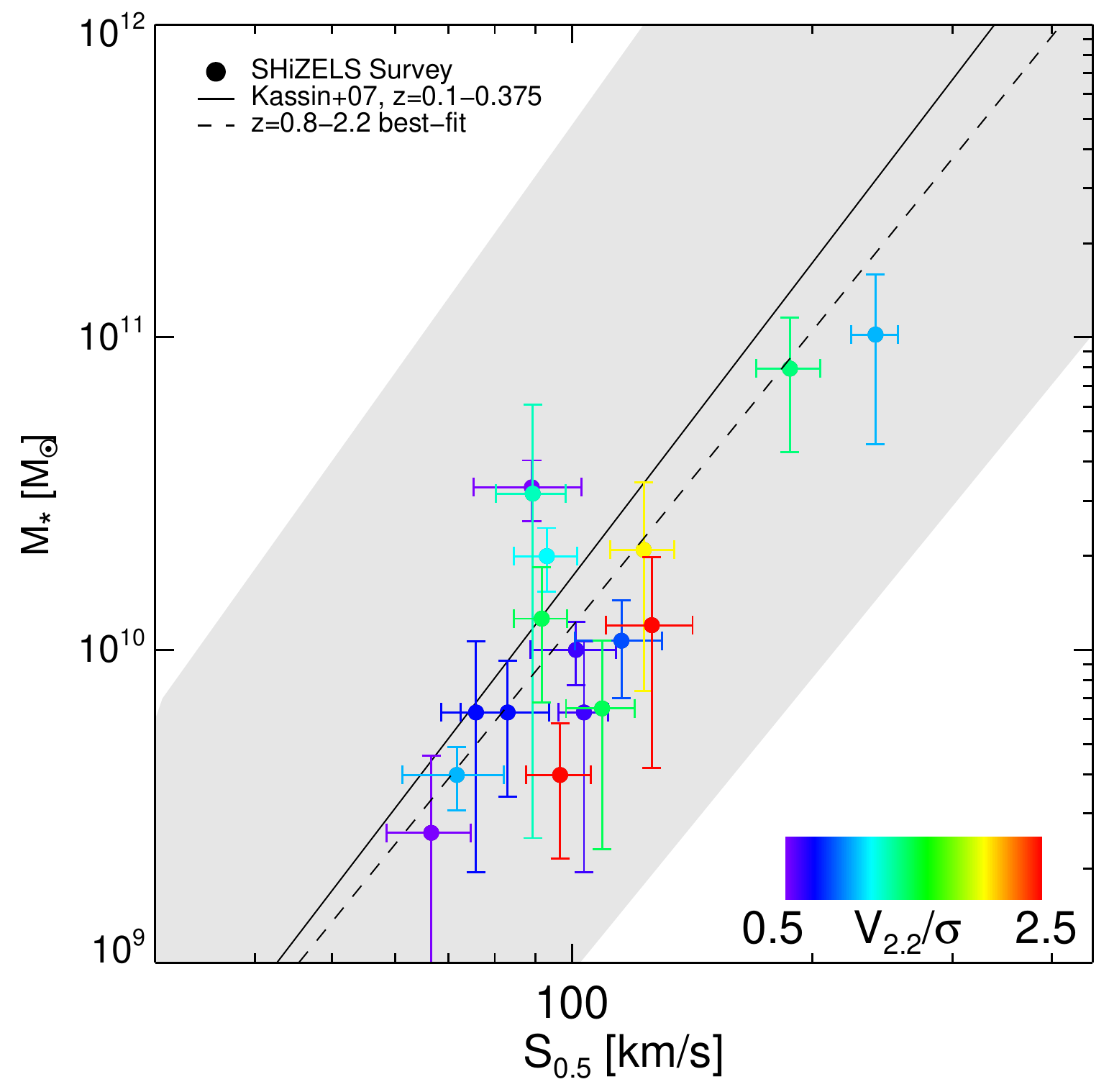}
 \caption{\label{fig:TF}
	  \textit{Top}: Evolution of the stellar mass TF relation measured from the SHiZELS survey at 
	  $z=0.8-2.23$ colour-coded using the V$_{2.2}/\sigma$ ratio. We only show our galaxies consistent 
	  with rotating systems together with their 1$\sigma$ velocity and stellar mass uncertainties. 
	  The solid line denotes the TFR at $z=0$ from \citet{Pizagno2005}. The dashed line represent the best-fit 
	  TF relation at $z=1-2$ from \citet{Swinbank2012a} based on the compilation of high-redshift points from 
	  \citet[][$z=0.6-1.3$]{Miller2011,Miller2012}; \citet[][$z=1$]{Swinbank2006}; \citet[][$z=1.5$]{Swinbank2012a}; 
	  \citet[][$z=2$]{Jones2010b}; \citet[][$z=2$]{Cresci2009} and \citet[][$z=3$]{Gnerucci2011b}. Galaxies with lower relative 
	  rotational support tend to be scattered to lower values along the velocity axis. This is consistent with the 
	  result found by \citet{Tiley2016}.
	  \textit{Bottom}: The M$_{\star}$-S$_{0.5}$ relationship measured from the SHiZELS survey at $z=0.8-2.23$.
	  The error bars show the 1$\sigma$ stellar mass and S$_{0.5}$ uncertainties. The data is colour-coded 
	  as in the image above. The solid line represents the relation at $z\sim0.2$ from \citet{Kassin2007} 
	  and the shaded area represents its 1$\sigma$ uncertainty. The dashed line corresponds to the best-linear-fit 
	  to our data. Our scatter is tighter than the intrinsic M$_{\star}$-S$_{0.5}$ scatter. The slope (0.32$\pm$0.2) 
	  and the intercept at $10^{10}$\,M$_\odot$(1.98$\pm$0.09) found from our 
	  best-fit are consistent with the uncertainties of the $z\sim0.2$ relation. This is consistent with no 
	  evolution of the M$_{\star}$-S$_{0.5}$ relation with redshift up to $z=2.23$.}
\end{figure}

The Tully-Fisher relation (TFR) is a fundamental scaling relation
describing the interdependence of luminosity or stellar mass
and the maximum rotational velocities (a dark matter mass tracer) in galaxies. 
It allows us to trace the evolution of the mass-to-luminosity ratio of 
populations of galaxies at different epochs. Recently the KROSS survey
(\citealt{Stott2016,Tiley2016}; Harrison et al. submitted) have provided 
a new perspective on TFR evolution by observing $\sim600$ galaxies at 
$z\sim0.9$. \citet{Tiley2016} derived an evolution of the stellar-mass 
TFR zero-point of $-$0.41$\pm$0.08 dex for rotationally 
supported galaxies defined with V/$\sigma$\,$>$\,3. However, when they analysed 
their data without this V/$\sigma$ constraint, they did not find any significant 
evolution of the M$_{\star}$-TFR zero-point. We note that the M$_{\star}$--TFR zero-point 
evolution found by \citet{Tiley2016} is contrary to some previous studies conducted at 
similar redshift \citep{Miller2011,Miller2012,Diteodoro2016}.

Similarly, \citet{Weiner2006a} and \citet{Kassin2007} introduced the kinematic
measure S$_{0.5} = (0.5$V$^2 + \sigma^2)^{0.5}$ which considers support by both rotational motions and dispersion 
arising from disordered motions \citep{Weiner2006a}. \citet{Kassin2007} computed the M$_{\star}$-TFR 
and M$_{\star}$--S$_{0.5}$ relations within 544 galaxies at 0.1\,$<z<$\,1.2. The M$_{\star}$--S$_{0.5}$ 
relationship was found to be a tighter relation compared with the M$_{\star}$-TFR relation, and
this relation also showed no evolution with redshift in either intercept or slope.

When measuring circular velocities, to be consistent with the previous \citet{Swinbank2012a} campaign, we use 
velocities observed at 2.2 times the disk scale length (V$_{2.2}$) corrected for inclination effects. The disk scale
length (r$_{\rm d}$) is defined as the radius at which the galaxy H$\alpha$ intensity has decreased to $e^{-1}(\sim0.37)$ 
times it's central value. 

\subsubsection{The Stellar Mass Tully-Fisher Relation}

In Fig.~\ref{fig:TF} we study the  M$_{\star}$-TFR at $z=0.8-2.23$ using 
SHiZELS survey galaxies classified as disky by our kinemetry analysis.  
The stellar masses and velocities from the comparison samples have been 
estimated in a fully consistent way, and these values (or corrections, 
where necessary) are presented in \citet{Swinbank2012a}. We also show
the TF relations at $z=0$ \citep{Pizagno2005} and the best-fit
relation at $z=1-2$ \citep{Swinbank2012a} from the literature. Even though 
we do not attempt to fit a relation to our data, we can see from 
Fig.~\ref{fig:TF} that apparently our sample is consistent with
no evolution in the zero-point of the M$_{\star}$-TFR out to $z=0.8-2.23$.

As suggested by \citet{Tiley2016} we estimated the rotational velocity
to dispersion velocity ratio. This is done by calculating the
V$_{2.2}$/$\sigma$ ratio. We show this parameter colour-coded 
in Fig.~\ref{fig:TF}. We find that galaxies with lower V$_{2.2}$/$\sigma$ ratio
(i.e. with greater pressure support) tend to be scattered to lower
values along the rotational velocity axis: this is consistent with
\citet{Tiley2016}, who found an evolution of the zero-point TFR at
$z=0.9$ when they select galaxies with V$_{80}$/$\sigma\geq3$ within
their sample (V$_{80}$ is the velocity observed at the radius which 
encloses the 80\% of the total H$\alpha$ intensity of the galaxy), although the complete 
sample is consistent with no evolution in the TFR zero-point.

This result suggests that the large scatter measured from the M$_{\star}$-TFR
at high-$z$ may be produced by galaxies which are supported by a combination
of rotational and disordered motions. If we do not take into account this effect then 
this could produce misleading conclusions. We note that galaxies which have greater 
rotational support within the SHiZELS survey tend to lie closer to the M$_{\star}$-TFR 
at $z=2$ derived by \citet{Swinbank2012a}, whilst galaxies with strong disordered motion support 
tend to be have a greater offset from this relationship. This trend perhaps implies that
galaxies may be moving onto the M$_{\star}$-TFR with time as the dynamics of the stars and gas 
in the central few kpc of the halos are yet to relax into a disk-like system.

\subsubsection{The M$_{\star}$--S$_{0.5}$ Relation}

The stellar mass TFR is found to be sensitive to which process dominates
the support of the galaxy. The scatter increases when galaxies with
pressure support equivalent to the rotational support
(V/$\sigma\sim1$) are included. Perhaps a more fundamental relation is
the M$_{\star}$-S$_{0.5}$ relationship \citep{Weiner2006a,Kassin2007}
which consider the support given by ordered and disordered motions
within the galaxy. In Fig.~\ref{fig:TF} we show the M$_{\star}$-S$_{0.5}$
relation for the SHiZELS survey using the inclination-corrected
speeds, colour-coded by V/$\sigma$ ratio. We also show the $z\sim0.2$
M$_{\star}$-S$_{0.5}$ relationship from \citet{Kassin2007} and the
best linear fit to the SHiZELS survey sample. We note that this relationship is 
fitted in the form $\log_{10}$(S$_{0.5})=$ a + b\,$\log_{10}($M$_{\star}\times10^{-10}$\,M$_\odot)$, 
where `a' is the interceptor. From Fig.~\ref{fig:TF} it can
be seen that our sample agrees with the $z\sim0.2$ M$_{\star}$-S$_{0.5}$ relationship
within 1$\sigma$ uncertainty: this is consistent with either no evolution of
interceptor or slope of the M$_{\star}$-S$_{0.5}$ relation with
redshift \citep{Kassin2007}. Despite the low number statistics, we do
not identify any dependency on V/$\sigma$ for this relation, 
contrasting with what was previously seen in the M$_{\star}$-TFR (Fig.~\ref{fig:TF}).

This relationship is consistent with a scenario of galaxy formation that begins with
matter assembling into a dark-matter halo with turbulent kinematics. The baryonic
component forms protodisks that are initially supported by a combination of rotational
and disordered motions. Finally, the material in these protodisks settles down,
unless they undergo major mergers \citep{Kassin2007}.

\subsection{Merger Fraction}
\label{sec:Merger_fraction}

\begin{figure}
 \centering
 \includegraphics[width=0.9\columnwidth]{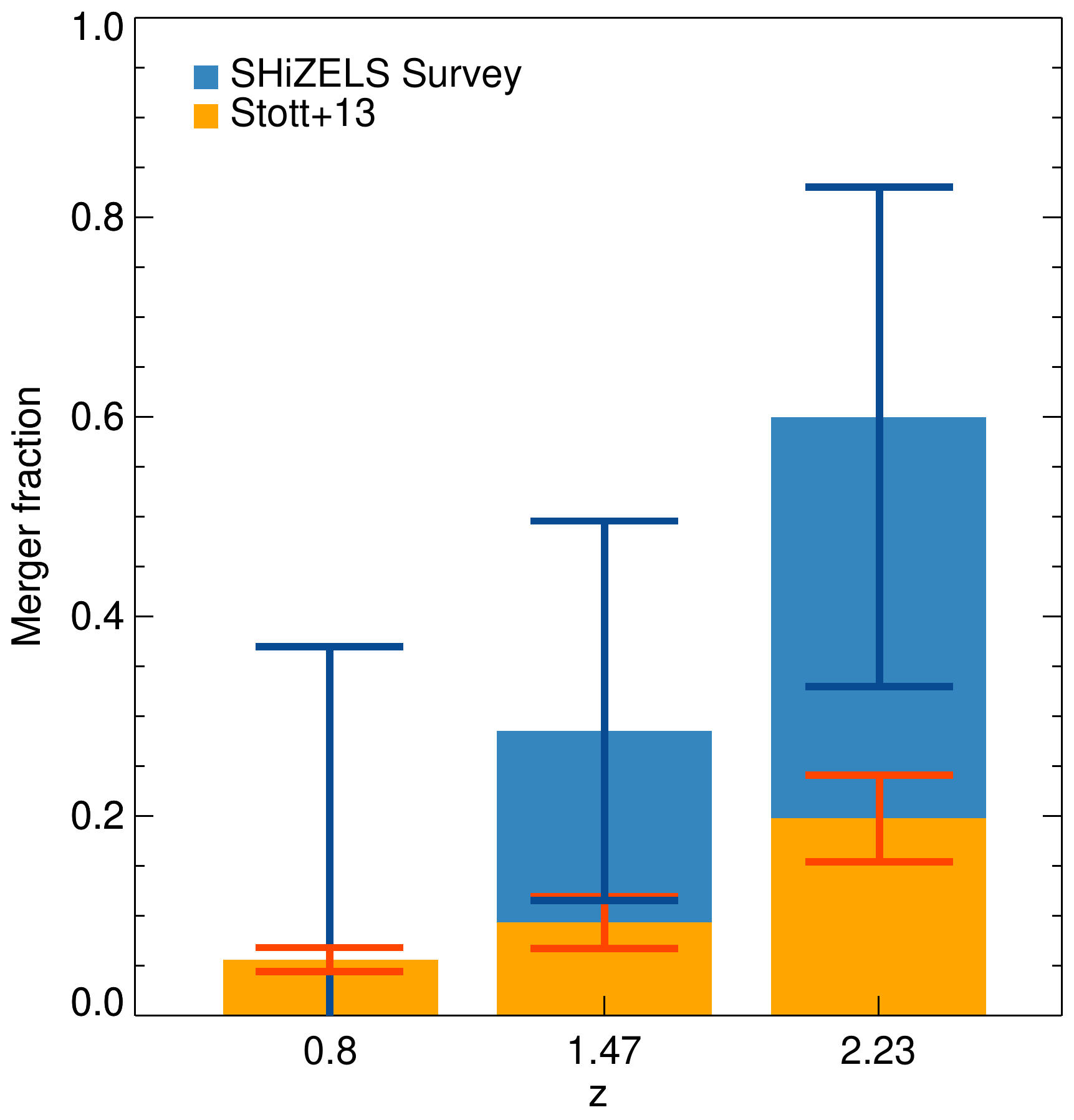}
 \caption{ \label{fig:MF}
	  In blue are represented the kinematically selected mayor merger fractions measured from the SHiZELS survey  
	  at $z=0.8-2.23$. The error bars for the SHiZELS survey are estimated by assuming binomial statistics 
	  (see \S\ref{sec:analysis}). The orange colour correspond to the data at similar sSFR ($\sim10^{-9}$\,yr$^{-1}$)
	  from \citet{Stott2013a}, who use the M$_{20}$ morphological criterion to classify mergers. We find higher 
	  merger fraction at each redshift slice.}
\end{figure}

To test whether it is galaxy mergers, secular processes or a combination of both 
that dominate and drive galaxy evolution at the peak era for star
formation, we need to measure the merger fraction ($f_{\rm merg}$) at
this epoch. From a theoretical perspective in the $\Lambda$CDM
paradigm, dark matter haloes merge hierarchically from the bottom up
\citep[e.g.][]{LaceyCole1993,Cole2000,Springel2005}. As baryonic
matter traces the underlying dark matter, we expect that galaxies merge
hierarchically as well.

\citet{Stott2013a} noted that the typical sSFR for galaxies increases with redshift 
within the HiZELS sample. They found greater merger fractions with increasing sSFR
suggesting that major mergers can lead to galaxies having unusually high sSFR. 
Although the targets within the SHiZELS survey have higher SFR at higher redshift, 
they also have higher stellar masses, maintaining the median sSFR per redshift slice 
roughly constant (see Fig.~\ref{fig:main_sequence}).

\begin{figure}
 \centering
 \includegraphics[width=1.0\columnwidth]{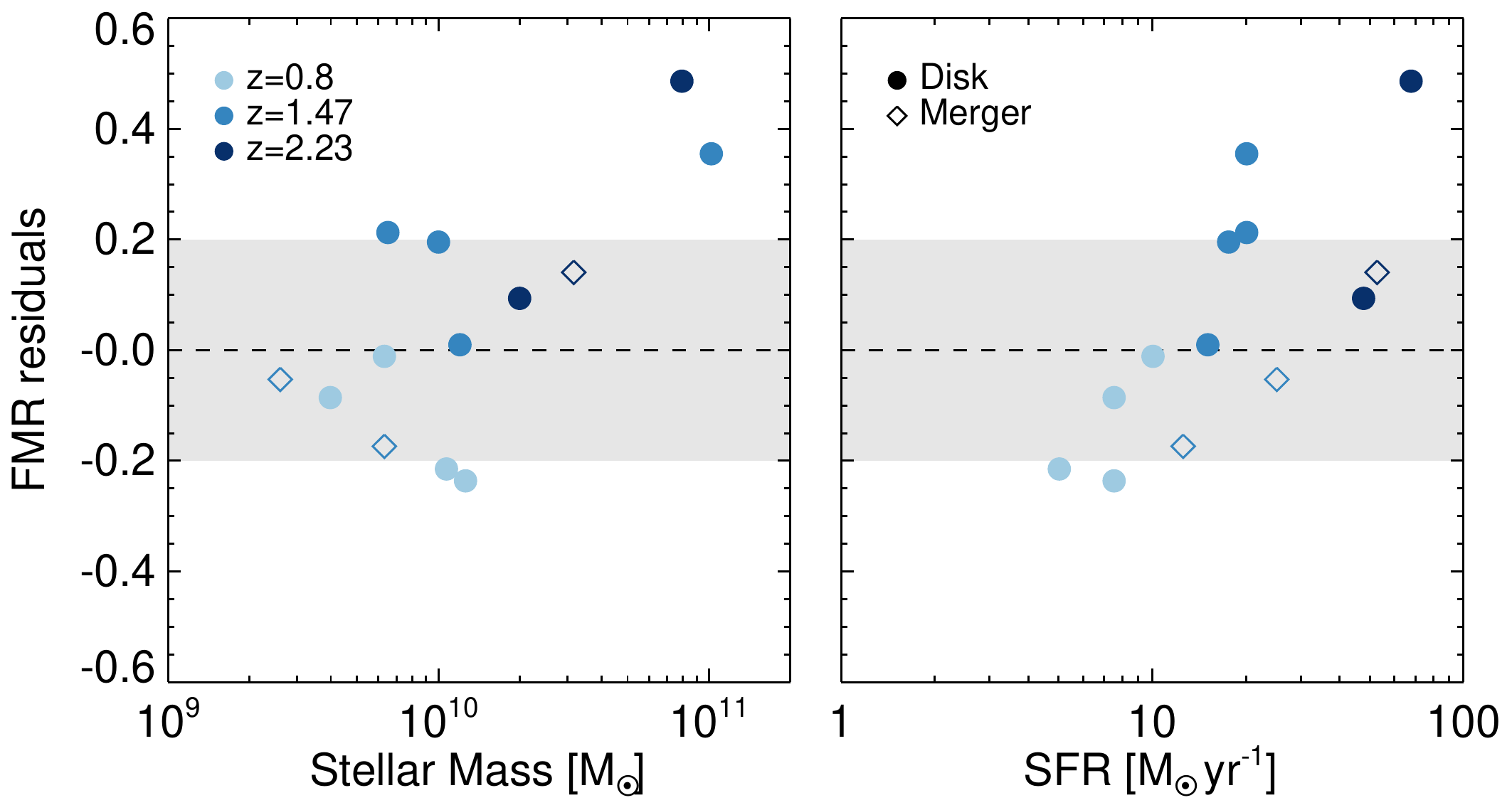}
 \caption{ \label{fig:fmr}
	 The metallicity residuals within the SHiZELS Survey calculated as the subtraction between the measured metallicities and 
	 the metallicities predicted by the FMR of \citet{Stott2013b}. We plot against the stellar mass (\textit{left}) and SFR (\textit{right}).
	 Negative values mean metallicities lower than expected by the FMR. The sky blue, blue and dark blue colours represent the 
	 sources at $z=0.8$, 1.47 and 2.23 respectively. Diamonds and circles show targets classified as Merger and Disks respectively from 
	 kinemetry criterion (see \S\ref{sec:analysis}). The shaded area corresponds to the scatter of the FMR of 0.2\,dex \citep{Stott2013b}. 
	 The measured residuals expected from the FMR is 0.23\,dex, which is consistent with the relationship. We find no trend 
	 between galaxy morphology and metallicity content within our sample.}
\end{figure}

\begin{figure*}
 \centering
 \includegraphics[width=2.0\columnwidth]{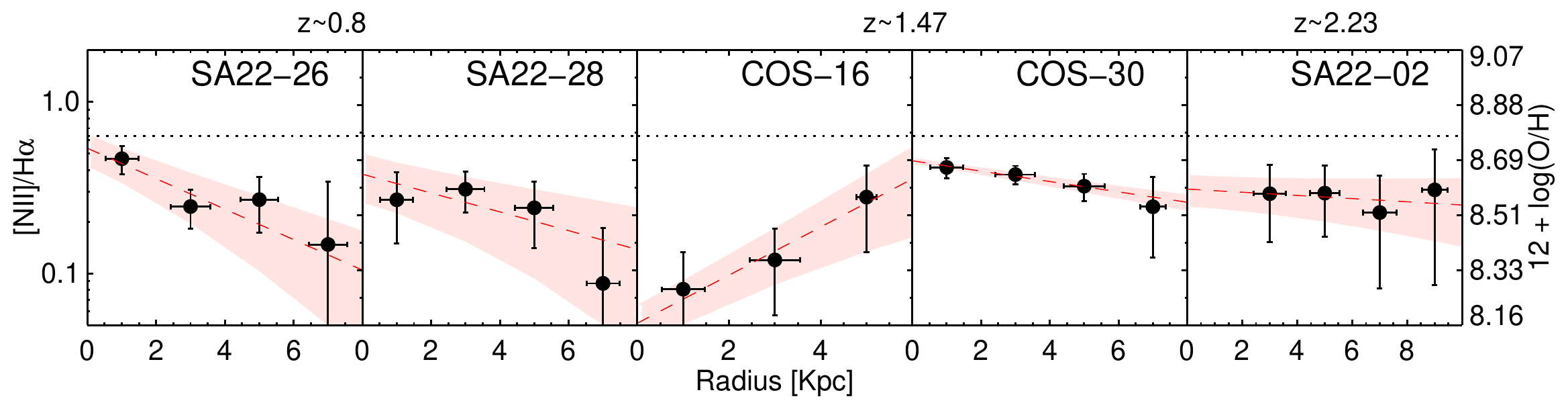}
 \caption{ \label{fig:metals} Metallicity gradients for five galaxies
   in our observed sample (Table~\ref{tab:table1}) from spatially resolved measurements as
   function of the physical radius derived from the best kinematical
   model. The red dashed line represents the best linear fit and the
   shaded region represents 1-dex uncertainties. The dotted 
   line shows the AGN limit of $\log_{10}$([N\,{\sc ii}]/H$\alpha$)$\approx-$0.2 at $z=0.8$ \citep{Kewley2013}. 
   This suggests the potential for a starting AGN activity within the central kpc in source SA22-26.
   Within our sample, SA22-02 has a gradient consistent with zero, SA22-26, SA22-28 and
   COS-30 have negative gradients and COS-16 source has positive metallicity gradient.}
\end{figure*}

We define `merger fraction' as the number of galaxies classified as
merger by the kinemetry criterion (see \S\ref{sec:analysis}) divided by
the total number of galaxies in the redshift slice (we do not consider
the unresolved galaxy classified as `compact' in this analysis).

In Fig.~\ref{fig:MF} we show the variation of the merger fraction as a
function redshift for the $z=0.8$, 1.47 and 2.23 redshift slices. From
the SHiZELS survey we find a merger fraction of 0.0$^{+0.4}$, 0.3$^{+0.2}_{-0.2}$
and 0.6$^{+0.2}_{-0.3}$ at $z=0.8$, 1.47 and 2.23, respectively. The error
in the merger fraction were estimated by assuming binomial statistics. 
These values are consistent with previous IFU surveys \citep[e.g.][]{Shapiro2008,Forster2009}.

We compare with \citet{Stott2013a} who used the M$_{20}$ morphological classification \citep{Lotz2004}
in the HiZELS sample but we only consider the merger fraction 
calculated for those galaxies with similar sSFR values ($\sim$10$^{-9}$\,yr$^{-1}$). 
From Fig.~\ref{fig:MF} we find a clear increase in merger fraction 
from $z=0.8$ to $z=2.23$ that seems to be stronger than the increase 
found by \citet{Stott2013a}. This suggests that 
methods based on surface brightness morphology classification may 
underestimate the number of major mergers at low sSFR and similar  
($0\farcs2-0\farcs7$) spatial resolution. We conclude that galaxy mergers may
have a dominant role in the evolution of `typical' star-forming galaxies at $z\geq1.5$,
but we caution regarding the low number of statistics.

\subsection{Metallicity Content}
\label{sec:chemical_abundances}

Measuring the internal enrichment and radial abundance gradients of
high redshift star-forming galaxies provides a tool for studying the
gas accretion and mass assembly process such as the gas exchange with the
intergalactic medium.  The [N\,{\sc ii}]/H$\alpha$ emission line
ratio can be used to determine the metallicity of high-z galaxies
using the conversion 12 + $\log_{10}$(O/H) = 8.9 + 0.57 $\log_{10}$([N\,{\sc
    ii}]/H$\alpha$) \citep{Pettini2004}. Our sample has a
median metallicity of 12 + $\log_{10}$(O/H) = 8.57$\pm$0.05, which is 
slightly below but still consistent with the solar value.  
Also our galaxies have metallicities consistent 
with previous studies that derive typical metallicities of 8.66$\pm$0.05 
and 8.58$\pm$0.07 for H$\alpha$-selected samples at $z\sim0.81$ 
and $z\sim0.84$--2.23 respectively \citep{Queyrel2012, Swinbank2012a, Sobral2013a}.

A relationship between mass, metallicity and SFR has been found in both 
the local and high-$z$ Universe \citep{Mannucci2010,Laralopez2010,Laralopez2013}
and measured by \citet{Stott2013b} for a sample drawn from the HiZELS survey at $z\sim1$. 
The shape of this `Fundamental Metallicity Relationship' (FMR) is, to first order, 
a manifestation of the positive correlation of the metallicity and stellar mass at 
fixed SFR and a negative correlation of the metallicity and SFR at fixed stellar mass.
The shape of the FMR can be explained as the result of the competing 
effects of chemical enrichment of the gas by the evolving stellar 
population, star-formation driven winds and the inflow of gas from 
the IGM. We test this relationship and its dependence on galaxy 
morphology using the SHiZELS Survey. In Fig.~\ref{fig:fmr} we show the 
difference between the measured metallicity of the SHiZELS Survey and 
the metallicity predicted by the FMR at $z=0.84-1.47$ \citep{Stott2013b}.
The measured scatter is consistent within 1$\sigma$ uncertainties from 
the FMR. On average, we find that the metallicity content in our 
star-forming galaxies is similar to galaxies of similar mass
and SFR at $z\sim0.1$ \citep{Stott2013b}. We note that this suggests no
evolution in the FMR up to $z\sim2.23$. 
Although most of our galaxies at $z\sim2.23$ are classified
as mergers, we find no trend between mergers and disks morphologies
within the residuals from this relationship.

In order to derive the chemical abundance gradients in our 
sample, we use the disk inclination and position angle (derived from
the best-fit dynamical model) to define $\sim$1\,kpc annuli centred at
the dynamical centre. Within each annulus we stack the spectrum
(considering emission line offsets in each annulus) to measure the
average [N\,{\sc ii}]/H$\alpha$ flux ratio by fitting a double gaussian 
profile with coupled gaussian widths. For a detection we enforce S/N\,$>5$ 
thresholds at each radius and a minimum of three radial 
detections per target. Then we fit a straight line as a function of 
galaxy radius in each case. From the eight galaxies with measured  
[N\,{\sc ii}]/H$\alpha$ flux ratio detected from their one-dimensional spectra 
(see \S~\ref{sec:ISM_properties}), five galaxies present reliable 
[N\,{\sc ii}]/H$\alpha$ gradients. We show this in Fig.~\ref{fig:metals} 
and the individual metallicity gradients values are reported in
Table~\ref{tab:table1}. We find a median of
$\Delta\log_{10}$(O/H)/$\Delta$R = $-$0.014$\pm$0.009\,dex\,kpc$^{-1}$,
i.e.\ a median gradient consistent with a negative gradient. In
comparison, \citet{Swinbank2012a} found a slightly steeper median
metallicity gradient ($\Delta\log_{10}$(O/H)/$\Delta$R = $-$0.027$\pm$0.006\,
dex\,kpc$^{-1}$) from their sample at similar redshift
range. Considering the combination of both studies (the full SHiZELS sample), we
find a median metallicity gradient of $\Delta\log_{10}$(O/H)/$\Delta$R =
$-$0.026$\pm$0.008\, dex\,kpc$^{-1}$. This result suggests 
that either low metallicity gas from the halo or IGM is accreted onto 
the outer disk, or metal enrichment is higher in the central region of the galaxy.

We do not have simultaneous access to [O\,{\sc iii}]5007, 
H$\beta$, H$\alpha$, and [N\,{\sc ii}]6583 or [S\,{\sc ii}]6717,6731 emission 
lines to distinguish any possible strong AGN contribution within our sample via a BPT diagram \citep{Baldwin1981}. 
Considering $\log_{10}$([N\,{\sc ii}]/H$\alpha$)$\approx-$0.2 flux ratio as a rough 
limit for identifying an AGN at $z\ge0.8$ \citep{Kewley2013}, only the central kpc 
of SA22-26 might be affected by low AGN contamination. This conclusion 
is also supported by the lack of broad recombination lines in the spectra.

Just one galaxy (COS-16) shows a positive metallicity gradient within
the SHiZELS survey -- a system which is also classified as a major
merger by the kinemetry criterion. From cosmological simulations, \citet{Tissera2016}
found that galaxies with positive metallicity gradients tend to exhibit
morphological perturbations and close neighbours. They suggest that those
galaxies have a high probability of interactions/mergers due to a high number 
of surrounding satellites. They analyse the evolution of a gas-rich equal-mass
merger and they found that both negative and positive metallicity gradients might
be produced during different stages of the merger evolution.

We note that we do not find any clear correlation between the asymmetries 
measured from the kinemetry coefficients (K$_{\rm V}$, K$_{\sigma}$ 
and K$_{\rm Tot}$) and the metallicity gradients observed within 
the SHiZELS survey. 

\subsection{Spatially Resolved Chemical Abundances}
\label{sec:Metallicity_Gradients}

\begin{figure*}
 \centering
 \includegraphics[width=2.0\columnwidth]{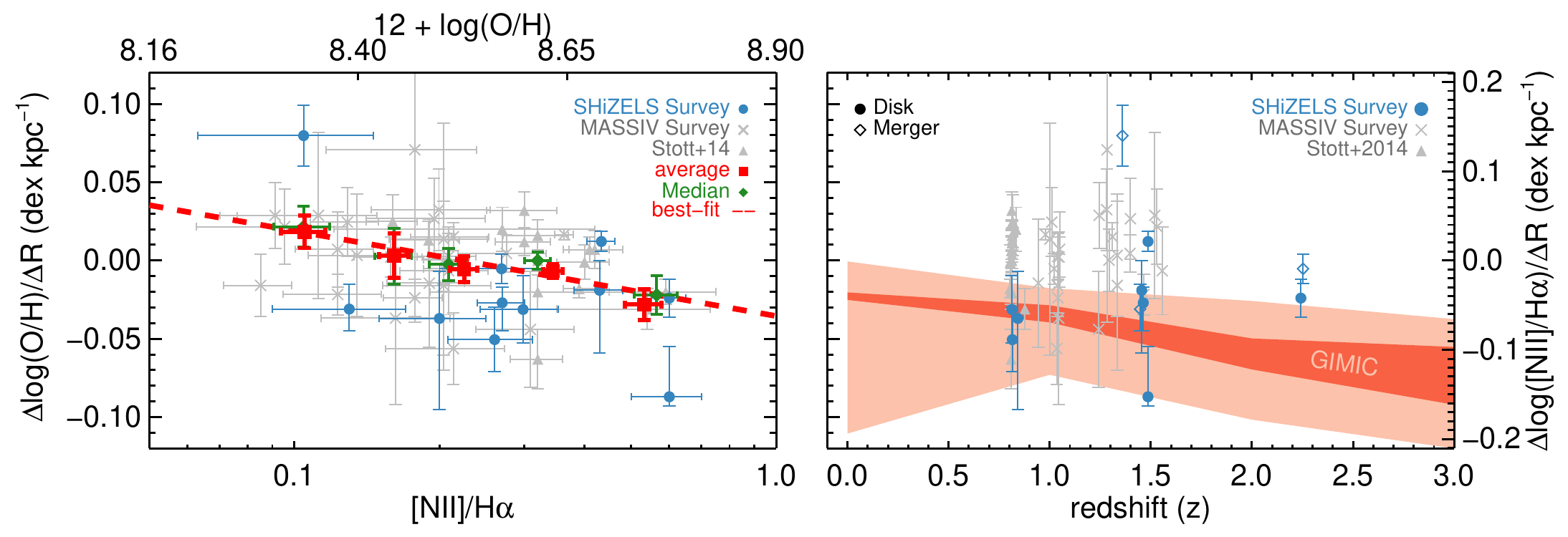}
 \caption{ \label{fig:grad_metals}
	  \textit{Left}: Metallicity gradient as a function of the [N\,{\sc ii}]/H$\alpha$ emission line ratio from the one-dimensional spectra
	  of each galaxy. The sky blue circles are individual galaxies within the SHiZELS Survey. The grey crosses denotes the sample 
	  from the MASSIV Survey \citep{Queyrel2012} corrected for a \citet{Pettini2004} metallicity calibration. The grey triangles denotes 
	  the sample from \citet{Stott2014}. The red squares and green diamonds show the average and median metallicity content and metallicity 
	  gradient per $\Delta\log_{10}$(O/H)=0.1 bin respectively. The error bars shows 1$\sigma$ uncertainties. Within the literature, we only 
	  consider [N\,{\sc ii}]/H$\alpha$ fluxes ratio above 1$\sigma$ detection threshold. The red-dashed line shows the best linear fit 
	  ($\Delta\log_{10}$(O/H)/$\Delta$R = a + b\,$\times\,\log_{10}$(O/H)) to the average values with a slope value of a$=-$0.10$\pm$0.04 and 
	  zero-point of b$=$0.89$\pm$0.30. The SHiZELS survey supports the anti-correlation such that metallicity gradients tend to be negative 
	  in metal-rich galaxies and positive in low-metallicity galaxies \citep[see also][]{Queyrel2012}. 
	  \textit{Right}:  Metallicity gradient as a function of redshift. We also show the theoretical 
	  evolution of the metallicity gradient with redshift from the GIMIC simulation \citep{Crain2009, McCarthy2012}. The shaded-red 
	  area shows the range of metallicity gradients for all disk galaxies in the simulation in the mass range 
	  9.5\,$<$\,$\log_{10}$(M/M$_\odot$)\,$<$\,11.5, whilst the shaded light-red denotes the 1$\sigma$ scatter at each epoch. 
	  Diamonds and circles show targets classified as Merger and Disk respectively from kinemetry criterion (see \S\ref{sec:analysis}).
	  The metallicity gradients measured from the SHiZELS survey do not support the evolution predicted from the GIMIC simulation 
	  from $z=0$ to $z=1.47$.}
\end{figure*}

A tentative anti-correlation between the metallicity gradient and the
global integrated metallicity was previously hinted at the `Mass Assemby Survey with SINFONI in VVDS'
(MASSIV), using star-forming galaxies at $z\sim1.2$ \citep{Queyrel2012}.
They suggest that metallicity gradients are more frequently negative
in metal-rich galaxies and more frequently positive in low-metallicity
galaxies. In Fig.~\ref{fig:grad_metals} we show the anti-correlation
suggested by \citet{Queyrel2012}, adding our SHiZELS galaxies and the sample
observed by \citet{Stott2014}. We calculate the average and median values per 
$\Delta\log_{10}=0.1$ bin. Then we calculate the best linear fit to the average 
and median values. The slope is 2.5$\sigma$ from being flat (2.0$\sigma$ by fitting the 
median values), but due to beam smearing effects and inclination angles the measured 
metallicity gradients are likely to be underestimated, especially those of \citet{Stott2014} 
and MASSIV samples which were measured on seeing-limited conditions ($\sim0\farcs7$). 
\citet{Stott2014} estimated that their observed metallicity gradients reflects only 
$\sim$70\% of the true values. This suggests that if we alleviate inclination angle effects 
with higher resolution IFU observations, these results will not change. Our results 
support the previous suggestion by \citet{Queyrel2012}.

We note that positive metallicity gradients could be explained by the infall of
metal-poor gas from the IGM into the centre of the galaxy, diluting
the gas and lowering its metallicity in the central regions. 
If the funnelling of metal-poor gas into the centres of galaxies 
is triggered by galaxy mergers, then from the merger fraction estimated for the SHiZELS Survey (Fig.~\ref{fig:MF}), 
we should expect to find more systems with positive/flat metallicity gradients at $z\sim2$.

Taking into account the metallicity gradients measured from
the SHiZELS Survey (Fig.~\ref{fig:grad_metals}), we compare its evolution with 
redshift with the prediction from the `Galaxies-Intergalactic Medium Interaction 
Calculation' simulation (GIMIC; \citealt{Crain2009,McCarthy2012,Swinbank2012a}), 
where the metallicity gradient evolution within disk galaxies is a consequence of a decrement of gas inflow 
rates from $z=2$ to $z=0$ and redistribution of gas within the galaxy disk.
The observed metallicity gradients for disk galaxies within the SHiZELS survey do not support the trend 
predicted by the GIMIC simulation between $z=0-2$. Nevertheless, we note that a much larger sample of disky galaxies at $z=2$
are needed to further test this.\\

\section{Conclusions}

We present new AO-aided SINFONI IFU observations of spatially resolved
H$\alpha$ kinematics of eleven mass-selected (M$_{\star}$ =
10$^{9.5-10.5}$\,M$_\odot$) `typical' star-forming galaxies from
the wide-field narrow-band HiZELS survey in three redshift slices,
$z=0.8$, 1.47 and 2.23. All galaxies lie within $<30''$ of bright
(\textit{R}<15.0) stars enabling natural guide star AO assisted
observations. Modelling the H$\alpha$ dynamics along the major kinematic 
axis of our galaxies, we derive a median dynamical-to-dispersion support of 
V$_{\rm max}$\,sin(i)/$\sigma$ = 1.6$\pm$0.3 (with a range of 1.1--3.8). We
classify the galaxies using a kinemetry analysis \citep{Shapiro2008}
finding six disk-like galaxies and four mergers. One galaxy is
unresolved. These new observations are combined with a previous similar
study (nine galaxies taken from \citealt{Swinbank2012a}) to create
a homogeneously selected sample of star-forming galaxies with dynamical 
characterisation at $\sim$\,kpc scales near the peak of the cosmic 
star-formation rate density.

We find a tentative increase of the merger fraction as a function of redshift 
($f_{\rm merg}\sim0.0^{+0.4}$, 0.3$^{+0.2}_{-0.2}$ and 0.6$^{+0.2}_{-0.3}$ at $z=0.8$, 
1.47, 2.23, respectively). Nevertheless, our results are consistent with previous IFU 
surveys \citep{Shapiro2008,Forster2009}, although we find higher merger fractions at a given sSFR 
in comparison to previous analyses by \citet{Stott2013a} who used a morphological 
classification from {\it HST} near-IR imaging.

We combine our observations with previous studies of intermediate and
high-redshift galaxies \citep{Swinbank2006,Cresci2009,Jones2010b,Gnerucci2011b,
  Miller2011,Miller2012,Swinbank2012a} to investigate the stellar mass
Tully-Fisher relation. We find that the scatter of this relation is
affected by the galaxy pressure support (V/$\sigma$) -- a result which
is consistent with \citet{Tiley2016}.  On the other hand, we also
investigate the M$_{\star}$-S$_{0.5}$ \citep{Kassin2007} relation within 
the SHiZELS survey at $z=0.8-2.23$. The kinematic measure S$_{0.5} = (0.5$V$^2 + \sigma^2)^{0.5}$ 
consider support by both rotational motions and dispersion arising from 
disordered motions \citep{Weiner2006a}. Our results are consistent 
(within 1$\sigma$) with the M$_{\star}$-S$_{0.5}$ relationship found at 
$z\sim0.2$, suggesting little or no evolution of this relation as 
function of redshift.

We measured the residuals from the `Fundamental metallicity relation' 
\citep{Stott2013b} at $z=0.84-1.47$, finding that the scatter is 
consistent with measurements errors, suggesting no variation in the
FMR up to $z=2.23$.

We measure metallicity gradients ($\Delta\log_{10}$(O/H)/$\Delta$R) using
the [N\,{\sc ii}]/H$\alpha$ ratio for 3, 7 and 2 galaxies at $z=$0.8, 1.47 and 2.23
within the SHiZELS Survey. These metallicity gradients ranges between $-$0.087 and 0.08\,dex\,kpc$^{-1}$,
with a median metallicity gradient of $\Delta\log_{10}$(O/H)/$\Delta$R = $-$0.027$\pm$0.008\,dex\,kpc$^{-1}$.
The evolution of metallicity gradients as a function of redshift in our modest sample
at $z\leq2$ do not exhibit any clear redshift trend such as the predicted by the GIMIC simulation
for galaxy disks, where gas inflow rate decreases with decreasing redshift progressively.
However, larger samples at $z=2$ are needed to further test this. 

We show that the metallicity gradient and global metallicity 
content are consistent with the anti-correlation suggested by
\citet{Queyrel2012}. This can be explained by the scenario in which
infall of metal-poor gas from the IGM into the central part of the
galaxy drives the positive gradients.

\section*{Acknowledgments}

We thank to the anonymous referee for his/her careful read of the manuscript and helpful comments and suggestions. 
This work is based on observations collected at the European Organization for Astronomical Research in
the Southern Hemisphere under ESO programme ID 092.A-0090(A).
This research was supported by CONICYT Chile (CONICYT-PCHA/Doctorado-Nacional/2014-21140483)
A.M.S. acknowledges an STFC Advanced fellowship, support from STFC(ST/L00075X/1) and the 
Leverhume foundation. I.R.S. acknowledges support from STFC(ST/L00075X/1),
the ERC Advanced Grant DUSTYGAL(321334) and a Royal Society/Wolfson Merit award.
D.S. acknowledges financial support from the Netherlands Organisation for Scientific research
(NWO) through a Veni fellowship. PNB is grateful for support from STFC via grant ST/M001229/1.








\appendix
\section{Serendipitous detection}
\label{sec:hizels} %

\begin{figure*}
 \centering
 \includegraphics[width=0.7\columnwidth]{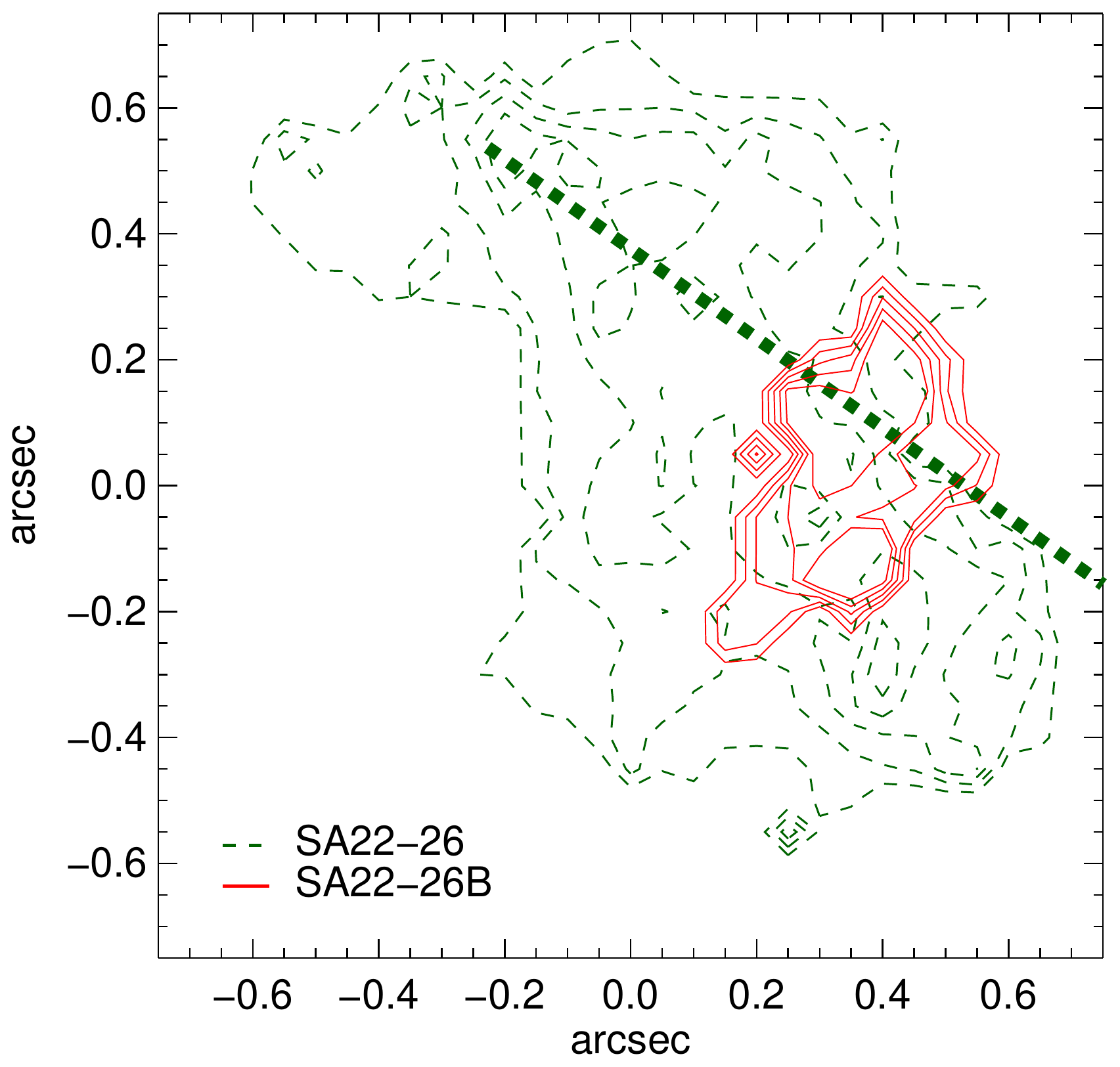}
 \includegraphics[width=1.0\columnwidth]{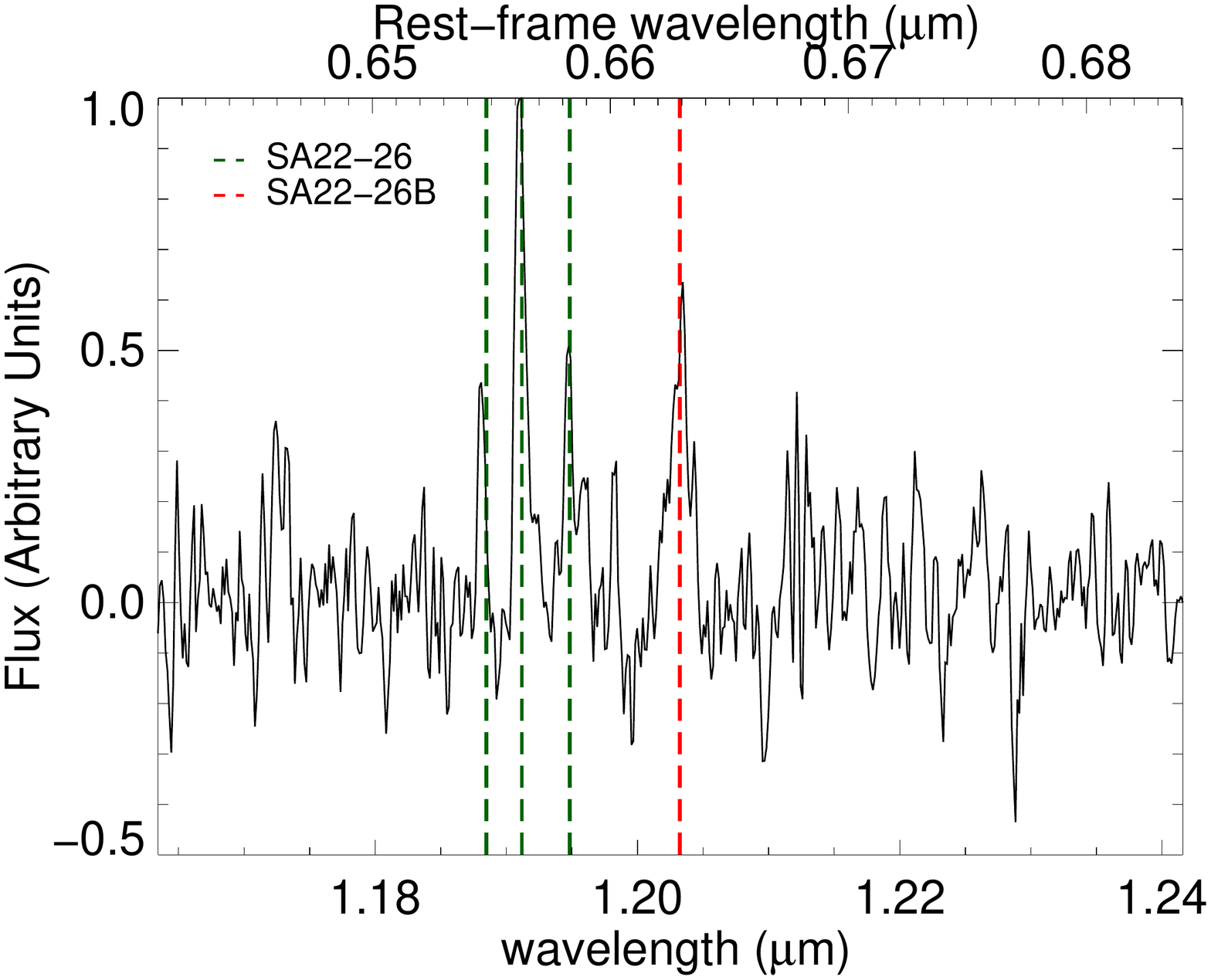}\\
 \includegraphics[width=0.343\columnwidth]{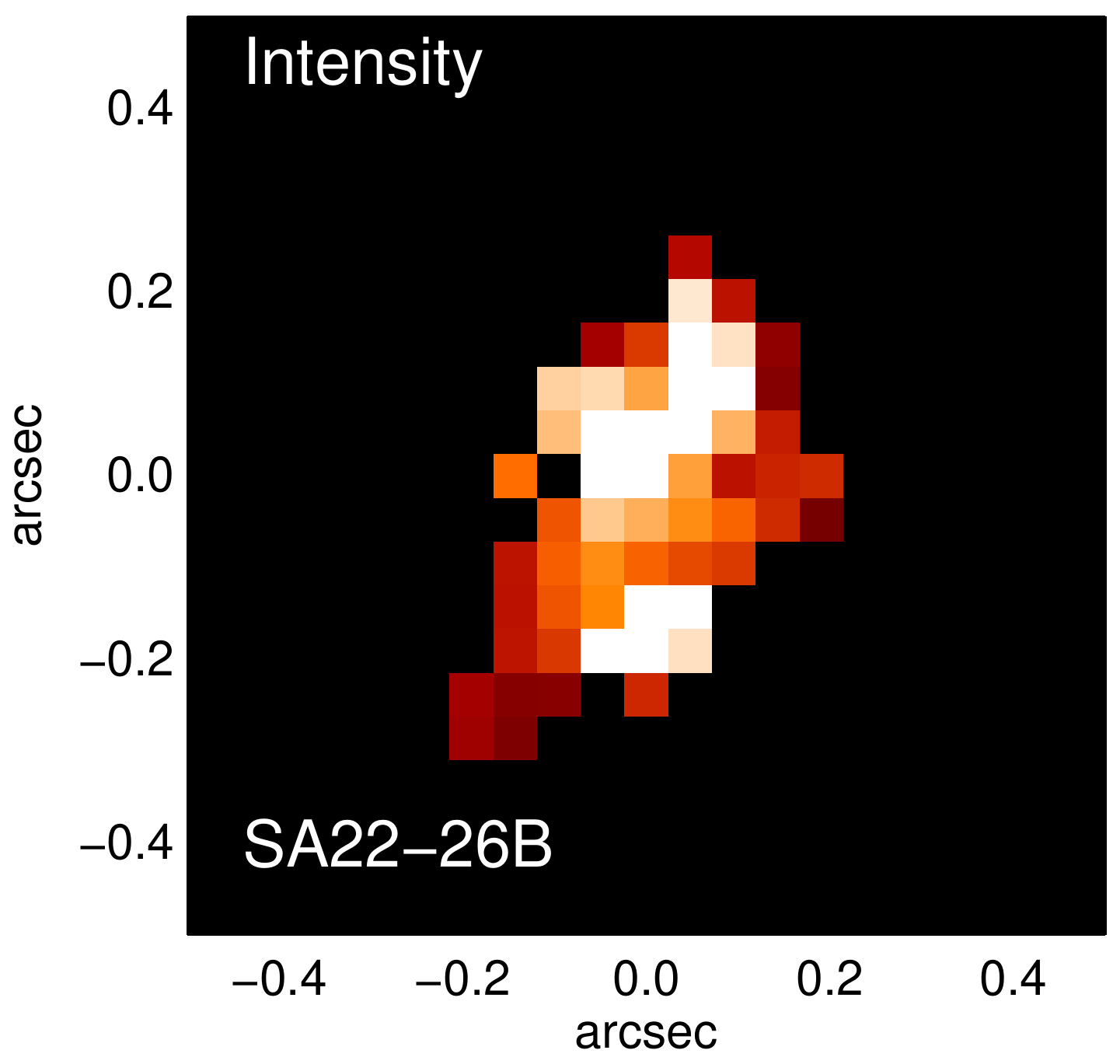}
 \includegraphics[width=0.32\columnwidth]{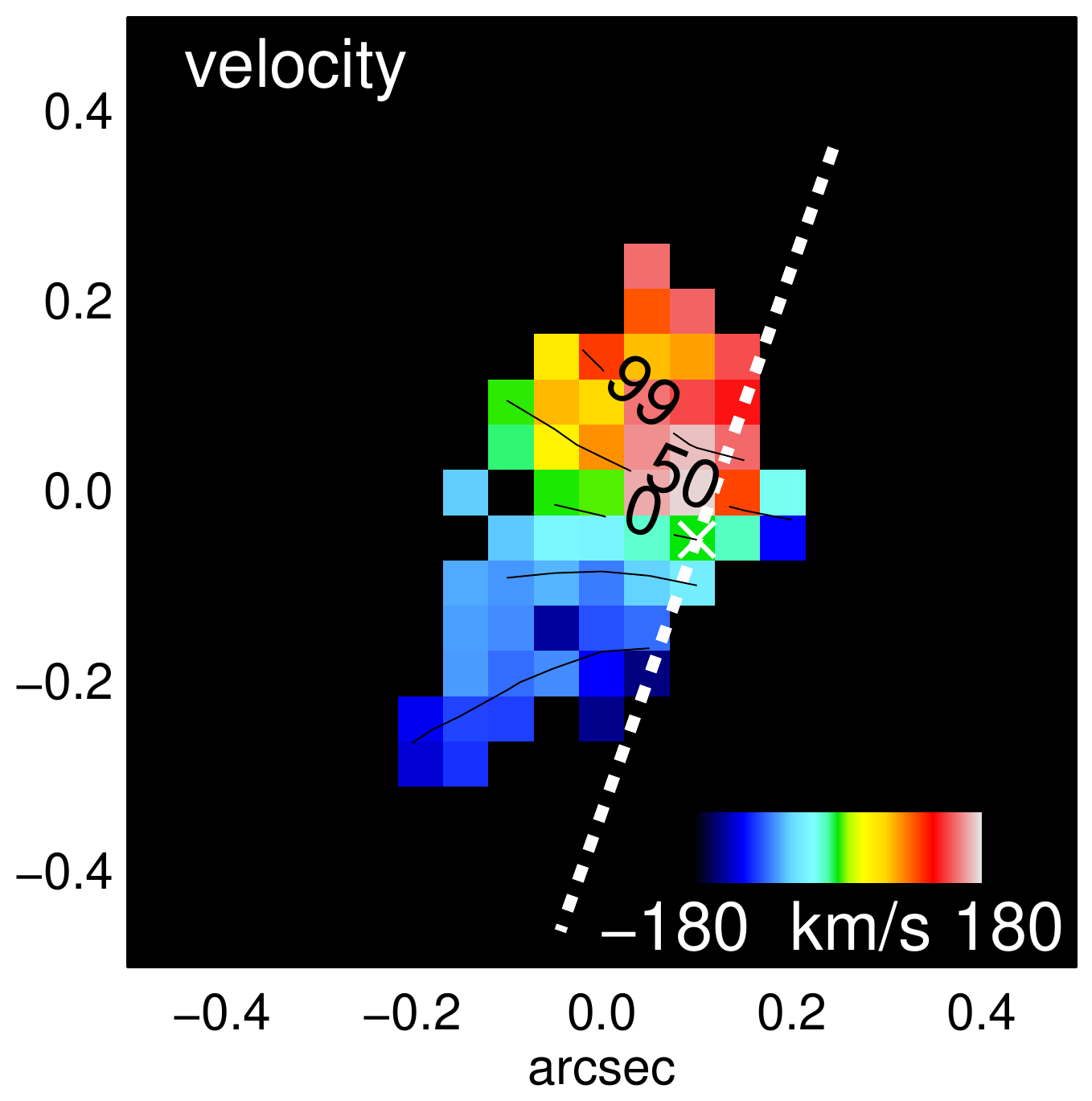}
 \includegraphics[width=0.32\columnwidth]{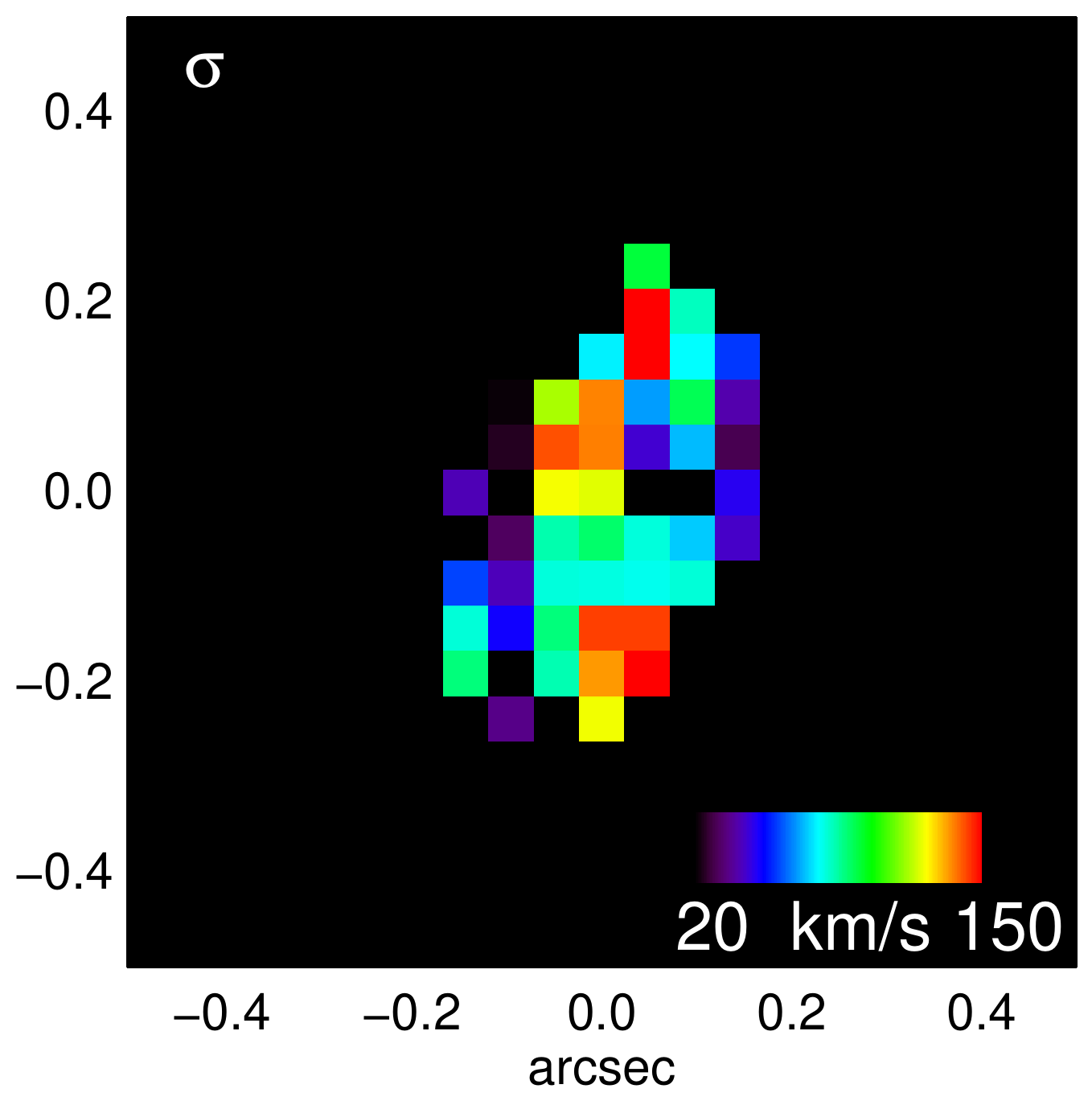}
 \includegraphics[width=0.32\columnwidth]{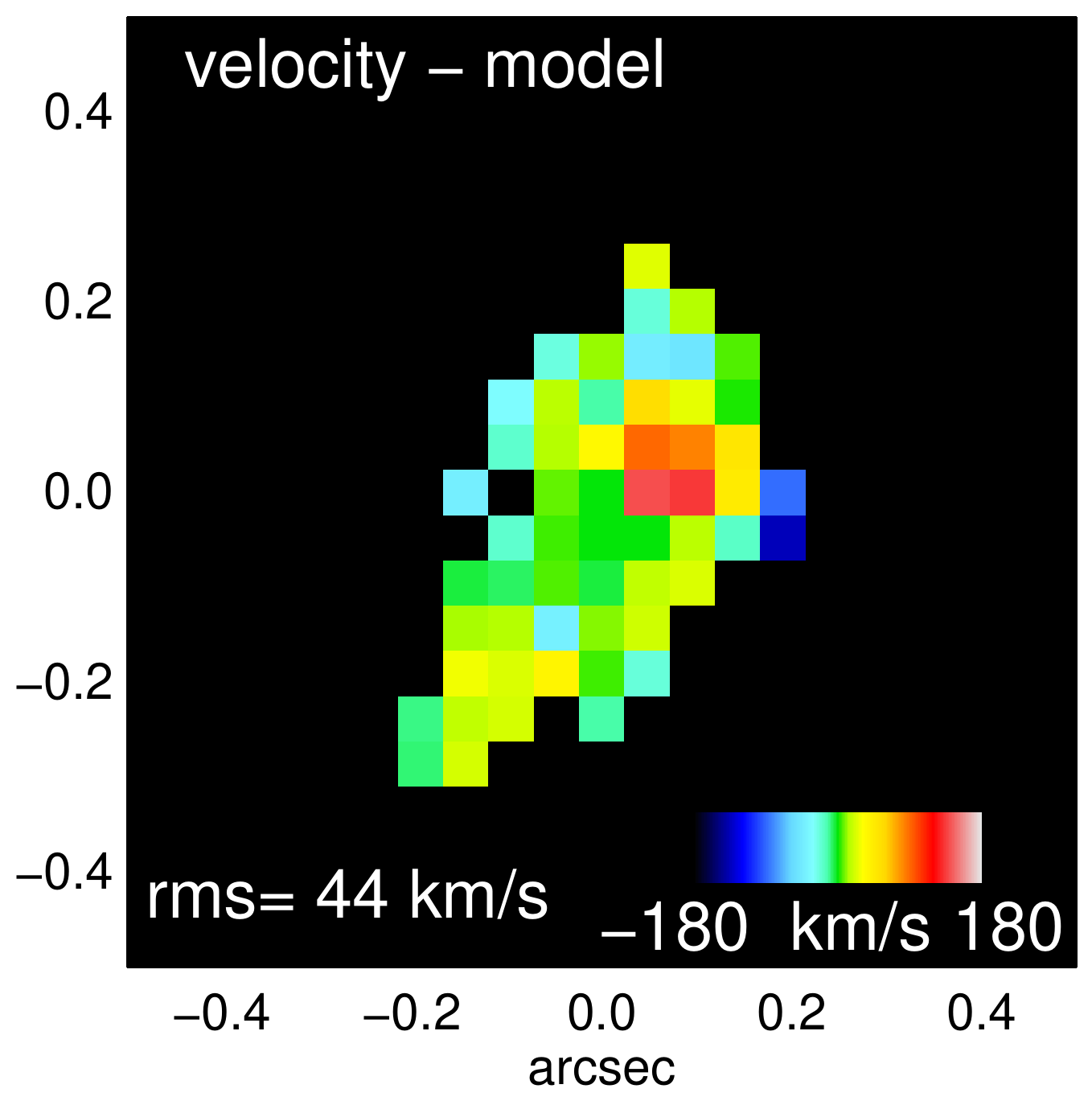}
 \includegraphics[width=0.345\columnwidth]{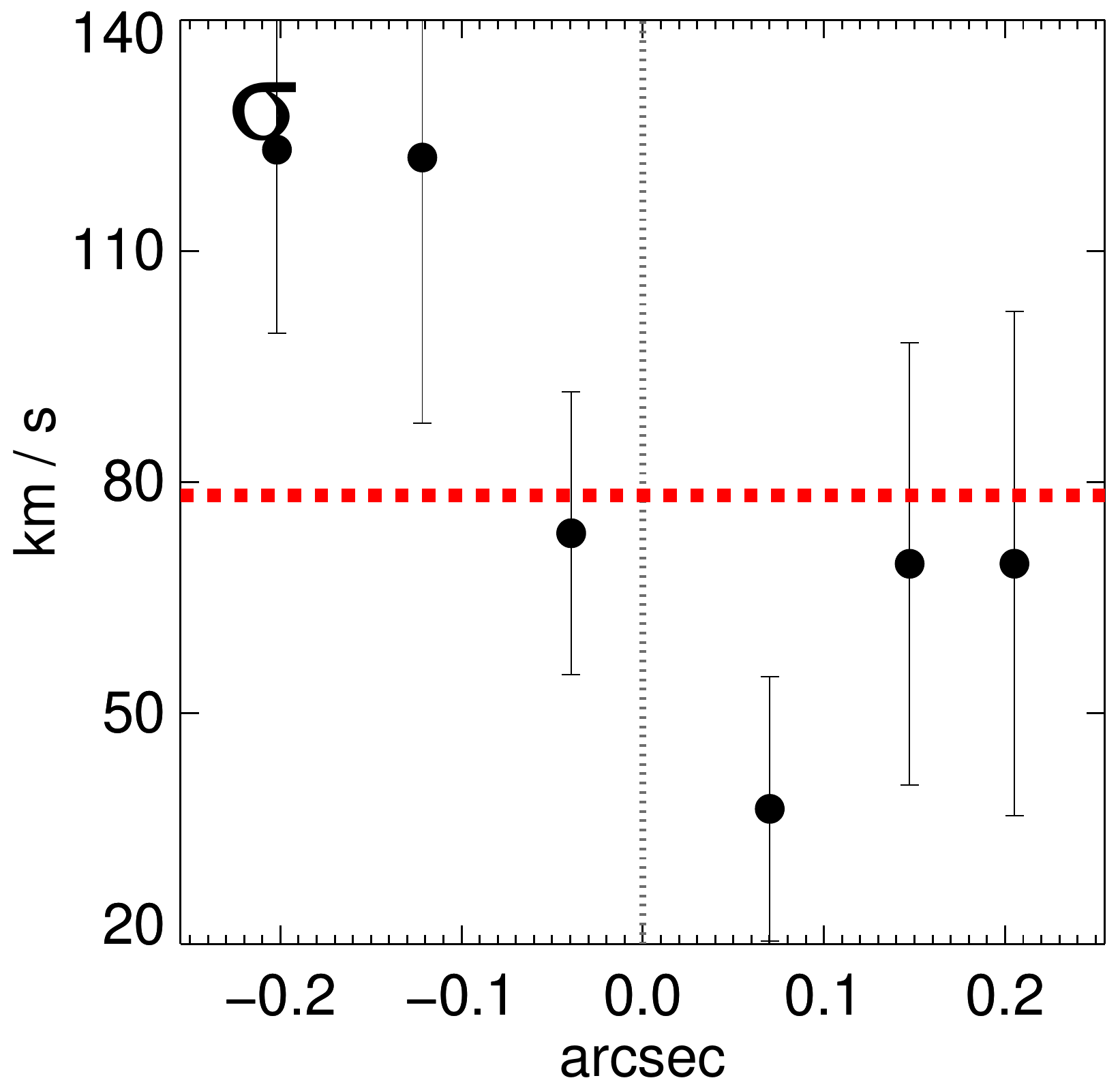}
 \includegraphics[width=0.373\columnwidth]{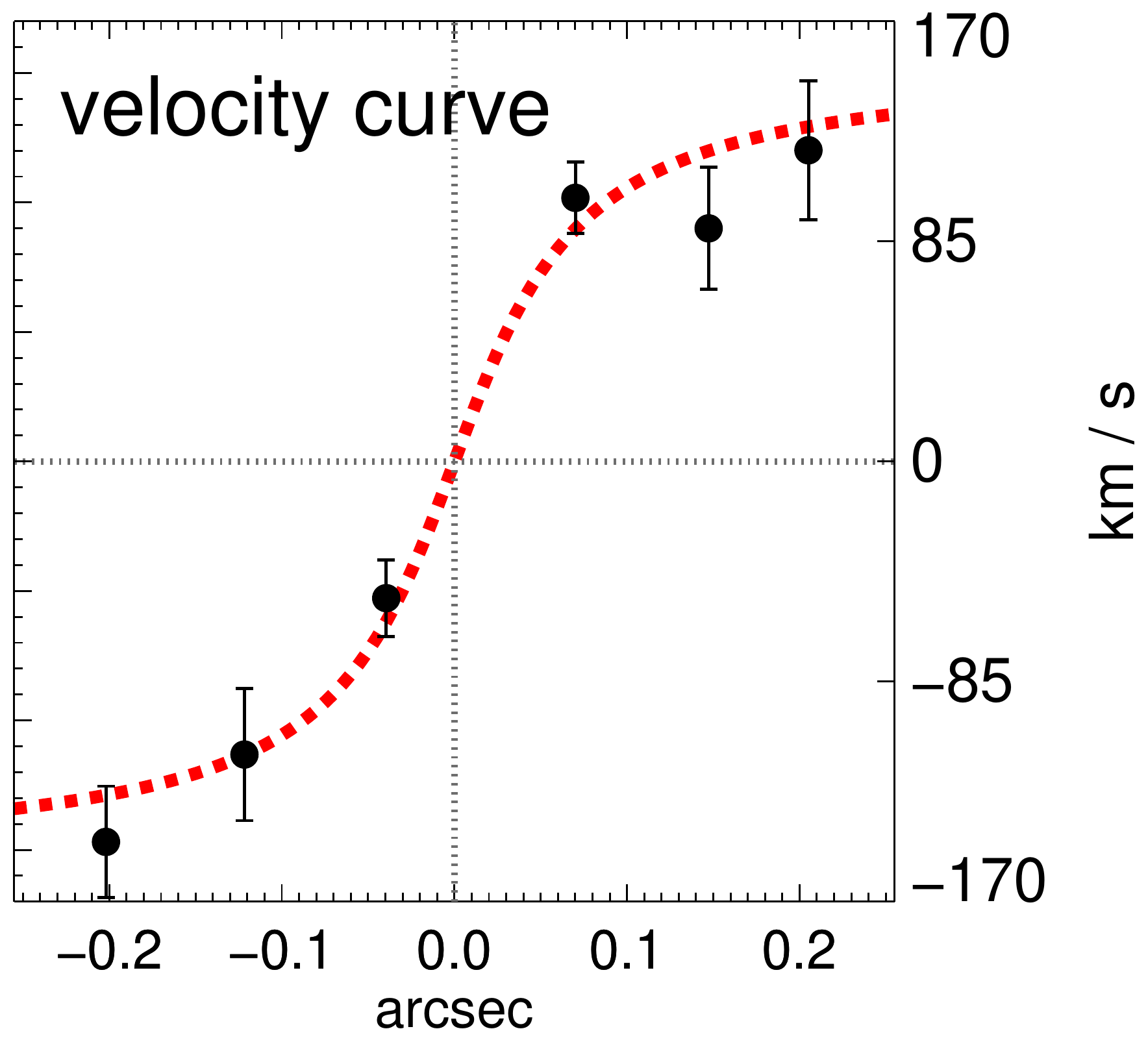}\\
 \caption{ \label{fig:Apendix_fig}
	  \textit{Top Left}: Intensity contours of our serendipitous detection (red) and SA22-26 (green-dashed) sources within the 
	  SINFONI data-cube. The dashed line shows the position angle derived for SA22-26 source. There is a clear spatially overlap between both 
	  emissions. \textit{Top Right}: Spatially integrated one-dimensional spectra of SA22-26B. The emission is integrated within the red contour 
	  showed in top-left figure. The red-dashed line shows the emission line detected at $\lambda\sim$1.203 $\mu$m. The green-dashed lines 
	  show the H$\alpha$, [N\,{\sc ii}]$\lambda\lambda$6583,6548 emission lines for the SA22-26 source extracted from the same area. 
	  \textit{Bottom}:  Intensity, velocity, line-of-sight velocity dispersion ($\sigma$), residual field, one-dimensional velocity dispersion 
	  profile and one-dimensional velocity profile (as in Fig.~\ref{fig:maps}) for our serendipitous detection.
	  The spatial scale is showed in arcseconds due to the uncertain redshift determination.}
\end{figure*}

Within the SA22-26 data-cube we found an unexpected emission line at $\lambda\sim1.203$\,$\mu$m. This emission line does not coincide with any 
expected emission line emitted from the SA22-26 source as it corresponds to $\lambda_{\rm rest}\sim0.663\,\mu$m in this galaxy's restframe. 
This unexpected emission line overlaps spatially with part of the H$\alpha$ emission from SA22-26 source (Fig.~\ref{fig:Apendix_fig}-\textit{top}). 
Furthermore, the kinematical behavior of this emission line region shows a position angle (Fig.~\ref{fig:Apendix_fig}-\textit{bottom}) which differs
from the position angle derived for the SA22-26 source (Fig.~\ref{fig:maps}). This suggest that this serendipitous emission does not comes from 
SA22-26. We did not find any other emission lines associated with this galaxy which could be used for determining the redshift of this 
possible new source. We call this source as SA22-26B. If we assume that this emission line is also a redshifted H$\alpha$ emission line, then 
SA22-26B is offset redwards by $\sim5500$\,km\,s$^{-1}$ (in the line-of-sight direction, $z\sim0.833$) from the SA22-26 source. The modest redshift
difference if the emission line is H$\alpha$ would be consistent with no lensing. We note that the emission from SA22-26 shows no sign of 
extinction by SA22-26B supporting the fact that this galaxy is background.

\bsp	
\label{lastpage}
\end{document}